\documentclass{article}

\usepackage{arxiv}

\usepackage[utf8]{inputenc} 
\usepackage[T1]{fontenc}    
\usepackage{hyperref}       
\usepackage{url}            
\usepackage{booktabs}       
\usepackage{amsfonts}       
\usepackage{nicefrac}       
\usepackage{microtype}      
\usepackage{graphicx}
\usepackage[numbers,sort]{natbib} 

\usepackage{amsmath,amssymb,amsfonts}

\newcommand{\vtheta}{\ensuremath{\boldsymbol{\theta}}}

\newcommand{\vx}{\ensuremath{\mathbf{x}}}
\newcommand{\vu}{\ensuremath{\mathbf{u}}}

\newcommand{\vy}{\ensuremath{\mathbf{y}}}

\newcommand{\vz}{\ensuremath{\mathbf{z}}}

\title{%
Systematic Evaluation of Generative Machine Learning Capability to Simulate Distributions of Observables at the Large Hadron Collider
}

\date{}

\usepackage{authblk}

\author[1,2]{%
Jan Gavranovi\v c\thanks{\href{mailto:jan.gavranovic@ijs.si}{\texttt{jan.gavranovic@ijs.si}}}%
}
\author[1,2]{%
Borut Paul Ker\v sevan\thanks{\href{mailto:borut.kersevan@ijs.si}{\texttt{borut.kersevan@ijs.si}}}%
}
\affil[1]{Jo\v zef Stefan Institute, Jamova 39, Ljubljana, 1000, Slovenia}
\affil[2]{Faculty of Mathematics and Physics, University of Ljubljana, Jadranska 19, Ljubljana, 1000, Slovenia}

\renewcommand{\undertitle}{}

\begin{document}
\maketitle

\begin{abstract}
    Monte Carlo simulations are a crucial component when analysing the Standard Model and New physics processes at the
    Large Hadron Collider. This paper aims to explore the performance of generative models for complementing the statistics of classical Monte Carlo
    simulations in the final stage of data analysis by generating additional synthetic data that follows the same kinematic distributions
    for a limited set of analysis-specific observables to a high precision. Several deep generative models are adapted
    for this task and their performance is systematically evaluated using a well-known benchmark sample containing the Higgs
    boson production beyond the Standard Model and the corresponding irreducible background. The paper evaluates the 
    autoregressive models and normalizing flows and the applicability of these models
    using different model configurations is investigated. The best performing model is chosen for a further evaluation using a
    set of statistical procedures and a simplified physics analysis. By implementing and performing a series of statistical
    tests and evaluations we show that a machine-learning-based generative procedure can be used to generate synthetic data
    that matches the original samples closely enough and that it can therefore be incorporated in the final stage of a
    physics analysis with some given systematic uncertainty.
\end{abstract}



\section{Introduction}

In data analysis of new physics searches at the Large Hadron Collider (LHC) \cite{Evans:2008zzb} experiments, the use of Monte Carlo (MC)
simulation is essential to accurately describe the kinematics of the known \emph{background} processes in order to determine
an eventual discrepancy with the measured data and attribute such a deviation to a certain new physics \emph{signal} hypothesis.
Describing LHC data precisely through MC simulations involves several key steps (see e.g. \cite{atlas-sim-infra2010}). First, particles are
generated based on a calculated differential cross-section, a process referred to as ``event generation''.
This generation is carried out using MC generator tools like Pythia \cite{Sjostrand:2014zea} or Sherpa \cite{Hoeche:2009rj}. Next, these particles are simulated as
they pass through the detector volume and interact with the detector's materials. This stage is typically performed using
the Geant4 toolkit \cite{Agostinelli:2002hh} and is known as ``detector simulation''. During this step, the model also incorporates various factors
that accurately represent the collision environment, such as the response to multiple simultaneous proton collisions in
the LHC, often referred to as ``pile-up'' events. These interactions in the detector are converted into simulated electronic
response of the detector electronics (``digitization'' step), and at this stage the MC simulated data matches the real
recorded collision data as closely as possible. Subsequently, the response of the detector trigger system is modelled,
and then the simulated data undergoes the same complex reconstruction procedure as the measured data, involving the reconstruction of basic analysis
objects, such as electrons or jets, followed by reconstruction of more involved event kinematics.

Eventually, the final data analysis optimizes the data selection (``filtering'') procedure to maximize measurement accuracy and new physics
discovery potential (statistical significance) and determines the potential presence of new physics using statistical
tests on the final data selection (for a nice overview see Ref. \cite{cowan2011asymptotic}). Both the filtering and final statistical analysis are based on comparing the MC signal and
background predictions with the real data by using several $\mathcal{O}(10)$ kinematic variables. Obviously, the
statistics of the simulated events limits the prediction accuracy of the background and signal events - ideally, the
number of simulated events would exceed the data predictions by several orders of magnitude to minimize the impact of
finite MC statistics on the systematic uncertainty of the final measurement. While the simulated background events can
typically be shared between several analyses at a LHC experiment, the simulation of a chosen signal process, and the subsequent choice of the relevant
kinematic observables is very analysis-specific.

\subsection{High-Luminosity LHC and the need for more computing power}

A fundamental problem with the standard MC simulation procedure described above is the need for large computational resources,
which restricts the physics analysis discovery process due to the speed and high cost of CPU and disk space needed to simulate
and store billions of MC events describing high energy particle collisions \cite{run2-computing-tdr2014}. The full procedure of MC event
simulation of one of the detectors at the LHC may take several minutes per event and produce $\mathcal{O}(1~\text{MB/event})$  of data of
which only $\mathcal{O}(1~\text{kB/event})$  of high-level features, i.e. reconstructed kinematic observables, is used in a
specific final analysis of a physics process of interest.

With the current Run 3 and, in the future, the High-Luminosity LHC (HL-LHC), it is expected that the experiments will
require even more computing power for MC simulation to match the precision requirements of physics analysis, which increase
proportionally to the size of the collected datasets that will come with the increase in integrated luminosity. Taken together, the data and MC
requirements of the HL-LHC physics programme are formidable, and if the computing costs are to be kept within feasible
levels, very significant improvements will need to be made both in terms of computing and storage, as stated in
Ref. \cite{atlas2020cdr} for the ATLAS detector \cite{PERF-2007-01}. The majority of resources will indeed be needed to
produce simulated MC events in the simulation chain from physics modelling (event generation) to detector simulation, reconstruction, and production of analysis formats.

\subsection{Machine learning for fast event generation}\label{sec: ml_gen}

To address the limitation posed by insufficient MC statistics in constraining the potential of physics analysis,
a promising approach is the utilization of Machine Learning (ML), specifically focusing on deep generative modelling.
For this purpose, a ``fast simulation'' approach is being investigated (see e.g.  \cite{aad2022atlfast3}), which aims
to replace the computationally intensive parts of the MC simulation chain with faster ML solutions.
The general idea is to create large numbers of events at a limited computing cost using a learning algorithm that was trained on a
comparatively smaller set of MC-simulated events. Several different approaches exist, trying to replace different MC stages
in simulation chain with faster ML solutions. Commonly used ML approaches include Generative Adversarial Networks (GAN),
Variational Autoencoders (VAE), normalizing flows, and diffusion models \cite{hashemi2023deep}.

he study presented in this paper focuses mainly on the autoregressive models and normalizing flows \cite{papamakarios2021normalizing}. The data
space in the final stage of a physics analysis is relatively low-dimensional, typically using $\mathcal{O}(10)$ observables, while
the precision requirements are high. This is a perfect setting for such models, which have a needed complexity and
necessary transformations to model such data samples in a computationally feasible way.

There have been many recent examples involving event generation \cite{gao2020event,Gao_2020,butter2021generative,verheyen2022event},
replacing the most computationally intensive parts of the detector simulation, such as the calorimeter response \cite{caloFlow,cresswell2022caloman,buckley2023inductive,Diefenbacher_2023},
jet modelling \cite{jetFlow}, anomaly detection \cite{Nachman_2020,golling2023interplay},
background estimation \cite{choi2020datadriven}, Bayesian uncertainties \cite{bayesFlows}, reweighting \cite{Nachman_2023},
end-to-end simulation \cite{vaselli2024endtoend} and many more.

The article focuses on the approach of developing a generative ML procedure for
a finite set of analysis-specific reconstructed kinematic observables. The generative ML algorithm is thus trained on a set
of MC-simulated and reconstructed events using the kinematic distributions used in the final analysis. The requirement is
to be able to extend the statistics of the existing MC using this procedure by several orders of magnitude with the
generation being fast enough that the events can be produced on-demand without the need for expensive data storage.
In other words, the ML algorithm will learn a to model the multi-dimensional distributions with a 'surrogate' model probability density $p(\vx)$ of
$\mathcal{O}(10)$ observables $\vx$ that can be used in the final stage of a given physics analysis. From a different perspective, this approach can also be interpreted as smoothing (``kriging'') of the multi-dimensional kinematic distributions derived from a finite learning sample.

The ML procedures used in particle physics are often built upon the most recent advances in ML tools used for commercial
purposes, whereby the goals and precision requirements in commercial applications are very
different from the ones used in a particle physics analysis. While ML approaches can achieve a \emph{visually impressive}
agreement between the training MC sample and the events generated using the derived ML procedure (for example with GAN
\cite{hashemi2019lhc} and VAE \cite{otten2021event}), one still needs to \emph{systematically evaluate} \cite{Kansal_2023,Das_2024}
whether the agreement is good enough in terms of the requirements of the corresponding physics data analysis.

The use of deep generative models for statistical amplification of MC samples in high-energy physics is a relatively new field, and
promises to be a powerful tool in the future \cite{Butter_2021,Matchev_2022,Bieringer_2022}. This paper aims to provide a
systematic evaluation of: what are the actual precision requirements in terms of statistical analysis of a new physics search,
and in view of these evaluates the performance of a few custom implementations of the recently trending ML models and a
set of statistical tools for their evaluation.


\section{The reference MC dataset}

The study in this paper uses the publicly available simulated LHC-like HIGGS dataset \cite{baldi2014searching} of a new physics beyond the
Standard model (BSM) Higgs boson production and a background process with identical decay products in the final state
and very  similar kinematic features,  to illustrate the performance of ML data generation in high-dimensional feature spaces.
The HIGGS dataset MC simulation uses the DELPHES toolkit \cite{delphes2009} to simulate the detector response of an LHC experiment.

The advantage of using this benchmark dataset is that it is publicly available and often used in evaluations of new ML tools by computer
science researchers and developers, and referenced even in groundbreaking works such as Ref. \cite{goodfellow2016}.

The signal process is the fusion of two gluons $gg$ into a heavy neutral Higgs boson $H^0$ that decays into heavy charged Higgs bosons $H^\pm$
and a $W^\mp$ boson. The $H^\pm$ then decays to a second $W^\pm$ boson and to a light Higgs boson $h^0$ that decays to $b$ quarks.
The whole signal process can be described as:
\begin{equation*}
    gg \rightarrow H^0 \rightarrow W^\mp H^\pm \rightarrow W^\mp W^\pm h^0 \rightarrow W^\mp W^\pm b \bar{b} \>.
\end{equation*}
The background process, which features the same intermediate state $W^\mp W^\pm b\bar{b}$ without the Higgs boson production
is the production and decay of a pair of top quarks, $gg \rightarrow  t \bar{t} \rightarrow W^\mp W^\pm b \bar{b}$, to a semi-leptonic final state.
Events were generated assuming 8 TeV collisions of protons at the LHC with masses set to $m_{H^0}=425$ GeV and $m_{H^\pm}=325$ GeV.

Observable final state decay products include electrically charged leptons $\ell$ (electrons or muons) and particle jets $j$.
The dataset contains semi-leptonic decay modes, where the first $W$ boson decays into $\ell \nu$ and the second into $jj$
giving us decay products $\ell \nu b$ and $jjb$. Further restrictions are also set on transverse momentum $p_{\text{T}}$,
pseudorapidity $\eta$, type and number of leptons, and number of jets.
Events that satisfy the above requirements are characterized by a set of observables (features) that describe the experimental
measurements using the detector data. These basic kinematic observables were labelled as ``low-level'' features and  are:
\begin{itemize}
    \item jet $p_{\text{T}}$, $\eta$ and azimuthal angle $\phi$, 
    \item jet $b$-tag, 
    \item lepton $p_{\text{T}}$, $\eta$ and $\phi$, 
    \item missing energy $E_m$, 
\end{itemize}
which gives us 21 features in total.
More complex derived kinematic observables can be obtained by reconstructing the invariant masses of the different intermediate
states of the two processes. These are the ``high-level'' features and are:
\begin{equation*}
    m_{jj},\> m_{jjj},\> m_{lv},\> m_{jlv},\> m_{b\bar{b}},\> m_{Wb\bar{b}},\> m_{WWb\bar{b}} \>.
\end{equation*}
Ignoring azimuthal angles $\phi$ due to the detector symmetry (giving a uniform distribution), and focusing only on continuous
features finally results in an 18-dimensional feature space.

\subsection{Data preprocessing}\label{sec: quantile}

Data preprocessing (also known as feature scaling) is a crucial step in training the ML surrogate models. It is used to transform the data
into a format that is more suitable for the ML algorithms. The data splitting into independent sets is presented in the
Appendix \ref{app: splits}. In this paper we use two different preprocessing methods: in the initial testing phase we use
the logit transformation followed by standardization, and in the final evaluation we use the quantile transformation.
This transformation (also known as a Gauss rank transformation) maps a variable's distribution to another probability
distribution \cite{scikit-learn,beasley2009rank}. The method transforms features to follow a normal distribution.
It reduces the impact of outliers making it a robust preprocessing scheme. For the final analysis we have chosen to
implement and use this transformation.


\section{Review of the explored ML methods}\label{sec: methods}

A short description of the evaluated normalizing flows and autoregressive models is presented in this section in order to give a relevant context to the ML
approaches evaluated in this paper.  The description closely follows the reviews from Refs. \cite{papamakarios2021normalizing, kobyzev2020normalizing}.
Preliminary results of the evaluation of the normalizing flows on the HIGGS dataset will be presented in this section.
We will further optimize the model performance and conduct additional evaluation using a set of statistical tools in the
following sections.

Let $\vu \in \mathbb{R}^D$ be a random vector with a known probability density function
$p_\text{u}(\vu): \mathbb{R}^D\rightarrow \mathbb{R}$. Distribution $p_\text{u}(\vu)$ is called a base distribution
and is usually chosen to be something simple, such as a normal distribution.
Given data $\vx \in \mathbb{R}^D$, one would like to know the distribution $p_\text{x}(\vx)$  it was drawn from. The solution is to express
$\vx$ as a transformation $T$ of a random variable $\vu$, distributed according to a distribution $p_\text{u}(\textbf{u})$, in such a way that
\begin{equation}
    \label{eq: sampling}
    \vx = T(\vu) \>, \quad \vu \sim p_\text{u}(\textbf{u}) \>,
\end{equation}
where $T$ is implemented using ML components, such as a neural network.
The transformation $T$ must be a diffeomorphism, meaning that it is invertible and both $T$ and $T^{-1}$ are differentiable.
Under these conditions, the density $p_\text{x}(\vx)$ is well-defined and can be calculated using the usual change-of-variables formula
\begin{equation}
    \label{eq: pdf}
    \begin{aligned}
        p_\text{x}(\vx) & = p_\text{u}\left(T^{-1}\left(\vx\right)\right) \left|\text{det}J_T\left(T^{-1}\left(\vx\right)\right)\right|^{-1}
        = p_\text{u}\left(T^{-1}\left(\vx\right)\right) \left| \text{det}J_{T^{-1}}\left(\vx\right) \right| \>,
    \end{aligned}
\end{equation}
where $J_T$ is a $D \times D$ Jacobian matrix of partial derivatives.

Invertible and differentiable transformations are composable, which allows one to construct a flow by chaining together different transformations.
This means that one can construct a complicated transformation $T$ with more expressive power by composing
many simpler transformations:
\begin{equation}
    T = T_K \circ \ldots \circ T_1 \quad \text{and} \quad T^{-1} = T_1^{-1} \circ \ldots \circ T_K^{-1} \>.
\end{equation}
A flow is thus referring to the trajectory of samples from the base distribution as they get sequentially transformed by each
transformation into the target distribution. This is known as forward or generating direction.
The word normalizing refers to the reverse direction, taking samples from data and transforming them to the base distribution, which
is usually normal. This direction is called inverse or normalizing direction and is the direction of the model training.
Flows in forward and inverse directions are then, respectively,
\begin{equation}
    \label{eq: dir}
    \begin{aligned}
         & \vz_k = T_{k}(\vz_{k-1}) \quad \text{for} \quad k=1,\ldots,K \>,    \\
         & \vz_{k-1} = T^{-1}_k(\vz_k) \quad \text{for} \quad k=K,\ldots,1 \>,
    \end{aligned}
\end{equation}
where $\vz_0 = \vu$ and $\vz_K = \vx$.

The log-determinant of a flow is given by
\begin{equation}
    \label{eq: det}
    \begin{aligned}
        \log \left|\operatorname{det} J_{T}\left(\mathbf{z}_{0}\right)\right| & =\log \left|\prod_{k=1}^{K} \operatorname{det}
        J_{T_{k}}\left(\mathbf{z}_{k-1}\right)\right|
        =\sum_{k=1}^{K} \log \left|\operatorname{det}
        J_{T_{k}}\left(\mathbf{z}_{k-1}\right)\right| \>.
    \end{aligned}
\end{equation}
A trained flow model provides event sampling capability by Eq. (\ref{eq: sampling}) and density estimation by Eq. (\ref{eq: pdf}).

The best description of the unknown probability density  $p_\text{x}(\vx)$ is obtained by fitting a parametric flow model
$p_\text{x}(\vx;\vtheta)$ with free parameters $\vtheta$ to a target distribution by using a maximum likelihood estimator
computing the average log-likelihood over $N$ data points
\begin{equation}
    \mathcal{L}(\vtheta) = -\frac{1}{N} \sum_{n=1}^N \log p_\text{x}(\vx_n; \vtheta) \>.
    \label{eq: loss}
\end{equation}
The latter defines the loss function of the ML algorithm and is thus the quantity optimized using gradient-based methods while training the ML procedure.
This can be done because the exact log-likelihood of input data is tractable in flow-based models. In order to keep the
computing load at a manageable level, averaging is performed over batches of data and not on the whole dataset, as is customary in practically all ML procedures.

Rewriting Eq. (\ref{eq: loss}) in terms of variables $\vu$ using Eq. (\ref{eq: pdf}) and introducing a parametric description
of the distribution $p_\text{u}(\textbf{u};\boldsymbol{\psi})$ from Eq. (\ref{eq: sampling}), one gets
\begin{equation}
    \label{eq: mle_loss}
    \begin{aligned}
        \mathcal{L}(\vtheta) = -\frac{1}{N} \sum_{n=1}^N [\log p_\text{u} \left( T^{-1}(\vx_n; \boldsymbol{\phi}); \boldsymbol{\psi} \right)
        - \log \left| \text{det} J_{T^{-1}}(\vx_n;\boldsymbol{\phi}) \right| ] \>,
    \end{aligned}
\end{equation}
where $\vtheta=\{\boldsymbol{\phi}, \boldsymbol{\psi} \}$ are the parameters of the transformation and base distribution, respectively.
The parameters $\boldsymbol{\psi}$ of the base distribution are usually fixed, for example, the zero mean and the unit variance
of a normal distribution.
From Eq. (\ref{eq: mle_loss}), one can see that in order to fit the flow model parameters one
needs to compute the inverse transformation $T^{-1}$, the Jacobian determinant, the density $p_\text{u}(\vu;\boldsymbol{\psi})$
and be able to differentiate through all of them. For sampling the flow model, one must also compute $T$ and be able to sample
from $p_\text{u}(\vu;\boldsymbol{\psi})$.

For applications of flow models, the Jacobian determinant should have at most $\mathcal{O}(D)$ complexity, which
limits the flow-model-based design.

\subsection{Coupling models}

The main principle of finding a set of transformations optimally suited to be used in flow-based generative ML models,
introduced by Ref. \cite{dinh2014nice}, is to focus on a class of transformations that produce Jacobian
matrices with determinants reduced to the product of diagonal terms. These classes of transformations are called
coupling layers.

A coupling layer splits the feature vector $\vz$ into two parts at index $d$ and transforms the second part as a function
of the first part, resulting in an output vector $\vz^\prime$. In the case of RealNVP model \cite{dinh2016density} the
implementation is a follows:
\begin{equation}
    \label{eq: affine}
    \begin{aligned}
         & \vz_{\leq d}^\prime = \vz_{\leq d} \>,                                                                                     \\
         & \vz_{> d}^\prime = \vz_{> d} \cdot \exp\left(\boldsymbol{\sigma}(\vz_{\leq d})\right) + \boldsymbol{\mu}(\vz_{\leq d}) \>.
    \end{aligned}
\end{equation}
This affine transformation of the form $\boldsymbol{s}\cdot\vz+\boldsymbol{t}$, consisting of separate scaling ($\boldsymbol{s}=\exp{(\boldsymbol{\sigma})}$) and translation ($\boldsymbol{t}=\boldsymbol{\mu}$)
operations, is implemented by distinct neural networks $f$. These operations depend on the variables $\vz_i$ in the other half
of the block ($i \leq d$), i.e. $\boldsymbol{\mu} = f_{\boldsymbol{\mu}}(\vz_{\leq d})$ and $\boldsymbol{\sigma} = f_{\boldsymbol{\sigma}}(\vz_{\leq d})$.
It is worth noting that this affine transformation possesses a straightforward inverse, eliminating the need to compute the
inverses of $\boldsymbol{s}$ and $\boldsymbol{t}$. Furthermore, it exhibits a lower triangular Jacobian with a block-like structure,
enabling the determinant to be computed in linear time.

When sequentially stacking coupling layers, the elements of $\vz$ need to be permuted between each of the two layers so that every
input has a chance to interact with every other input. This can be done either by using binary checkerboard masks, or
with a trained permutation matrix
\begin{equation}
    \vz^\prime = \mathbf{W}\vz
\end{equation}
as in the Glow model \cite{kingma2018glow}. The idea is thus to alternate these invertible linear transformations with coupling layers.

\begin{figure}[ht!]
    \centering
    \includegraphics[width=0.9\textwidth]{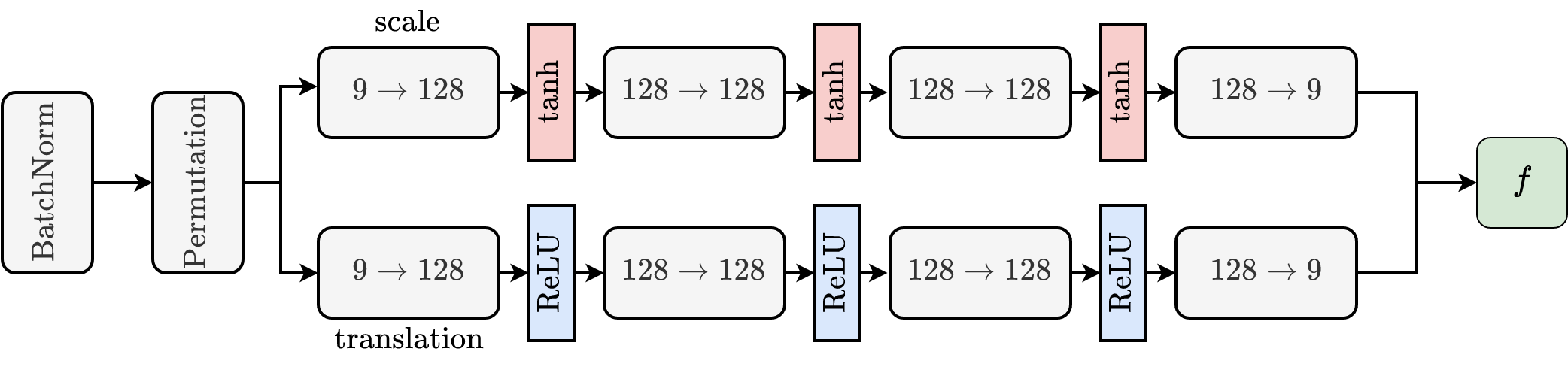}
    \caption{The building block of a coupling layer in a flow. The block consists of a coupling layer with batch normalisation
        and learned permutations. The scaling and translation networks have the same architecture but differ in activation functions.
        Scaling network uses tanh functions, whereas the translation network uses ReLU functions.}
    \label{fig: glow}
\end{figure}

In order to further speed up and stabilize flow training, batch normalisation is introduced at the
start of each coupling layer as described in Ref. \cite{dinh2016density}. One block of such a flow is schematically
presented in Fig. \ref{fig: glow}.

\begin{figure}[ht!]
    \centering
    \includegraphics[width=0.48\textwidth]{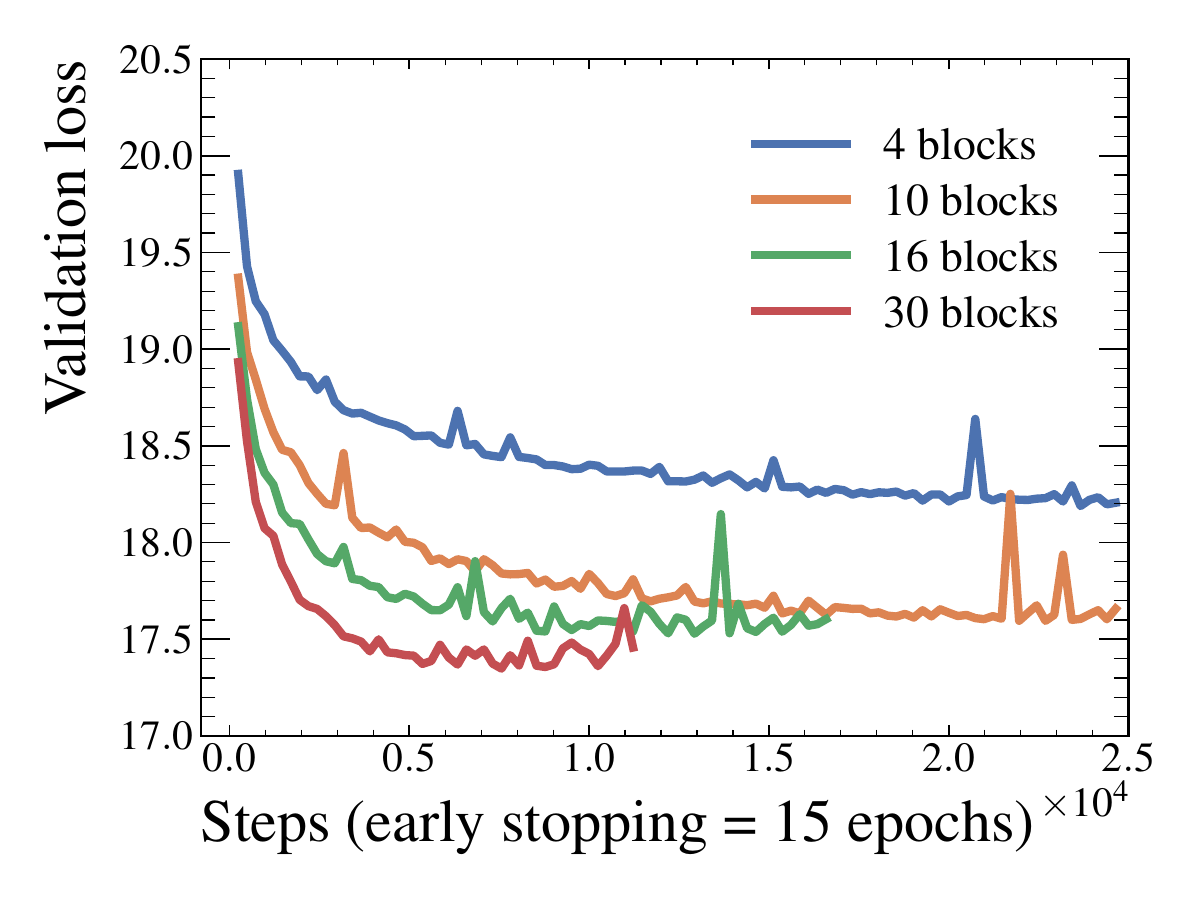}
    \includegraphics[width=0.48\textwidth]{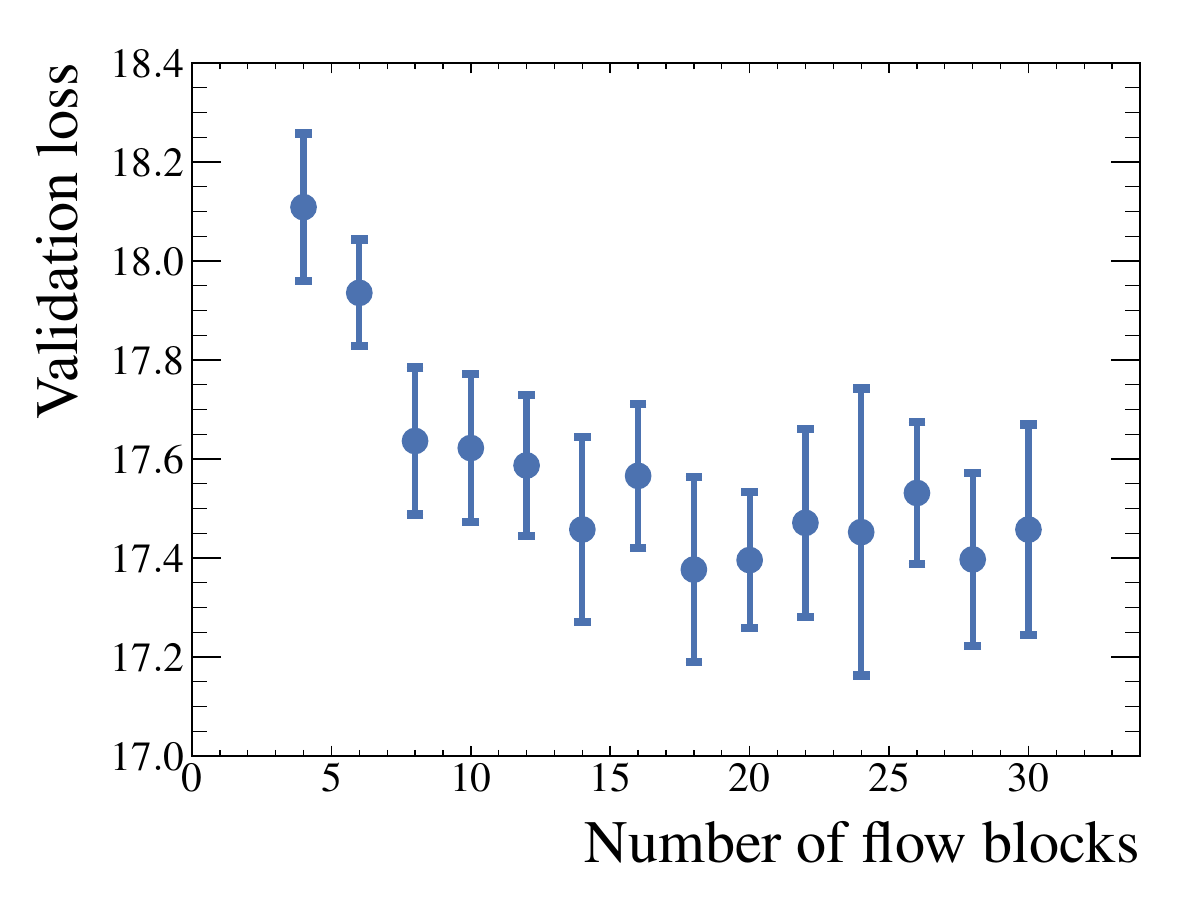}
    \caption{Validation loss as a function of training steps (epochs) and the number of flow blocks. Early stopping was
        employed to account for over-fitting. All other hyper-parameters were kept constant. Models were trained on $2.5\times 10^5$
        events. The uncertainties presented in the right plot were estimated by repeated training and validation using random
        sampling of events. One can observe rapidly diminishing gains by using more than 10 blocks.}
    \label{fig: glow_losses}
\end{figure}

The expressive power of a flow can be increased by composing multiple blocks of coupling layers, batch normalisations and
permutations. The number of blocks in the flow is the main hyper-parameter of such a model. The Fig. \ref{fig: glow_losses} shows
the dependence of validation loss, i.e. the value of the loss function from Eq. (\ref{eq: loss}) obtained on an independent
validation sub-sample of the HIGGS dataset, on the number of blocks in a flow. Models with more blocks are harder to train,
and show signs of over-fitting earlier. A list of all other hyper-parameters is given in Table \ref{tab: hyper1} in the
Appendix \ref{app: hyper}.

\subsection{Autoregressive models}

\begin{figure}[ht!]
    \centering
    \includegraphics[width=0.55\textwidth]{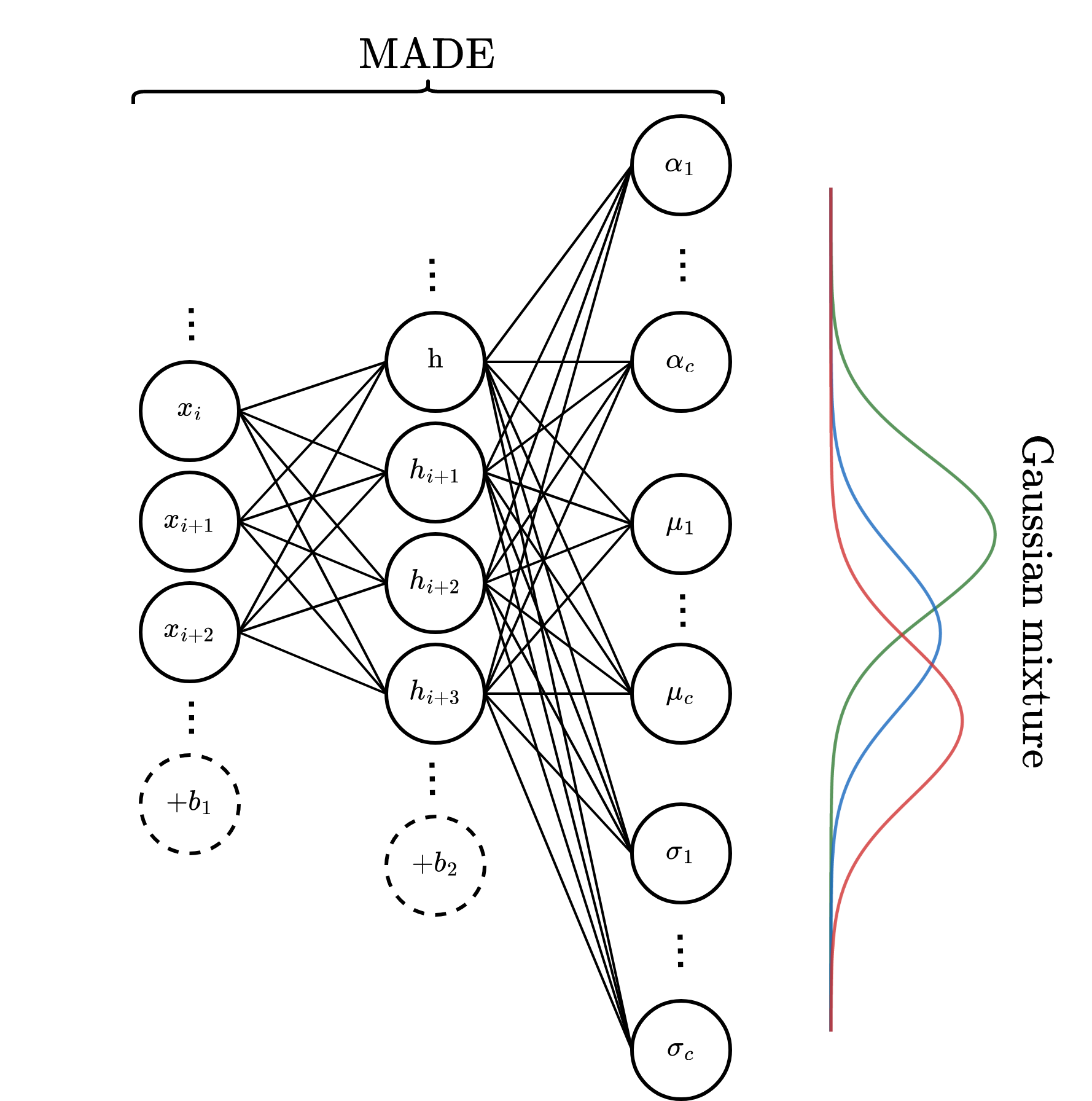}
    \caption{The mixture of Gaussian terms (MOG) output of a neural network implemented with MADE.}
    \label{fig: gmm}
\end{figure}

Using the chain rule of probability, one can rewrite any joint distribution over $D$ variables (as discussed in Ref. \cite{pml2Book})
in the form of a product of conditional probabilities
\begin{equation}
    \label{eq: auto}
    p(\vz) = p(z_1)p(z_2|z_1)p(z_3|z_2,z_1)p(z_4|z_3,z_2,z_1) \ldots = \prod_{i=1}^D p\left(\vz_i|\vz_{<i}\right) \>,
\end{equation}
where each conditional $p(\vz_i|\vx_{<i})=p(\vz_i;c_i(\vz_{<i}))$ is modelled by some parametric distribution with
parameters $c_i$. If $p\left(\vz_i;c_i(\vz_{<i})\right)$ is conditioned on a mixture of Gaussian terms (MOG), one gets a
RNADE model from Ref. \cite{uria2013rnade}:
\begin{equation}
    \label{eq: mixtures}
    p(\vz_i|\vz_{<i}) = \sum_{c=1}^C \boldsymbol{\alpha}_{i, c}\, \mathcal{N}(\vz_i;\boldsymbol{\mu}_{i, c},\boldsymbol{\sigma}^2_{i, c}) \>,
\end{equation}
Where $C$ are the number of components in the mixture.
The mixture model parameters are calculated using a neural network
that returns a vector of outputs $(\boldsymbol{\mu}_{i}, \boldsymbol{\sigma}_{i}, \boldsymbol{\alpha}_{i})=f(\vz_{<i};\vtheta_i)$,
as illustrated on Fig. \ref{fig: gmm}.

Specifically, the mixture of Gaussian parameters for the conditionals is calculated in the following sequence:

\begin{equation}
    \begin{aligned}
        \mathbf{h}(\vz)          & = \text{ReLU}(\mathbf{W}^\top \vz + \mathbf{b}) \>,                                            \\
        \boldsymbol{\alpha}(\vz) & = \text{softmax}(\mathbf{W}^\top_\alpha \mathbf{h}(\vz) + \mathbf{b}_\alpha) \>,               \\
        \boldsymbol{\mu}(\vz)    & = \mathbf{W}^\top_\mu \mathbf{h}(\vz) + \textbf{b}_\mu \>,                                     \\
        \boldsymbol{\sigma}(\vz) & = \text{ELU}(\mathbf{W}^\top_\sigma \mathbf{h}(\vz) + \mathbf{b}_\sigma) + 1 + \varepsilon \>,
    \end{aligned}
\end{equation}

where  $\mathbf{W}$ are the weight matrices, and $\mathbf{b}$ the bias vectors. ReLU, softmax and ELU are the activation functions.
The event sampling generative step is performed simply as sampling from a Gaussian mixture. As a substantial simplification,
a single neural network with $D$ inputs and $D$ outputs for each parameter vector can be used instead of using $D$ separate
neural networks ($f_i$) for each parameter. This is done with a MADE network \cite{germain2015made} that uses a masking
strategy to ensure the autoregressive property from Eq. (\ref{eq: auto}). Adding Gaussian components to a MADE network
increases its flexibility.  The model was proposed in Ref. \cite{papamakarios2017masked} and  it is stated in this reference that it can fact be shown that the MADE with the mix of Gaussian components (MoG) is a universal density approximator; with sufficiently many hidden units and Gaussian components, it can approximate any continuous density arbitrarily well. In our paper we call it  MADEMOG, as it is commonly tagged. The MADE
network was implemented using residual connections from Ref. \cite{he2016identity} as described in Ref. \cite{nash2019autoregressive}.

\clearpage

\begin{figure}[ht!]
    \centering
    \includegraphics[width=0.99\textwidth]{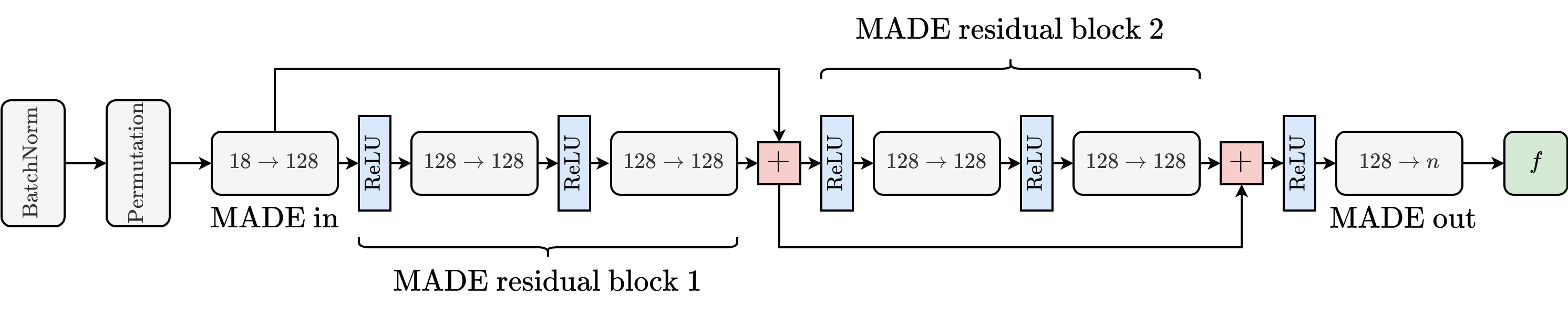}
    \caption{\centering Block of a used MAF model. The architecture follows the one outlined in Ref. \cite{durkan2019neural}.}
    \label{fig: maf}
\end{figure}

MADE networks can also be used as building blocks to construct a Masked Autoregressive Flow or MAF \cite{papamakarios2017masked}
by sequentially stacking MADE networks. For notational simplicity we will write all the following equations for the first
block of the flow only. For all the next blocks the same equations are used with the only difference that the input to the block $\vz_k$
is the output of the previous block $\vz_{k-1}$ (for the first block this is the base distribution $\vu$ from Eq. (\ref{eq: dir})).
In this case, the $i$-th conditional is given by a single Gaussian
\begin{equation}
    \label{eq: maf_gauss}
    p\left(\vz_i;c_i(\vz_{<i})\right) = \mathcal{N}\left(\vz_i;\boldsymbol{\mu}_i, (\exp(\boldsymbol{\sigma}_i))^2\right) \>,
\end{equation}
where $\boldsymbol{\mu}_i = f_{\boldsymbol{\mu}_i}(\vz_{<i})$ and $\boldsymbol{\sigma}_i = f_{\boldsymbol{\sigma}_i}(\vz_{<i})$ are MADE networks.
The generative sampling in this model is straightforward:
\begin{equation}
    \label{eq: maf_forward}
    \vz_i = \vu_i \cdot \exp(\boldsymbol{\sigma}_i) + \boldsymbol{\mu}_i \>, \quad \text{where} \quad \vu_i \sim \mathcal{N}(0, 1)
\end{equation}
with a simple inverse
\begin{equation}
    \label{eq: maf_inverse}
    \vu_i = (\vz_i - \boldsymbol{\mu}_i)\exp (-\boldsymbol{\sigma}_i) \>,
\end{equation}
which is the flow model training direction.
Due to the autoregressive structure, the Jacobian is triangular (the partial derivatives $\partial x_i/\partial u_j$
are identically zero when $j>i$), hence its determinant is simply the product of its diagonal entries.
One block of such a flow is presented in Fig. \ref{fig: maf}.
A further option is to stack a MADEMOG on top of a MAF (and train it jointly), which is then labelled as a MAFMADEMOG.

\begin{figure}[ht!]
    \centering
    \includegraphics[width=0.49\textwidth]{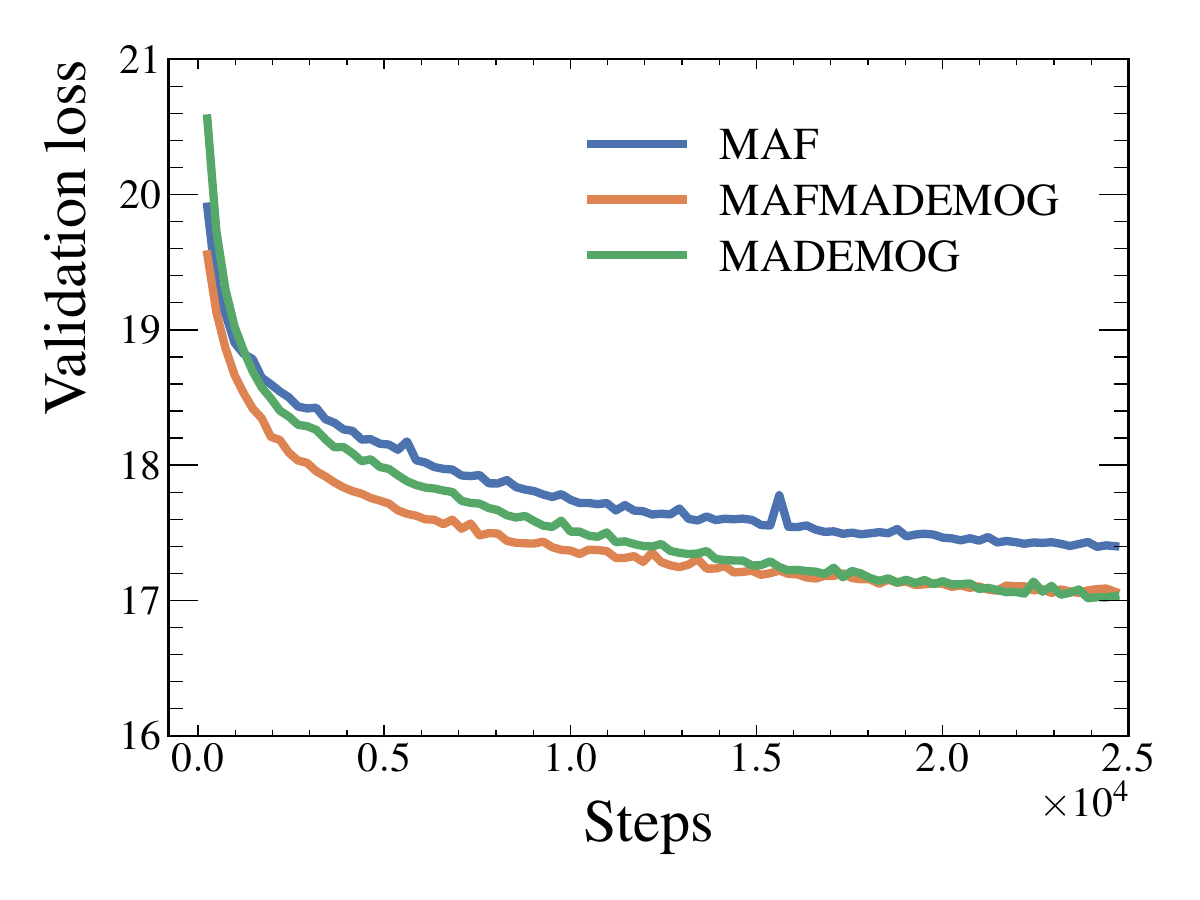}
    \includegraphics[width=0.49\textwidth]{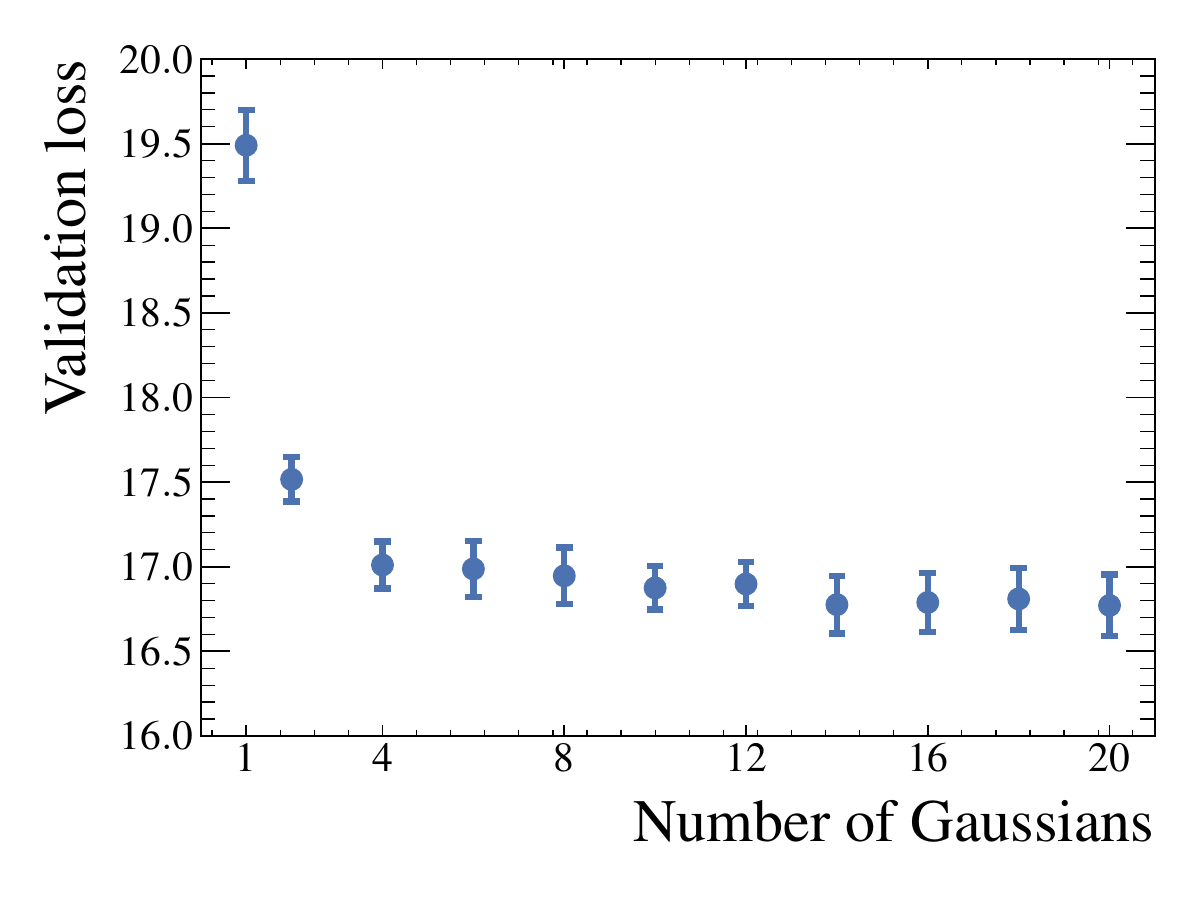}
    \caption{Validation loss as a function of training steps (epochs) for different types of models and the dependence of validation
        loss on the number of Gaussian components for the MAFMADEMOG model. Models were trained on $2.5\times 10^5$ events. The uncertainties presented in the right plot were estimated by repeated training and validation using random
        sampling of events. One can observe rapidly diminishing gains by using more than 6 Gaussian components.}
    \label{fig: mades_losses}
\end{figure}

Training curves for all three types of models are shown in Fig. \ref{fig: mades_losses}, where one can see that the MADEMOG and
MAFMADEMOG achieve similar performance, in both cases significantly better than the simpler MAF. The dependence on the number
of Gaussian terms is also shown, where one can see that the model performance saturates at a relatively low number of
chosen Gaussian mixtures. The baseline ML model was eventually selected to be MADEMOG, by measuring the quality of
generated samples, evaluated using statistical measures presented in Sections \ref{sec: distances} and \ref{sec: c2st}.

To maximize the physics potential we have further trained the MADEMOG model, which has shown promising performance with
a simple architecture at the good sampling speed (see appendix \ref{app: times} for ML event generation times). To combat
the problems introduced in this section (mainly training instability and over-fitting) we have used a GELU
activation function \cite{hendrycks2016gaussian} and cosine learning scheduler with warm restarts \cite{loshchilov2016sgdr}.
We have also used Gaussian feature scaling introduced in section \ref{sec: quantile}. The network has 20M trainable parameters
(see Table \ref{tab: hyper_comp} in Appendix \ref{app: hyper}).
This setup was eventually shown to be the most performant and was used in the final studies in this paper.
The dependence on the number of used parameters for the MADEMOG network is shown in Appendix \ref{app: paramas}.

\subsection{Spline transformations}

At the end of the flow block, one needs to choose a specific transformation $f$. So far, affine transformations of
Eq. (\ref{eq: affine}) and Eq. (\ref{eq: maf_forward}) were used, which were combined with sampling from Gaussian
conditionals as in Eq. (\ref{eq: mixtures}). Rather than relying on basic affine transformations, this approach can be
extended to incorporate spline-based transformations.

A spline is defined as a monotonic piecewise function consisting of $K$ segments (bins).
Each segment is a simple invertible function (e.g. linear or quadratic polynomial, as in Ref. \cite{muller2019neural}).
In this paper, rational quadratic splines are considered, as proposed in Ref. \cite{durkan2019neural} and used by Ref. \cite{Gao_2020}.
Rational quadratic functions are differentiable and are also analytically invertible in the case of monotonic segments.
The spline parameters are then determined from neural networks in an autoregressive way. The resulting autoregressive
model, replacing affine transformations of Eq. (\ref{eq: maf_forward}) and its inverse Eq. (\ref{eq: maf_inverse}) with
equivalent expressions for splines, is labelled as RQS in the studies of this paper. Symbolically, the generative step is thus:
\begin{equation}
    \label{eq: rqs_forward}
    \vz_i = \text{RQS}_i\left(\vu_i,K_i,\boldsymbol{\theta}_i\right), \quad \text{where} \quad \vu_i \sim \mathcal{N}(0, 1)
\end{equation}
with an inverse
\begin{equation}
    \label{eq: rqs_inverse}
    \vu_i = \text{RQS}^{-1}_i\left(\vz_i,\vu_i,K_i,\boldsymbol{\theta}_i\right),
\end{equation}
where the $K$ represents the bin parameters and $\boldsymbol{\theta}$ denotes the remaining model (spline) parameters.

\begin{figure}[ht!]
    \centering
    \includegraphics[width=0.49\textwidth]{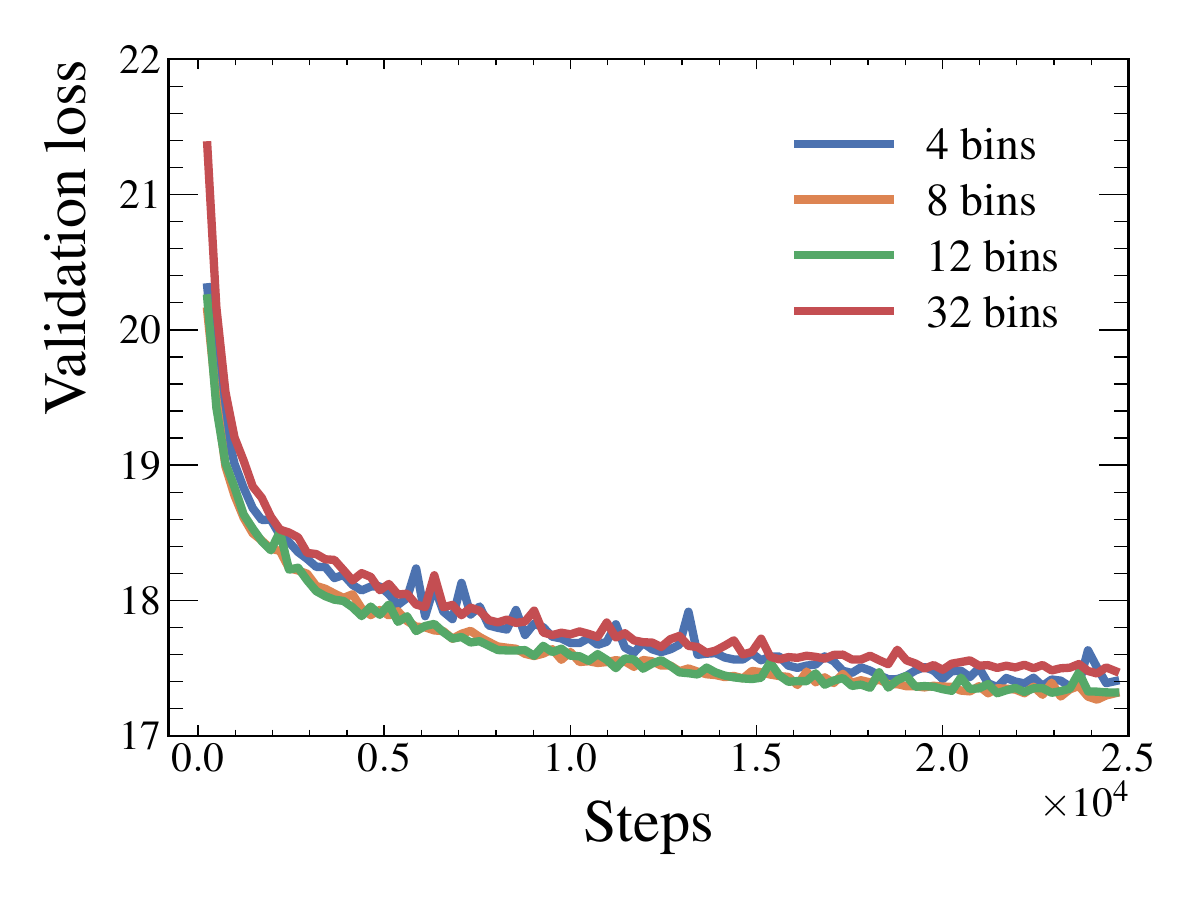}
    \caption{Validation loss as a function of training steps and number of spline bins.}
    \label{fig: splines_losses}
\end{figure}

Fig. \ref{fig: splines_losses} shows the validation loss  for different number of spline bins.
The validation loss does not seem to decrease substantially with the increasing number of bins, however, the quality of
generated samples, evaluated using statistical methods presented in Sec. \ref{sec: distances} and Sec. \ref{sec: c2st}. does, in fact, improve with more bins. For the studies performed in this paper, rational
splines in 32 bins were eventually chosen. As one can observe from the Fig. \ref{fig: mades_losses} and Fig. \ref{fig: splines_losses}, the performance of this algorithm
is comparable to the one of the MADEMOG, however the computing requirements are noticeably higher. This was one further
reason why the MADEMOG was chosen as the baseline algorithm for further studies, as already stated.

\clearpage

\section{Performance evaluation of the ML techniques}\label{sec: performance}

After defining the representative MC dataset and generative ML procedures, one can focus on the main objective of
this paper, which is to systematically evaluate the performance of such tools with a finite-sized training sample in
terms of requirements based on a physics analysis at the LHC and HL-LHC.

As described in the introduction, a typical particle physics data analysis with the final selection and the statistical evaluation
procedure uses an order of $\mathcal{O}(10)$ reconstructed kinematic observables (which are very often used to construct the ``ultimate''
selection variable using ML techniques). The corresponding available statistics of MC signal and background events used
to construct the predicted distributions of these kinematic observables, is however often of the order $\mathcal{O}(10^6)$
events or less due to the computing resource constraints and severe filtering requirements of a new physics search. Consequently, the ultimate
kinematic region, where one searches for new physics signal, can even contain orders of magnitude fewer MC events, since one
is looking for unknown new processes at the limits of achievable kinematics, which translates to ``tails'' of kinematic distributions.

\subsection{ML-generated distributions of observables}

\begin{figure*}
    \centering
    \includegraphics[width=0.8\textwidth]{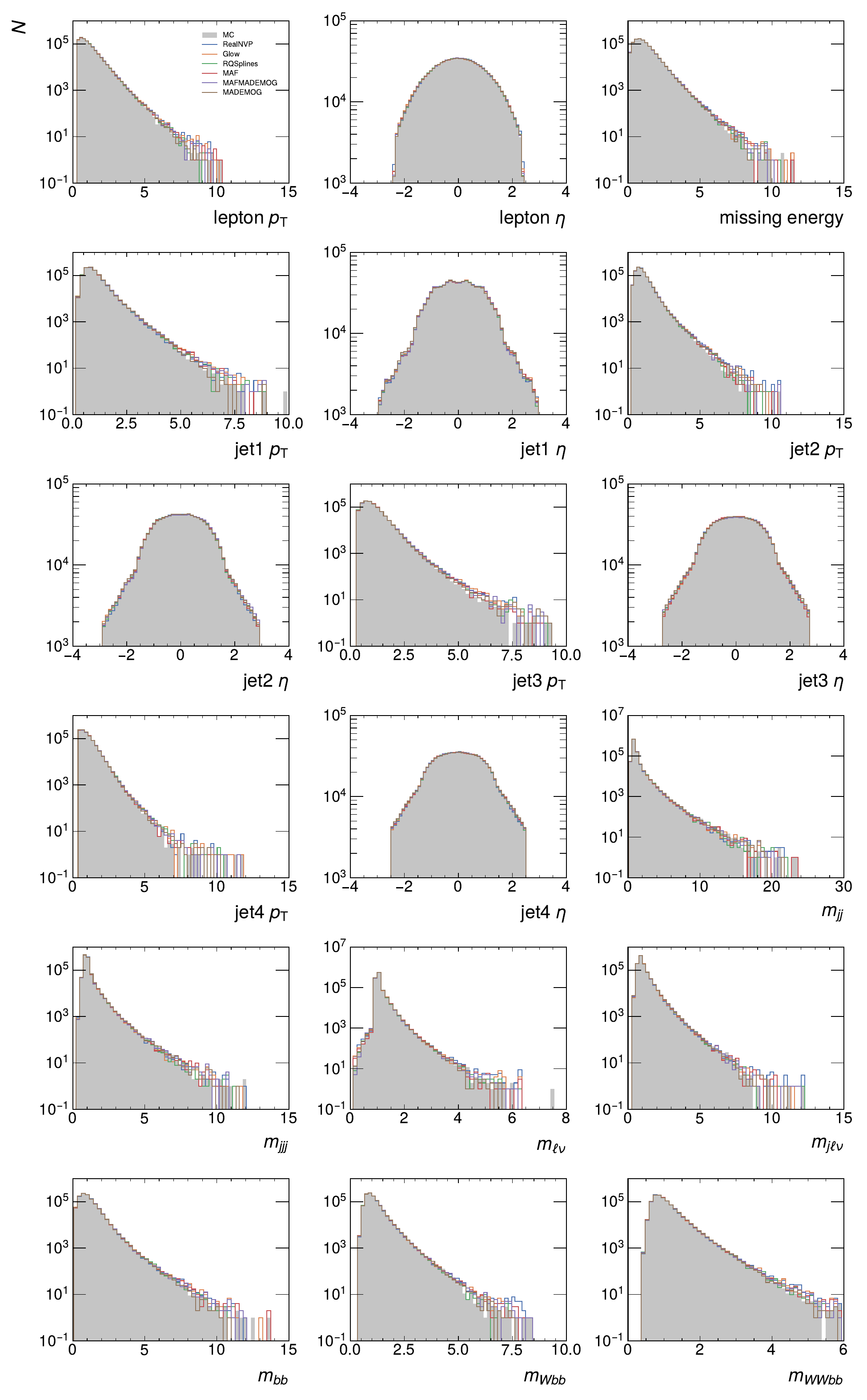}
    \caption{Distributions of ML-generated events using the described models, trained on the HIGGS dataset on its kinematic observables. The
        original MC distribution from this dataset is shown in grey. Visually, the quality of reproduction is very good.}
    \label{fig: dists}
\end{figure*}

Histograms of  event distributions of kinematic observables used in the HIGGS sample, for which the corresponding events were ML-generated,
are shown in Fig. \ref{fig: dists}, with the detailed model configuration given in Table \ref{tab: hyper_comp}.
ML-generated distributions are compared with distributions of the corresponding MC events from a subset which was not used in training to avoid any possible bias, as detailed in the Appendix \ref{app: splits}. One can observe that the
models reliably reproduce the original distributions at least on this (visual) level of comparison.
Further plots are shown in the Appendices \ref{app: nolog} and \ref{app: corner}.

\begin{figure*}
    \centering
    \includegraphics[width=0.6\textwidth]{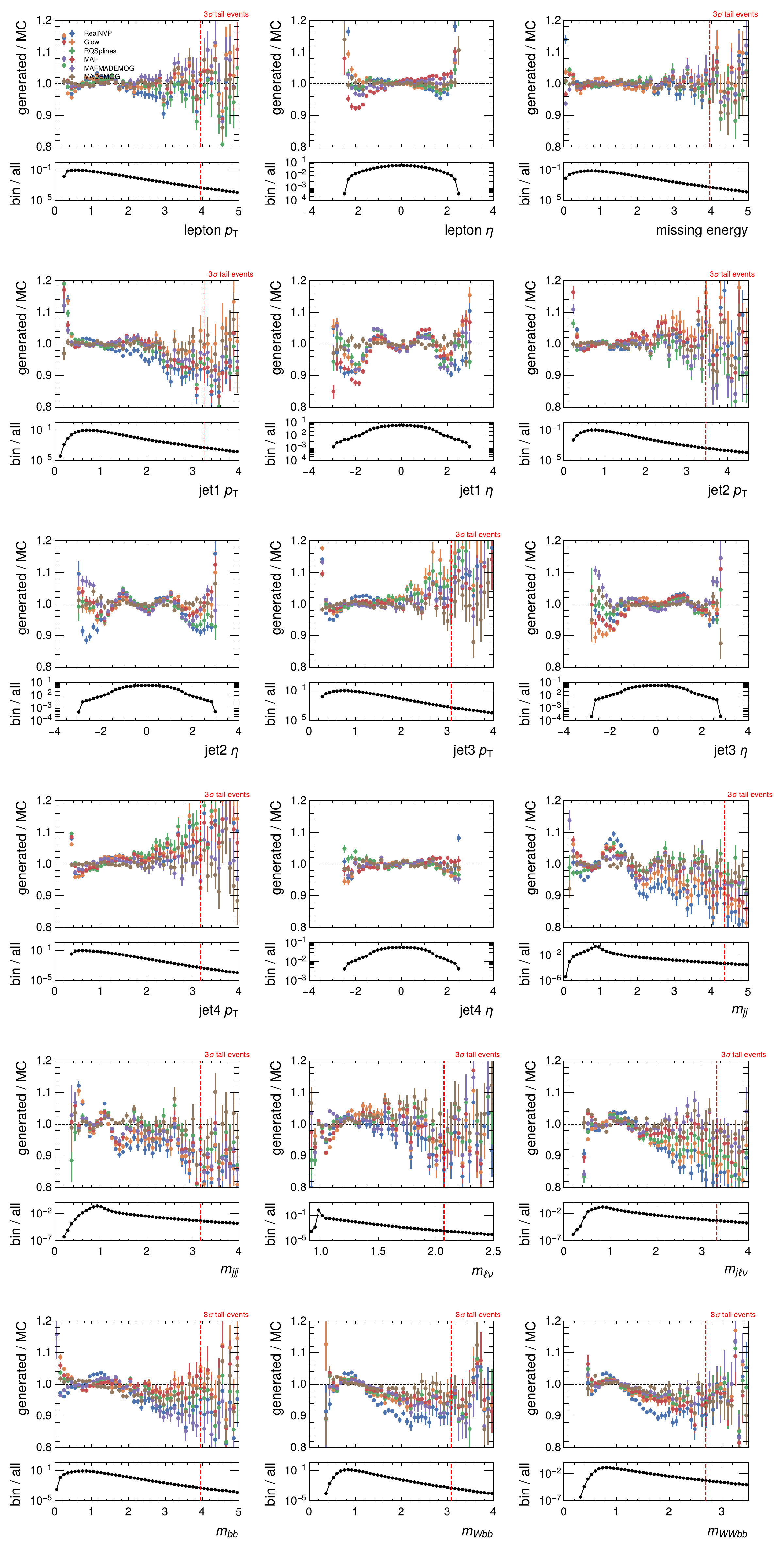}
    \caption{Histograms containing binned ratios of ML-generated events and MC-simulated events using the described models. Smaller plots containing
        the normalized original distribution are given under each ratio plot for easier comparison.
        Discrepancy between bins for MC events and ML-generated events increases in the tails of the distributions with poor
        statistics, to give a representative impact of the model uncertainties in the  error bars due to finite statistics, 
        the numbers of ML-generated and MC-simulated events are set to be the same. 
        The vertical dashed red line indicates the region containing the ``3$\sigma$'' tail
        (i.e. approximately the $10^{-3}$ fraction of MC events), indicating very poor  statistics of MC events available
        for training the ML algorithm.}
    \label{fig: ratios}
\end{figure*}

In order to give a more detailed insight in the quality of reproduction, histograms containing binned ratios of these ML-generated events and MC-simulated
events are shown in Fig. \ref{fig: ratios}. It is evident that the overall agreement is visually very good - the deviations between the ML and MC are for all cases most pronounced in the tails where the event count becomes very low. Performance in these tail regions diminishes due to the fact that the learning algorithm did not see enough rare events in the tails of distributions and could not reproduce a reliable surrogate distribution in that region.

\begin{figure*}
    \centering
    \includegraphics[width=0.8\textwidth]{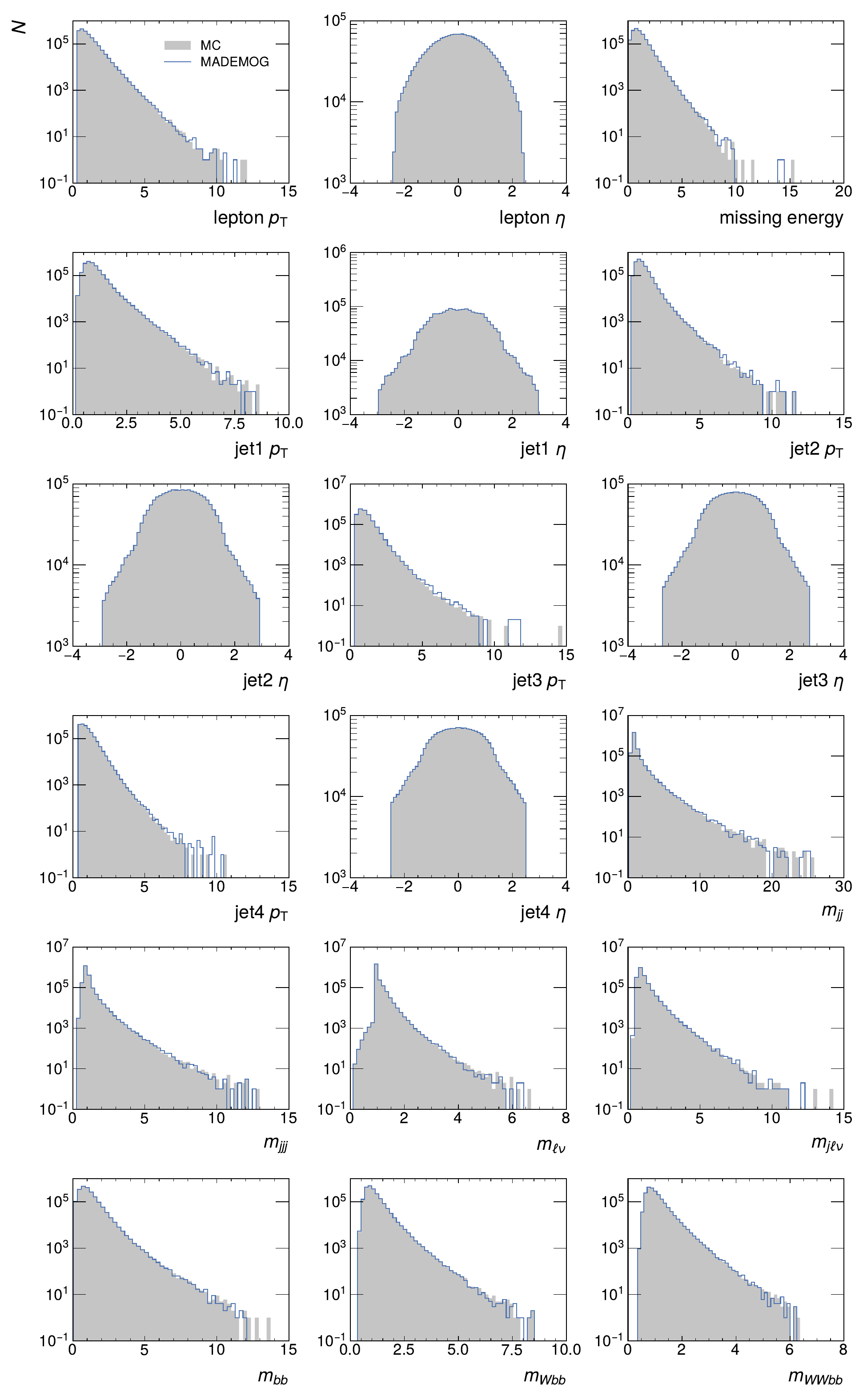}
    \caption{Distributions of ML-generated events using the selected MADEMOG model, trained on the HIGGS dataset on its kinematic observables. The
        original MC distribution from this dataset is shown in grey. Visually, the quality of reproduction is very good.}
    \label{fig: dists_best}
\end{figure*}

\begin{figure*}
    \centering
    \includegraphics[width=0.6\textwidth]{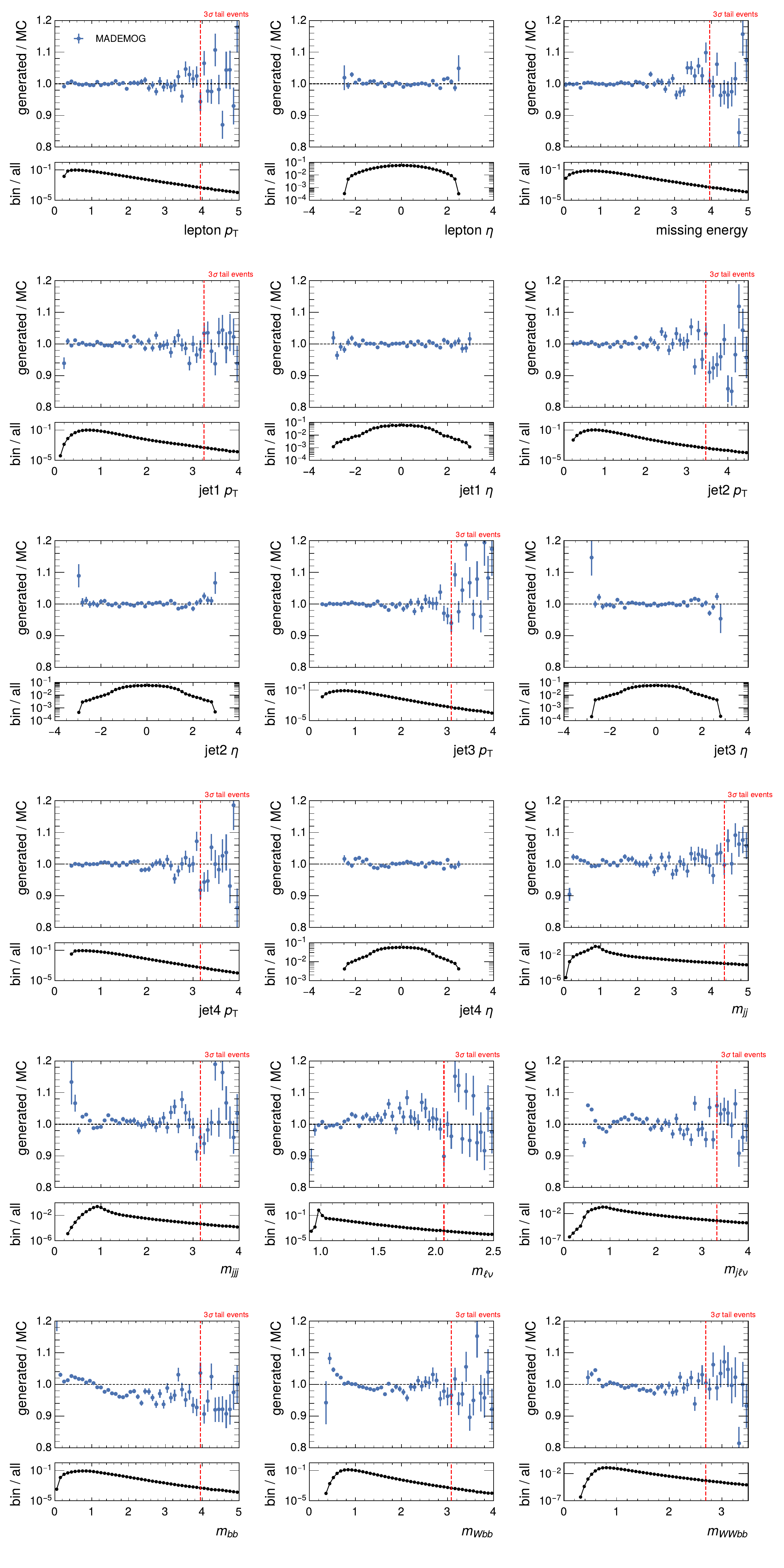}
    \caption{Histograms containing binned ratios of ML-generated events and MC-simulated events using the selected MADEMOG model. Smaller plots containing
        the normalized original distribution are given under each ratio plot for easier comparison.
        Discrepancy between bins for MC events and ML-generated events increases in the tails of the distributions with poor
        statistics, to give a representative impact of the model uncertainties in the  error bars due to finite statistics, 
        the numbers of ML-generated and MC-simulated events are set to be the same. 
        The vertical dashed red line indicates the region containing the ``3$\sigma$'' tail
        (i.e. approximately the $10^{-3}$ fraction of MC events), indicating very poor  statistics of MC events available
        for training the ML algorithm.}
    \label{fig: ratios_best}
\end{figure*}

The dedicated comparisons for the MADEMOG model, which was selected to be used in the statistical studies presented in this paper, are shown in figures Fig. \ref{fig: dists_best} and Fig. \ref{fig: ratios_best}. 

\clearpage

\clearpage

\clearpage

\clearpage

\clearpage

\clearpage

\subsection{Evaluating the ML generation using divergence measures between probability distributions in one dimension}\label{sec: distances}

The probability distributions we can use in statistical tests can, in this study, be extracted from the event samples. We have the ML-generated events
$\{\vx_1,\ldots,\vx_N \}\sim P_{\text{ML}}$ and the MC events $\{\tilde{\vx}_1,\ldots,\tilde{\vx}_M \}\sim P_{\text{MC}}$.
Determining if the samples come from the same distribution (i.e. if $P_{\text{MC}}=P_{\text{ML}}$)
is known as a two-sample test. Such a test can be computed by defining some suitable divergence metric
and comparing it to a baseline reference. For this purpose, we have used several well known divergence measures (or ``tests''): the $\chi^2$ distance, the Kullback-Leibler divergence,
the Hellinger distance and the Wasserstein distance \cite{pml2Book}. To present a successful test and define the baseline reference, we have split the 
MC event sample in two subsets and performed the distance calculation on these two parts. This then represents the ideal result of a successful test in the presence of statistical uncertainty (that we use as the baseline reference in what follows). 

\begin{figure}[ht!]
    \centering
    \begin{minipage}[t]{0.46\textwidth}
        \centering
        \includegraphics[width=0.99\textwidth]{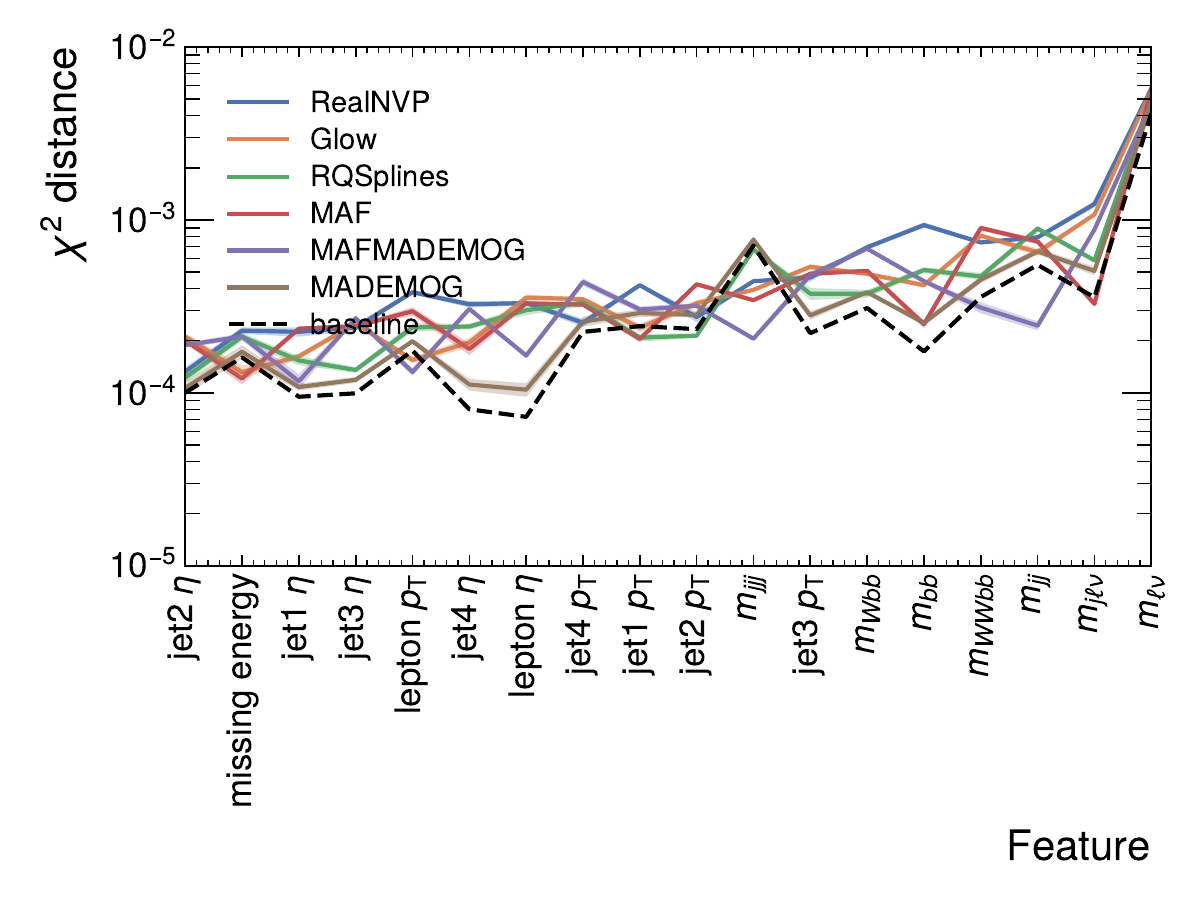}
    \end{minipage}
    \hfill
    \begin{minipage}[t]{0.46\textwidth}
        \centering
        \includegraphics[width=0.99\textwidth]{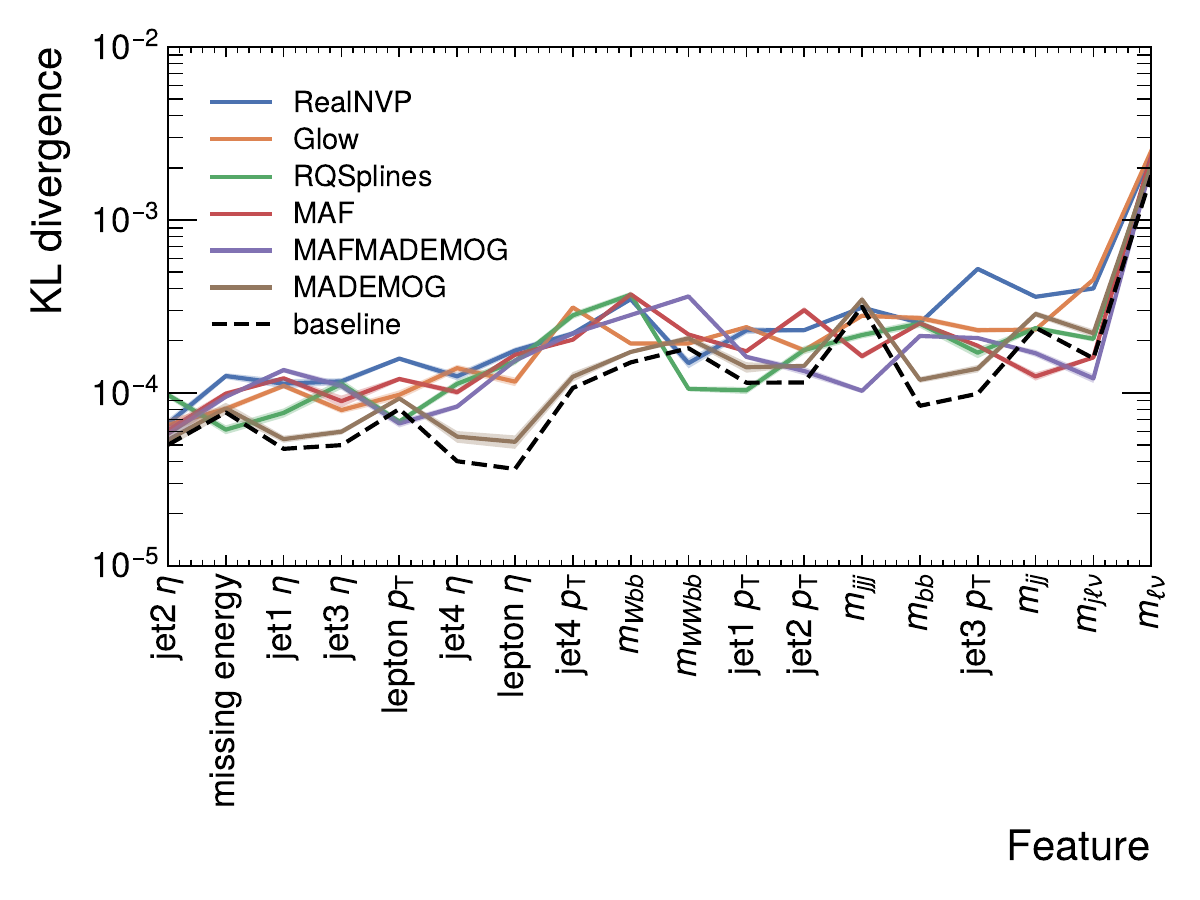}
    \end{minipage}
    \hfill
    \begin{minipage}[t]{0.46\textwidth}
        \centering
        \includegraphics[width=0.99\textwidth]{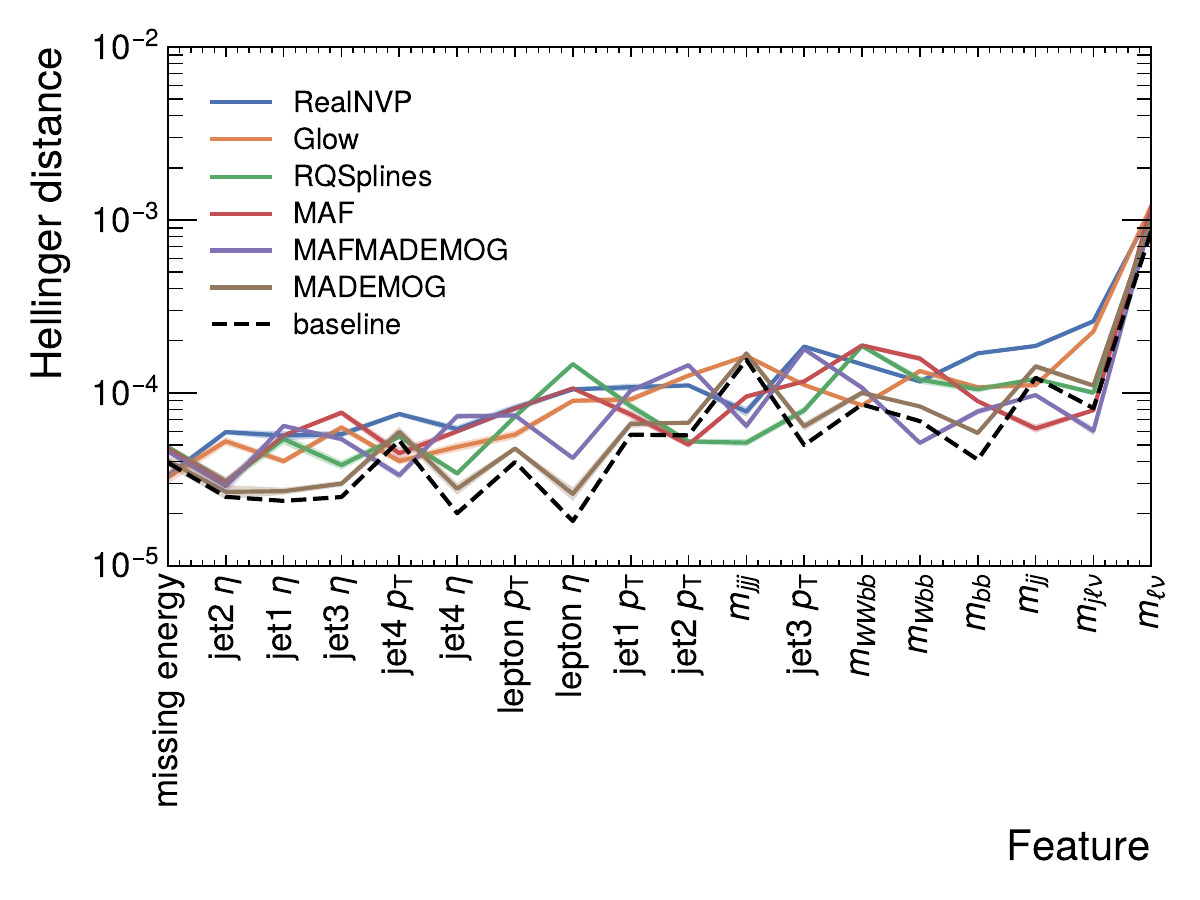}
    \end{minipage}
    \hfill
    \begin{minipage}[t]{0.46\textwidth}
        \centering
        \includegraphics[width=0.99\textwidth]{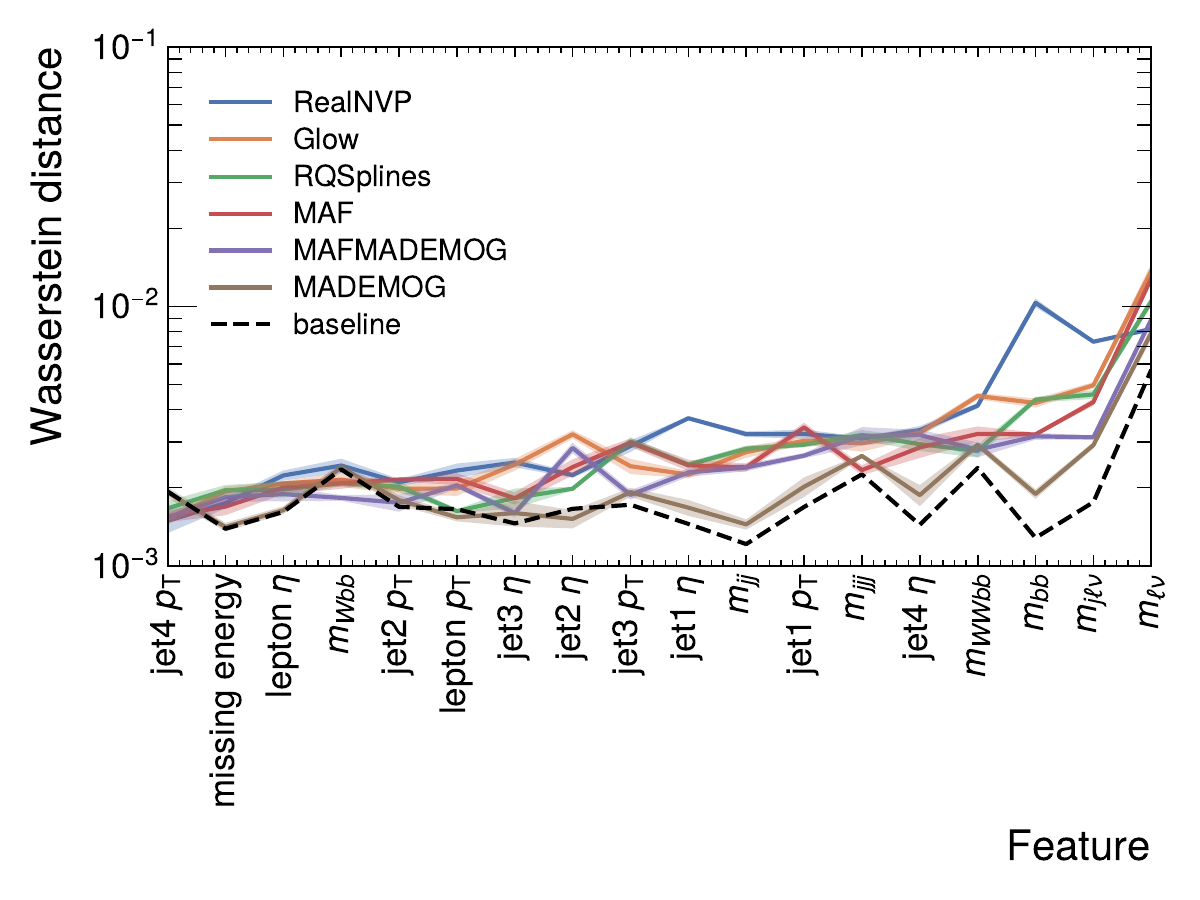}
    \end{minipage}
    \caption{Evaluation of the performance of the generative ML algorithms using different divergence measures. The plots show
        the statistical ``distance'' between the generated and MC distributions. The baseline reference (successful test of `MC vs MC') is shown as the dashed line. The width of the test results is the estimated variance of the test, obtained by resampling, i.e. repeating the ML-generation procedure.}
    \label{fig: distances}
\end{figure}

All these tests are in practice performed as one-dimensional, i.e.
their values are calculated per-feature. The results are shown in Fig. \ref{fig: distances}, with the baseline reference (successful test of `MC vs MC') shown as the dashed line. The width of the presented test results is the  variance of the test, estimated by repeated ML-generation using random sampling of events (i.e. resampling). 
The results show that the similarity of the ML-generated events is consistent across the different tests and is indeed close to the baseline MC events, which is now confirmed
using these objective measures. Comparing the obtained results to  different generative methods presented in the Ref
\cite{Das_2024}, the values of these tests show that the applied methods are in fact quite competitive to other approaches,
confirming the representative value of these studies.

\subsection{Classifier two sample test (C2ST) as a multi-dimensional distribution comparison}\label{sec: c2st}

As stated above, we would like to asses if samples from $P_{\text{ML}}$ and $P_{\text{MC}}$ are equivalent. In other words
we want to accept or reject the hypothesis $P_{\text{ML}}=P_{\text{MC}}$. A relatively new procedure is to use ML tools for this purpose and
train a classifier to distinguish between the two samples. If the classifier is unable to distinguish between the two samples, we can conclude that the two samples
are equivalent \cite{lopezpaz2016revisiting}. The goal is thus to train a `maximally confused' classifier with a 50\% accuracy
and its AUC equal to 0.5. To achieve this, we construct a joint dataset with label 0 for MC events and label 1 for
ML events:
\begin{equation}
    \mathcal{D} = \{ (\vx^\prime_i, 0) \}^n_{i=1} \cup \{ (\vx^{\prime\prime}_i, 1) \}^n_{i=1} = \{ \vx_i, \vy_i \}^{2n}_{i=1} \>,
    \label{eq: c2st}
\end{equation}
where we have assumed same sample sizes for simplicity.
We then shuffle the dataset and split it into a training and a holdout (validation) set. We train a classifier
$f(\vx; \vtheta)\approx p(\vy=1|\vx)$ on the training set, the score of which we can interpret as the probability measure for the label
$\vy=1$ based on inputs $\vx$.
We can then use classification accuracy or AUC on the validation set as a measure of the similarity of the two distributions: Ideally, with 
the two MC and ML distributions (hypotheses) being indistinguishable, this would result in the classification failing to distinguish 
the two, resulting in the AUC being 0.5 (and the accuracy of 50\%), i.e. the classifier would be `ideally confused'.
This procedure is favoured compared to the classical divergence measures we evaluated in the previous section,
because it scales very well to multi-dimensional distributions, both in terms of simplicity and as well as computing requirements.

\begin{figure}[h!]
    \centering
    \begin{minipage}[t]{0.44\textwidth}
        \centering
        \includegraphics[width=0.99\textwidth]{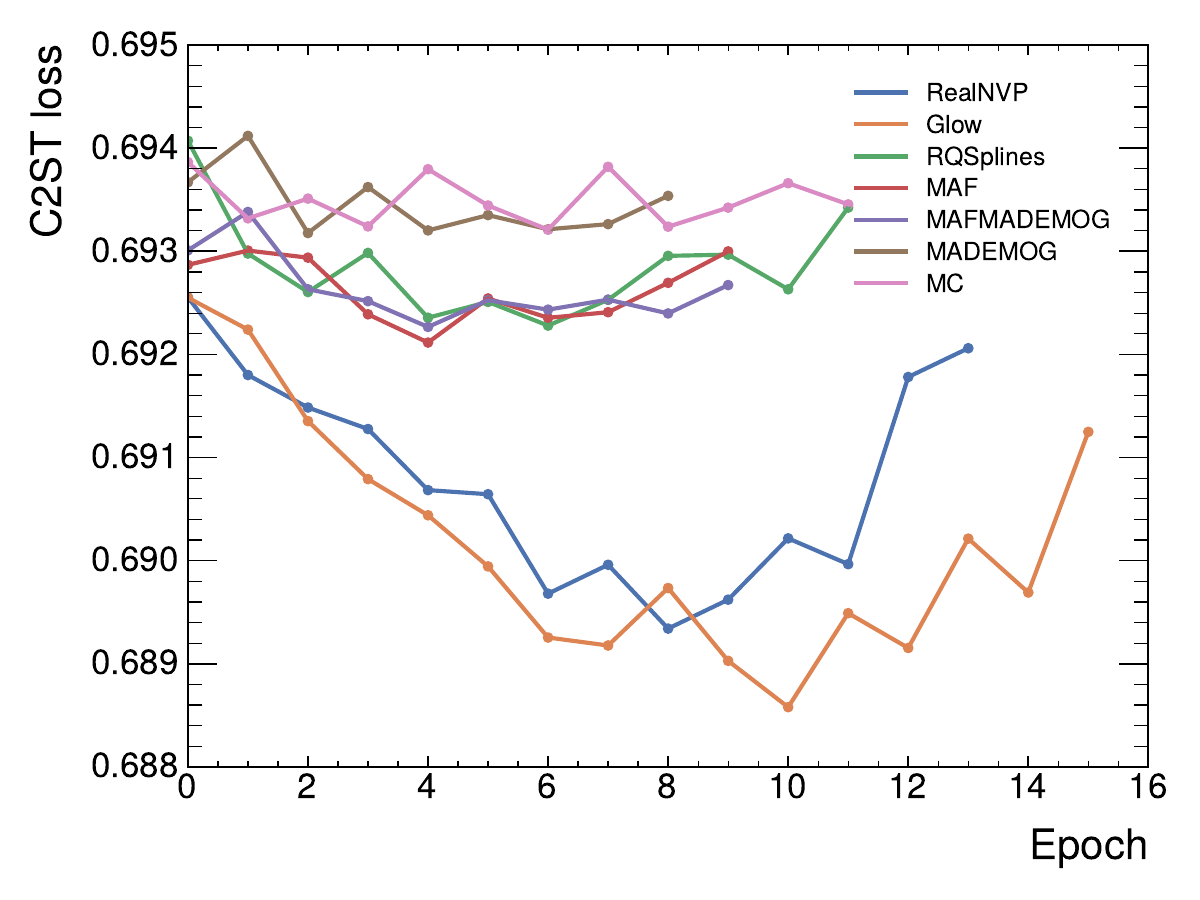}
    \end{minipage}
    \hfill
    \begin{minipage}[t]{0.44\textwidth}
        \centering
        \includegraphics[width=0.99\textwidth]{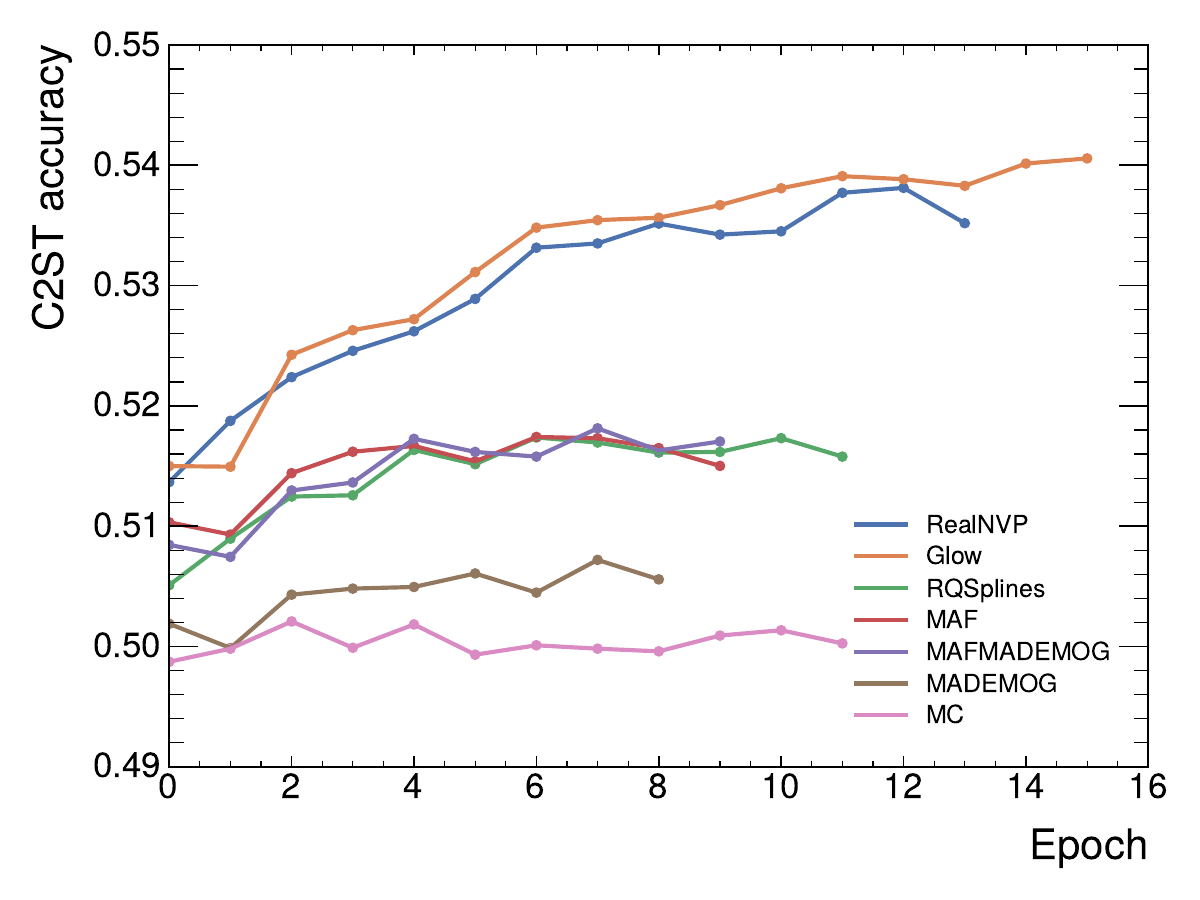}
    \end{minipage}
    \caption{The validation loss and accuracy while training the (C2ST) classifier using a dataset given by Eq. (\ref{eq: c2st}). The curves end at different points due to the early stopping algorithm, used to prevent over-training and optimize the training, as is the common best practice.}
    \label{fig: c2st_train}
\end{figure}

\begin{figure}[h!]
    \centering
    \begin{minipage}[t]{0.44\textwidth}
        \centering
        \includegraphics[width=0.99\textwidth]{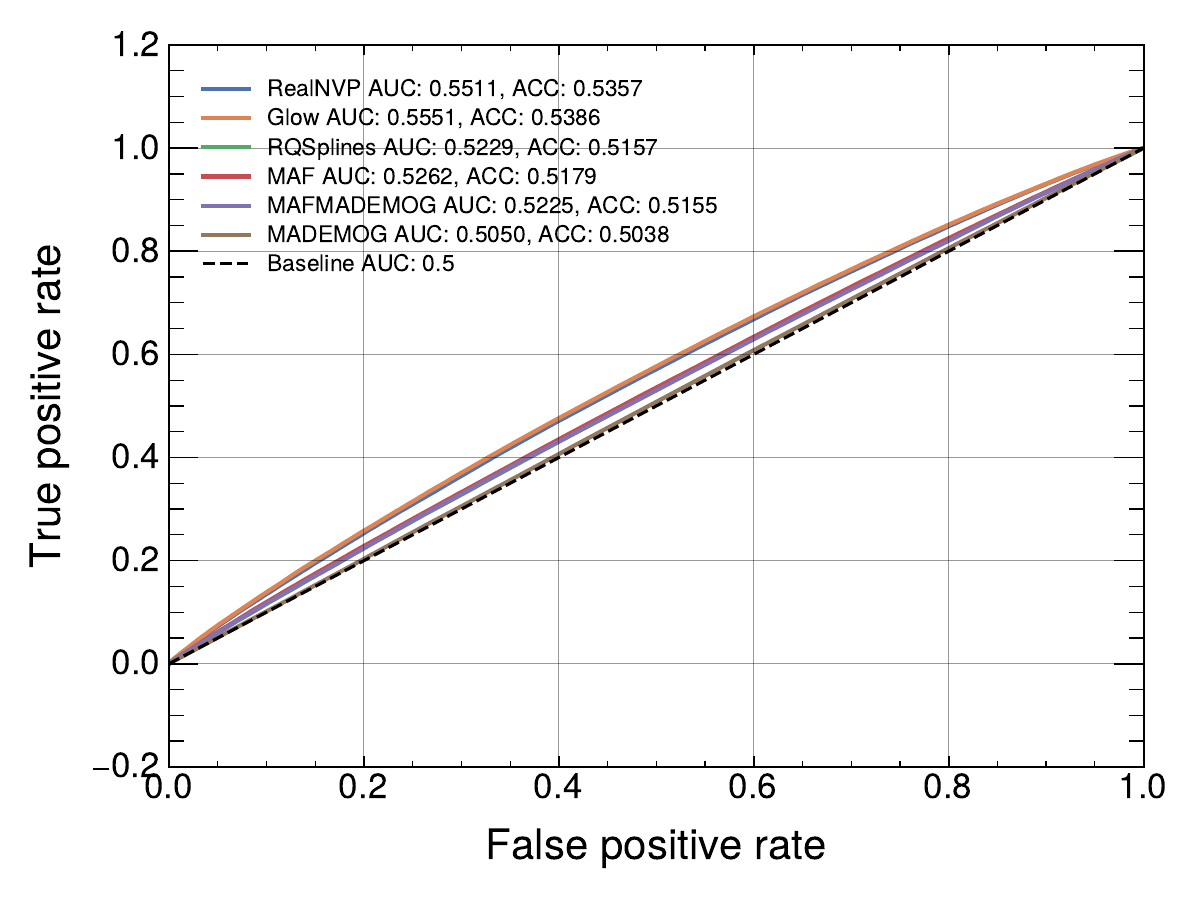}
        \caption{ROC curve for the C2ST classifier for all the presented models. The MADEMOG model is the best performant one, with the
        AUC and accuracy (ACC) very close to 0.5, leading to an `ideally confused' classifier, which (almost) cannot separate the MC from ML-generated events.}
        \label{fig: c2st_roc}
    \end{minipage}
    \hfill
    \begin{minipage}[t]{0.44\textwidth}
        \centering
        \includegraphics[width=0.99\textwidth]{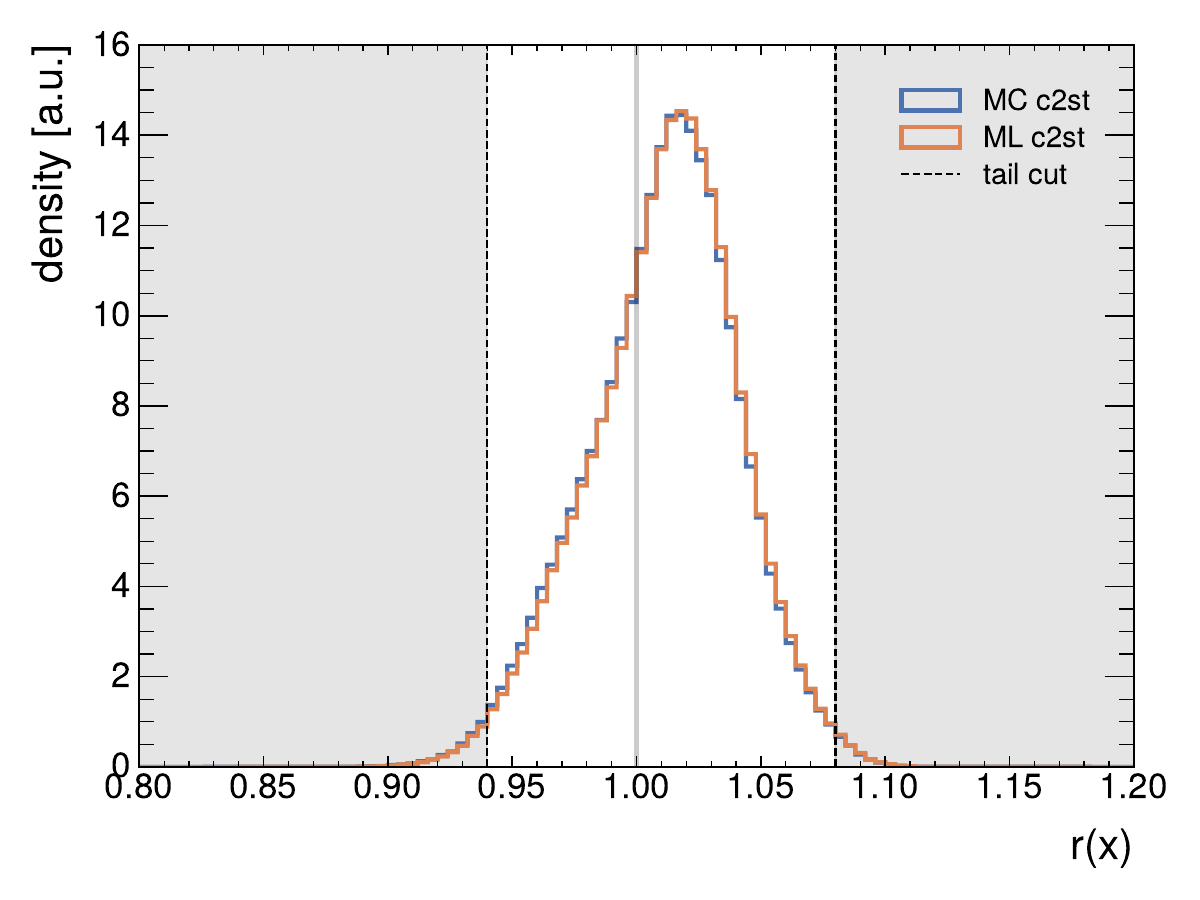}
        \caption{Event distribution of the density ratio $r(\vx)$ for the selected MADEMOG generative ML algorithm. Events in the
        tails of the distribution are expected to have the worst agreement between MC and ML generated events and are further scrutinized
        for differences in Fig. \ref{fig: c2st_fail}}
        \label{fig: c2st_w}
    \end{minipage}
\end{figure}

The obtained training curves, validation loss and validation accuracy for this (C2ST) classifier are shown in Fig. \ref{fig: c2st_train} and the
resulting  ROC curve is shown in
Fig. \ref{fig: c2st_roc}.  The classifier achieves an AUC close to 0.5 and an accuracy close to 50\%, just slightly above,
meaning that it can barely distinguish between the two samples. The same can be said looking at the validation
loss and accuracy while training. We can identify a deviation from the ideally confused classifier, but this is expected due
to the finite training sample size used in training the generative model, which translates to poor modelling of the events
in the tails of the distributions that the classifier can pick up to distinguish between the two samples.

\begin{figure*}
    \centering
    \includegraphics[width=0.86\textwidth]{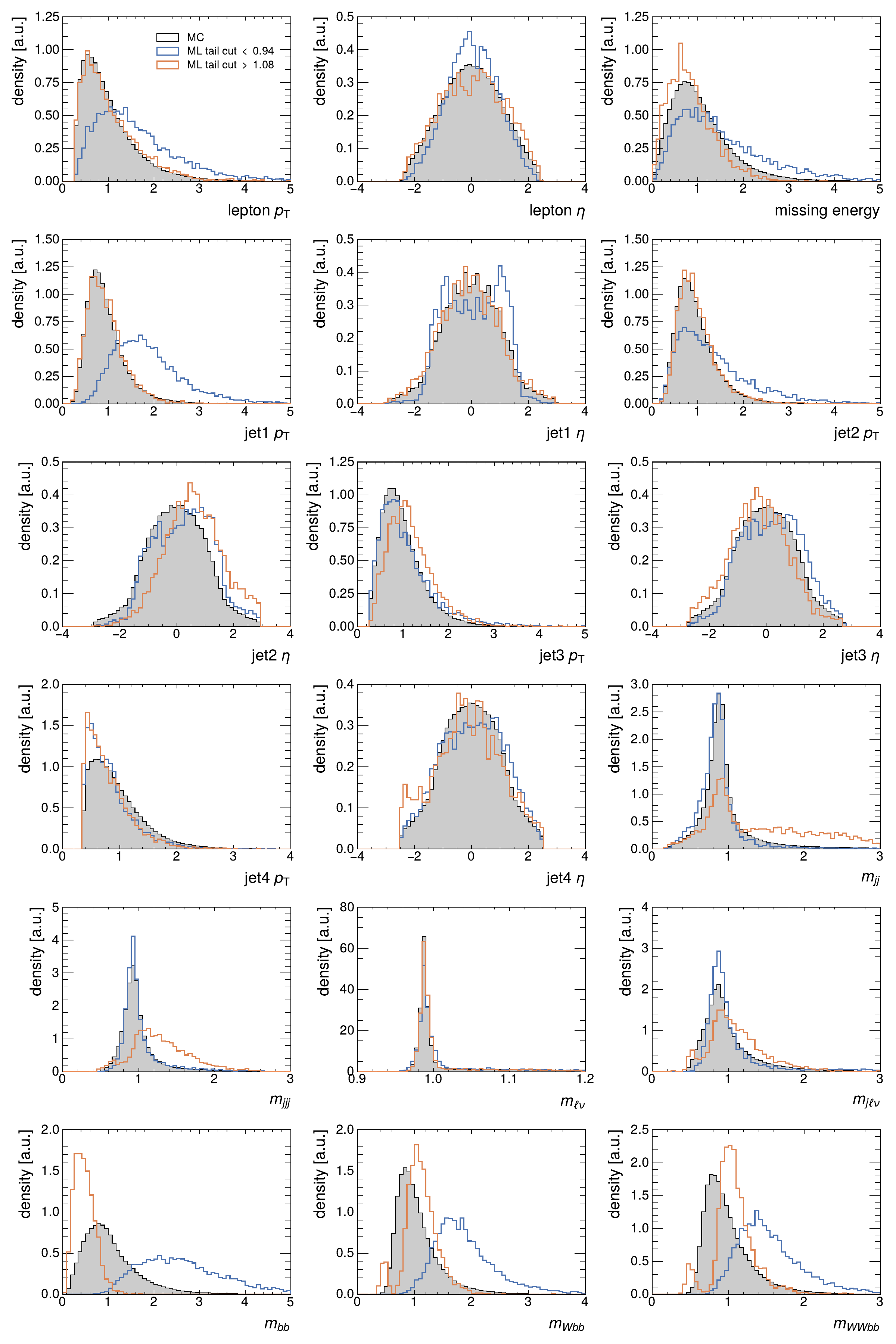}
    \caption{Event distributions of MADEMOG generated events from the tails of the density ratio and comparison to the actual MC shape. One can
    observe that the events that worst match the MC sample deviate especially in the tails of kinematic distributions, as predicted.}
    \label{fig: c2st_fail}
\end{figure*}

To gain an even more detailed insight into the strengths and weaknesses of our generative procedure, we can extract the
density ratio $r(\vx)$  from the same C2ST procedure, details are given in the Appendix \ref{app: c2st}.
This density ratio can then be used to search for the failure modes of the generative ML model \cite{Das_2024}, and thus
has further advantages w.r.t. to the previously described tests, since the events in the tails of the $r(\vx)$ ratio
value can be identified as the ones where the generative ML model fails to reproduce the MC events. The obtained distributions
of ML-generated events w.r.t. the $r(\vx)$ ratio value, along with and independent MC sample for reference, are shown
in Fig. \ref{fig: c2st_w} and the kinematic  distributions corresponding to events in the tails of the $r(\vx)$ ratio distribution
 are shown in Fig. \ref{fig: c2st_fail}. The plots in this Figure again confirm the insight that the discrepancy
between MC and ML events is most pronounced in the tails of the distributions, where the training event count becomes
very low.

\clearpage

\clearpage

\section{ML performance evaluation in a physics analysis}\label{sec: analysis}

In order to provide a relevant quantitative evaluation of the impact of replacing MC-simulated distributions with ML-generated ones
in an analysis environment, a simplified analysis setup was constructed, that
matches a typical analysis of a new physics search in an LHC experiment and its statistical evaluation. In this setup, the HIGGS sample was used
for training the generative ML algorithms for the background modelling, with subsets being reserved for validation and testing, as described in 
the Appendix \ref{app: splits}. In the following studies, the MADEMOG model
trained on the HIGGS dataset was used to generate the new background samples as described in the previous sections.

In order to emulate a typical analysis procedure in an LHC experiment, a further step was introduced, namely the 
generative ML algorithm was trained and applied on background events in the HIGGS sample without any kinematic cuts on its observables,
while the statistical analysis studies were done after an additional cut on a new ML classifier, trained to separate the signal from background events. In the training stage of this classifier, the MC signal events from the HIGGS sample and the ML-generated background events were used in equal fractions. This new classifier is a standard (simple) feed-forward neural network with 3 layers of 512 neurons each, producing an output score using a sigmoid function.

\begin{figure}[h!]
    \centering
    \begin{minipage}[t]{0.4\textwidth}
        \centering
        \includegraphics[width=0.99\textwidth]{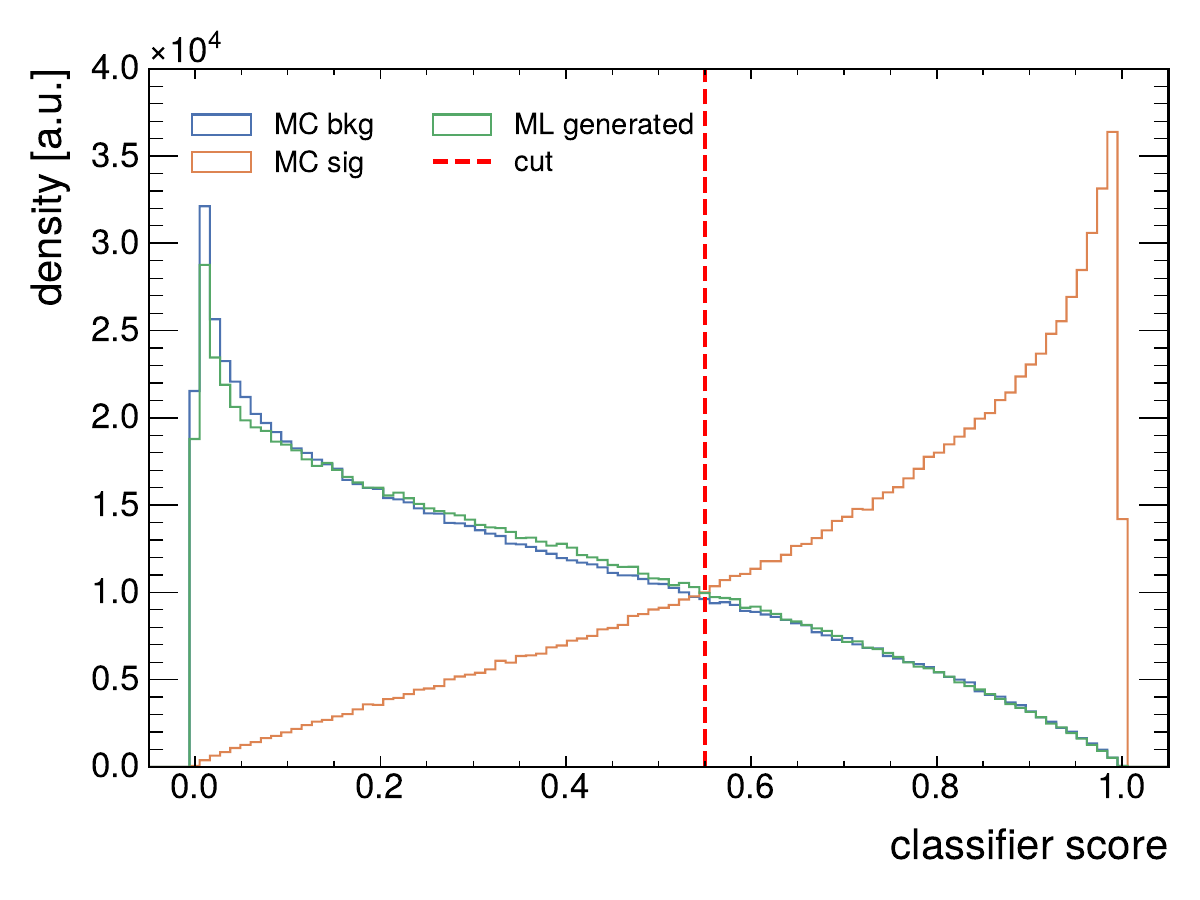}
        \caption{MC signal, MC background and ML background score, as given by the selection classifier, described in the text. A
        good signal and background separation is achieved, the score distribution in the region of interest shows a good agreement between
        the MC and ML-generated background samples.}
        \label{fig: class_out}
    \end{minipage}
    \hfill
    \begin{minipage}[t]{0.4\textwidth}
        \centering
        \includegraphics[width=0.99\textwidth]{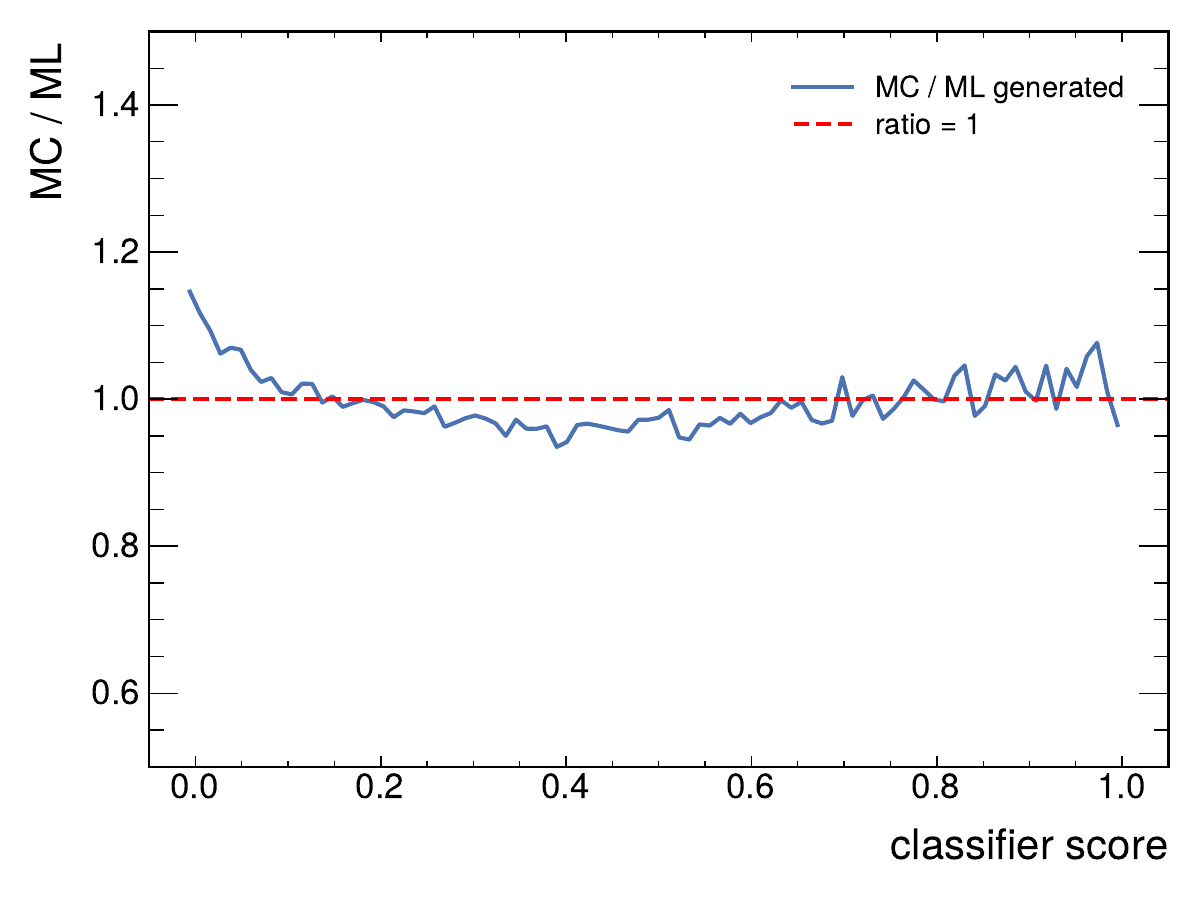}
        \caption{The ratio of classifier scores for background MC and generated ML events is shown; moderate deviations from the ideal value of 
        one are visible, confirming good performance of ML-sampling and the adequacy of the trained selection classifier.}
        \label{fig: miss_class}
    \end{minipage}
\end{figure}

The performance of this classifier is shown in  Fig. \ref{fig: class_out}, with the final signal selection cut at the value of the classifier (output) score of 0.55.
The visual agreement between the classifier scores of the independent MC-simulated and  ML-generated samples is quite good in the finally selected samples, with the discrepancy between the two on the level of up to 10\% and is shown as a ratio between the two score distributions in Fig. \ref{fig: miss_class} for clarity.

This additional selection step is introduced to match the standard analysis approach, where first a relatively simple baseline selection is applied on the
data and MC samples as the first stage of analysis optimisation, at which the agreement of kinematics
of the signal and background predictions is generally good enough and the statistics of the MC samples is sufficiently large to train a
ML classifier (as well as a generative algorithm).  A representative LHC analysis then uses this ``ultimate'' classifier to perform a final data selection to achieve an optimal signal-to-background separation in terms of discovery significance. However, the MC-simulated samples in new physics searches tend to have quite low statistics after this final selection, which is then too low to be usable to effectively train a generative ML procedure. Consequently, as also done in this paper, the training of the generative ML algorithm needs to happen before this ultimate selection and the impact of the classifier cut on the agreement between the MC and ML-generated samples needs to remain adequate after the selection step. At this point it is worth stressing that one needs to be careful in training this classifier that the discrepancies between the MC and ML-generated samples does not propagate (and even amplify) in the final selection, thus in a real analysis careful comparison checks should be made to validate such a classifier - in practice, much the same as is already regularly done in the LHC experiment analyses when checking for discrepancies between the real data and  the standard MC simulation and how it propagates through a ML-based classifier.

To sum up, implementing the required analysis procedure involving a classifier cut on the ML generated samples is also an innovative
implementation of the study presented in this paper.
The procedure can be summarised as follows:
\begin{enumerate}
    \item train a generative model on  MC background samples and generate large datasets of new events,
    \item use the MC and ML-generated events to train a neural network classifier and validate the procedure,
    \item apply the final selection on the ML-generated samples using this classifier,
    \item normalise the ML-generated samples to match the integrated luminosity of the data sample after the final selection,
    \item perform the statistical analysis.
\end{enumerate}

\begin{figure*}
    \centering
    \includegraphics[width=0.85\textwidth]{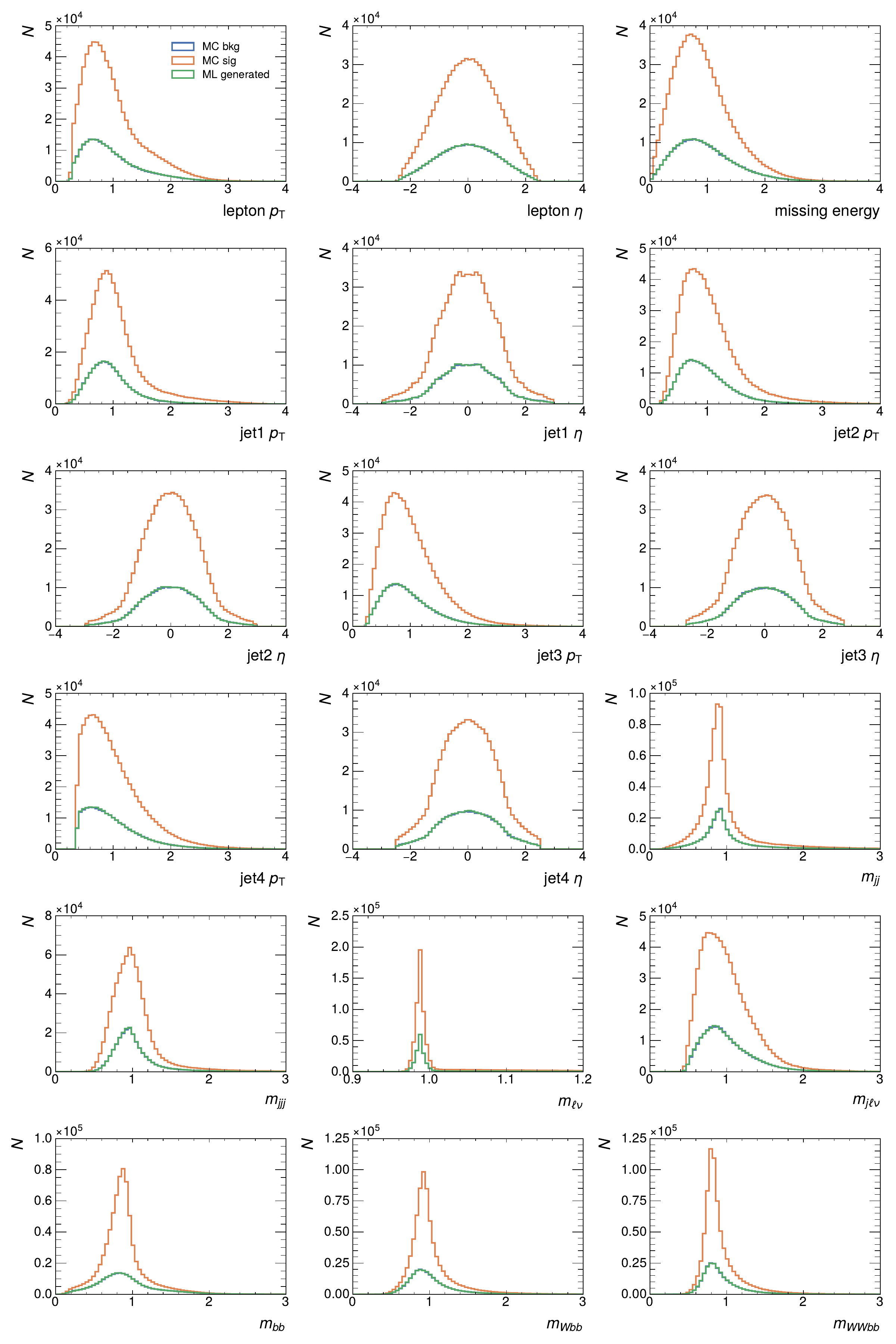}
    \caption{Kinematic distributions of ML (green) and MC-generated (blue) background events after the cut on
        the classifier from Fig. \ref{fig: class_out} at the classifier score value of 0.55. The MC-generated signal events are
        also shown for completeness (orange).}
    \label{fig: cut_dists}
\end{figure*}

By introducing a (ML) classifier cut, one can also see how well the multi-dimensional generation of correlated variables
is in fact performing. As a further insight from a different perspective: without subsequent (final) kinematic cuts, each relevant kinematic
distribution can just be modelled (smoothed) independently, without a need for a generative procedure.
As one can observe in Fig. \ref{fig: cut_dists}, a good agreement is preserved between the MC and ML samples,
demonstrating that the variable correlations are adequately modelled and that the generative ML approach is thus
preforming well enough for such an analysis approach.

In this study the background yield is chosen and then the integrated luminosity with a given cross section is calculated. The evaluation was done
in the background range $B \in [10^4,10^6]$, corresponding to a luminosity increase up to three orders of magnitude
($L \in [10~\text{fb}^{-1}, 1000 ~\text{fb}^{-1}]$). The signal yield ($S$) fraction with respect to the background(B), i.e. $S/B$, was set to have
a value of $S/B$=5\% of the background yield in the performed tests. An inclusive relative systematic uncertainty ($\beta$), which in a real
analysis would comprise both theoretical and experimental uncertainties, was set to $\beta=$10\%, with the relevant subset of the studies performed also at two further values of  $\beta=$5\% and $\beta=$20\% to cover the representative range of typical LHC analyses. This systematic uncertainty  excludes the systematic uncertainty due to finite simulated sample statistics, which is the focus of this study and is being treated separately
and has the expected total Poisson-like $\sim \sqrt{N}$ dependence on the number of ML-generated events for the  background
prediction, translating into the appropriate multinomial uncertainties in the normalised (and re-scaled) binned distributions.

The data prediction was constructed by adding the MC-simulated samples of background and signal with a chosen signal content,
giving a so-called Asimov dataset, which perfectly matches the MC-samples and is often used in the performance validation
of a statistical analysis in an LHC experiment. For the purpose of the studies in this paper, the MC-simulated background
is then replaced with the corresponding ML-generated samples, the production of which was previously described.

The signal prediction
is retained as the MC one because in LHC analyses, the signal simulation, which generally needs a comparatively small number
of generated events, is typically not as much of an issue as the background. The final analysis selection is centred around
the signal prediction, retaining most of the signal statistics while only selecting the low-statistics tails of the background
samples.

This setup aims to provide a good measure of the quality of the ML approach in a LHC analysis environment - ideally
the results after this replacement would still perfectly match the results of the original MC setup on the Asimov dataset,
however, an agreement well within the joint statistical and systematic uncertainty is still deemed as acceptable, and thus a
successful  implementation of the generative ML procedure.

The achieved statistical agreement certainly depends on the integrated luminosity, i.e. data statistics, assuming a constant presence
of a fixed total systematic uncertainty. The dependence on integrated luminosity then translates into a requirement on the quality of ML-generated samples,
assuming that an arbitrarily large statistics (size) of these ML-generated samples is easily achievable.

With the aim of closely matching a typical statistical analysis, as done in LHC experiments, the HistFactory model from the
{\tt pyhf} \cite{pyhf_joss, pyhf} statistical tool was used, and different standard procedures of evaluation of the agreement
between data and simulation predictions were implemented (profile likelihood calculation, upper limit estimation, background-only
hypothesis probability etc.). The statistical analysis was performed using two different possibilities for choosing
the optimal variable, resulting in a binned distribution w.r.t. the $m_{bb}$ in the first case and classifier score
in the second case, which aims to give an optimal separation between the shape of the background and signal predictions.
In the statistical analysis, the signal presence is evaluated by using a scaling factor $\mu$ (signal strength) of the
predicted signal normalisation. Statistical error scale factors $\boldsymbol{\gamma}=\{\gamma_i\}$  are used to model
the uncertainty in the bins due to the limited statistics of (ML-)simulated samples. By using the fast ML event generation,
one aims to push this uncertainty to negligible values, as is indeed done in the study presented below. As already stated,
the simulated predictions are given an additional  relative uncertainty of $\beta=$10\% in each bin of the distribution as $\boldsymbol{\beta}=\{\beta_i\}$ to model the systematic
uncertainty contribution on the predicted background shape and normalisation. In the likelihood calculation, the parameters $\boldsymbol{\gamma}$
and $\boldsymbol{\beta}$ are modelled as nuisance parameters in addition to $\mu$ as the main parameter of the likelihood fit.

\begin{figure}[ht!]
    \centering
    \includegraphics[width=0.49\textwidth]{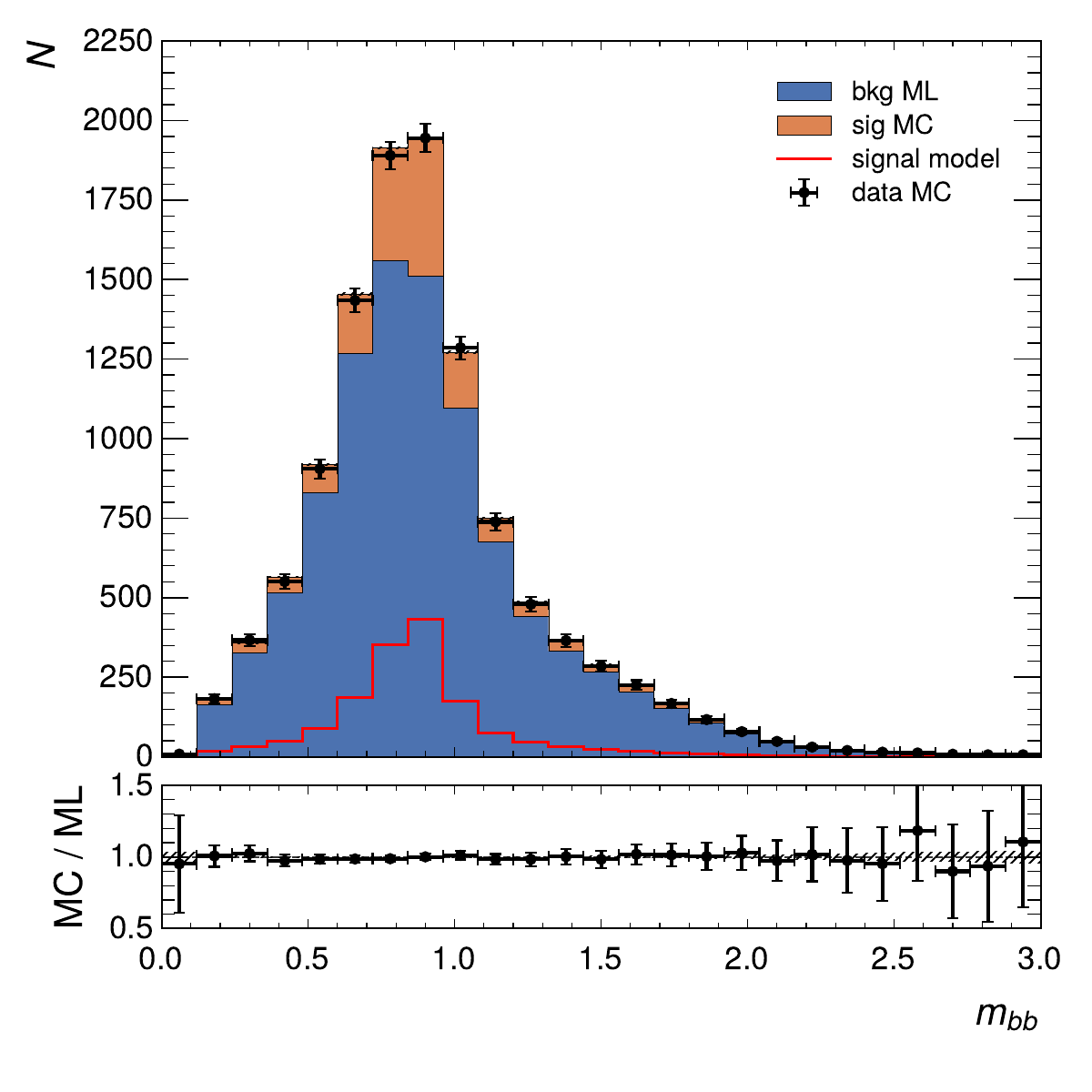}
    \includegraphics[width=0.49\textwidth]{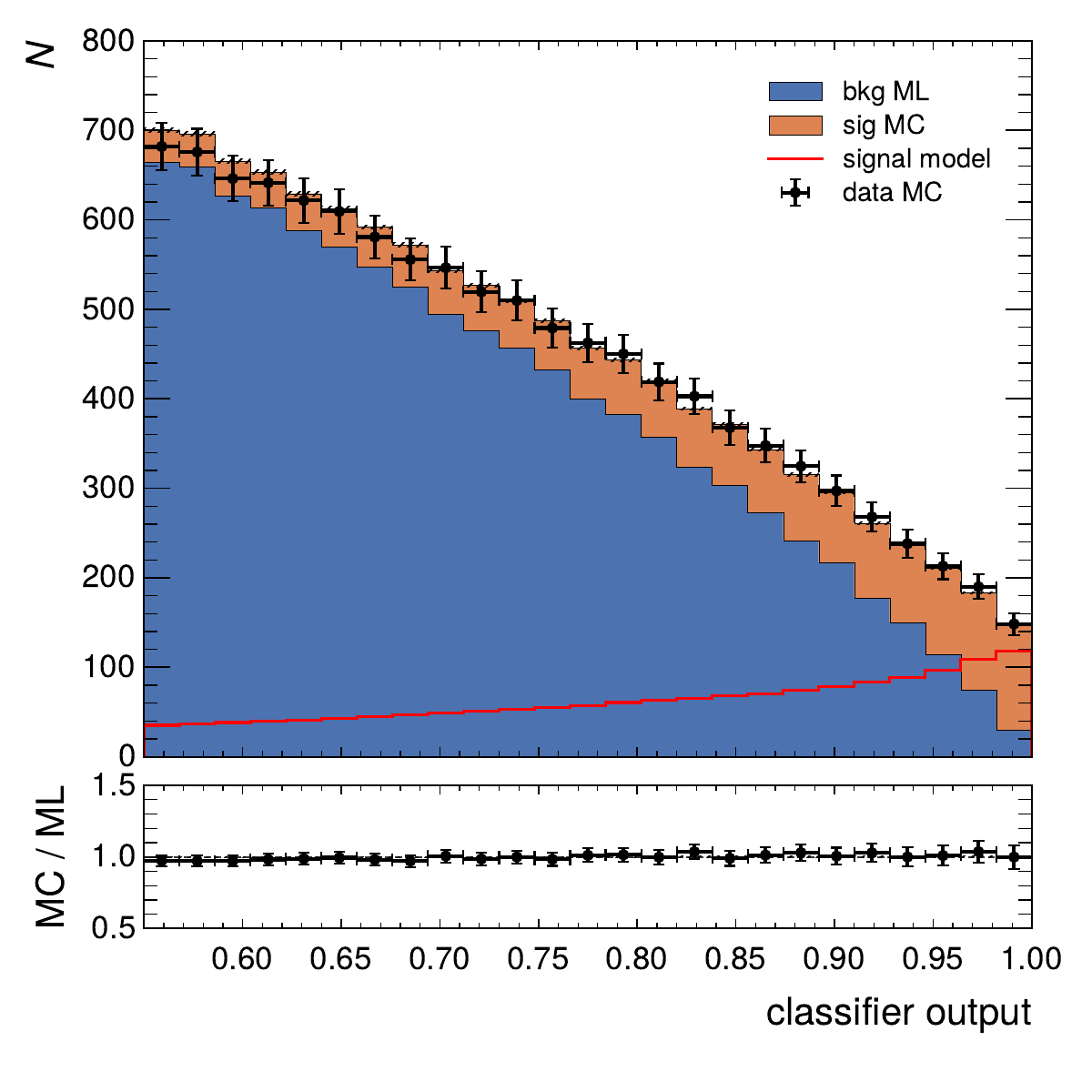}
    \caption{The $m_{bb}$ and classifier score distributions after the classifier cut. The ``MC data'' (crosses) represents the Asimov
        data set composed from MC signal and background prediction and matches quite well the combined MC signal (orange) and
        ML background (blue) prediction. The red histogram shows the separate shape of the signal prediction.}
    \label{fig: hists_after_cut_rescale}
\end{figure}

The binned distributions of samples used in this statistical analysis are shown in Fig. \ref{fig: hists_after_cut_rescale}
for the two relevant observables. One can see that the agreement between the Asimov data and the simulation prediction
using the ML-generated background sample seems to be good, which is an encouraging starting point for a detailed analysis.

In more detail (see e.g. Ref \cite{cowan2011asymptotic}), the likelihood is defined as the product over all bins of the Poisson probability to observe \(N_i\)
events in a particular bin \(i\):
\begin{equation}
    \begin{aligned}
        L_{\text{phys}}(\text{data} \, | \, \mu) = L_{\text{phys}}(\mu)
        = \prod_{i \in \text{bins}} \text{Pois}(N_i \, | \, \mu S_i + B_i)
        = \prod_{i \in \text{bins}} \frac{(\mu S_i + B_i)^{N_i}}{N_i!} e^{-(\mu S_i + B_i)},
    \end{aligned}
\end{equation}
where \(S_i\) and \(B_i\) are the expected signal and background yields, respectively. As already stated, the main parameter
of interest is the signal strength, denoted as \(\mu\).


The systematic uncertainties are included in the likelihood via a set of nuisance parameters (NP), denoted as
\(\boldsymbol{\theta} = (\boldsymbol{\beta}, \boldsymbol{\gamma})\), that modify the expected background yield, i.e.
\(\{B_i\} \to \{B_i(\boldsymbol{\theta})\}\). The overall relative systematic
uncertainties $\beta_i$ can be encoded into Gaussian functions and subsequently into an auxiliary
likelihood function \(L_{\text{aux}}(\boldsymbol{\beta})\), while the uncertainties on the background predictions due to the limited
number of simulated events are accounted for in the likelihood function considering Poisson terms
\begin{equation}
    L_{\text{stat}}(\boldsymbol{\gamma}) = \prod_{i \in \text{bins}} \frac{(\gamma_i B_i)^{B_i} e^{- \gamma_i B_i}}{\Gamma(B_i)},
\end{equation}
where \(\Gamma\) is the gamma function.

The profile likelihood function can finally be defined as a product of three likelihoods
\begin{equation}
    L(\mu, \boldsymbol{\theta}) = L_{\text{phys}}(\mu) \cdot L_{\text{aux}}(\boldsymbol{\beta})
    \cdot L_{\text{stat}}(\boldsymbol{\gamma}).
\end{equation}
A likelihood fit using this likelihood is then performed to determine the value of $\mu$ and its uncertainty,
as well as the nuisance parameters. The estimates of $\mu$ and  $\boldsymbol{\theta}$  are obtained as the values of
the parameters that maximise the likelihood function \(L(\mu, \boldsymbol{\theta})\) or, equivalently,
minimise \(-\!\ln L(\mu, \boldsymbol{\theta})\).

This profile likelihood function is then used to construct the test statistics commonly used in the LHC analyses,
nicely described in Ref. \cite{cowan2011asymptotic}, namely the $t_{\mu},\tilde{t}_{\mu},q_{\mu}$ and $\tilde{q}_{\mu}$. 
In particular, the compatibility of the data with the background-only hypothesis (with \(\mu = 0\)) is performed using the 
special test statistic $q_0=\tilde{t}_{\mu=0}$, constructed for the positive signal discovery estimation.

The test statistic is
used to calculate a $p$-value ($p_0$) that quantifies the agreement of the data with the \(\mu = 0\) background-only hypothesis as:
\begin{equation}
    p_0 = \int_{q_0^{\text{obs}}}^{\infty} f(q_0 \, | \, \mu=0) \, \text{d}q_0,
\end{equation}
where \(q_0^{\text{obs}}\) is the value of test statistics observed in the data, and \(f(q_{0} \, | \, \mu=0)\) is the
probability density function of the test statistic \(q_{0}\) under the \(\mu=0\) signal strength assumption.
The \(p\)-value can be expressed in units of Gaussian standard deviations \(Z = \Phi^{-1}(1 - p_0)\), where \(\Phi^{-1}\)
is the inverse Gaussian CDF. The rejection of the background-only hypothesis ($\mu=0$) with a significance of at least
\(Z = 5\) (corresponding to \(p_0 = 2.87 \times 10^{-7}\)) is considered a discovery.

The upper limits on the signal strength are derived at a $\text{CL}=95\%$ confidence
level using the \(\text{CL}_\text{s}\) method~\cite{CLsRead}, for which both the signal plus background, \(p_{S+B}\),
and background-only, \(p_B\), values need to be calculated. The final confidence level \(\text{CL}_\text{s}\) is
computed as the ratio $\text{CL}_\text{s} \equiv \frac{p_{S+B}}{1-p_B}$, which excludes a signal hypothesis at
$\text{CL}=95\%$ when giving a value below  5{\%}.

\subsection{Results of likelihood tests}

\begin{figure}[ht!]
    \centering
    \includegraphics[width=0.47\textwidth]{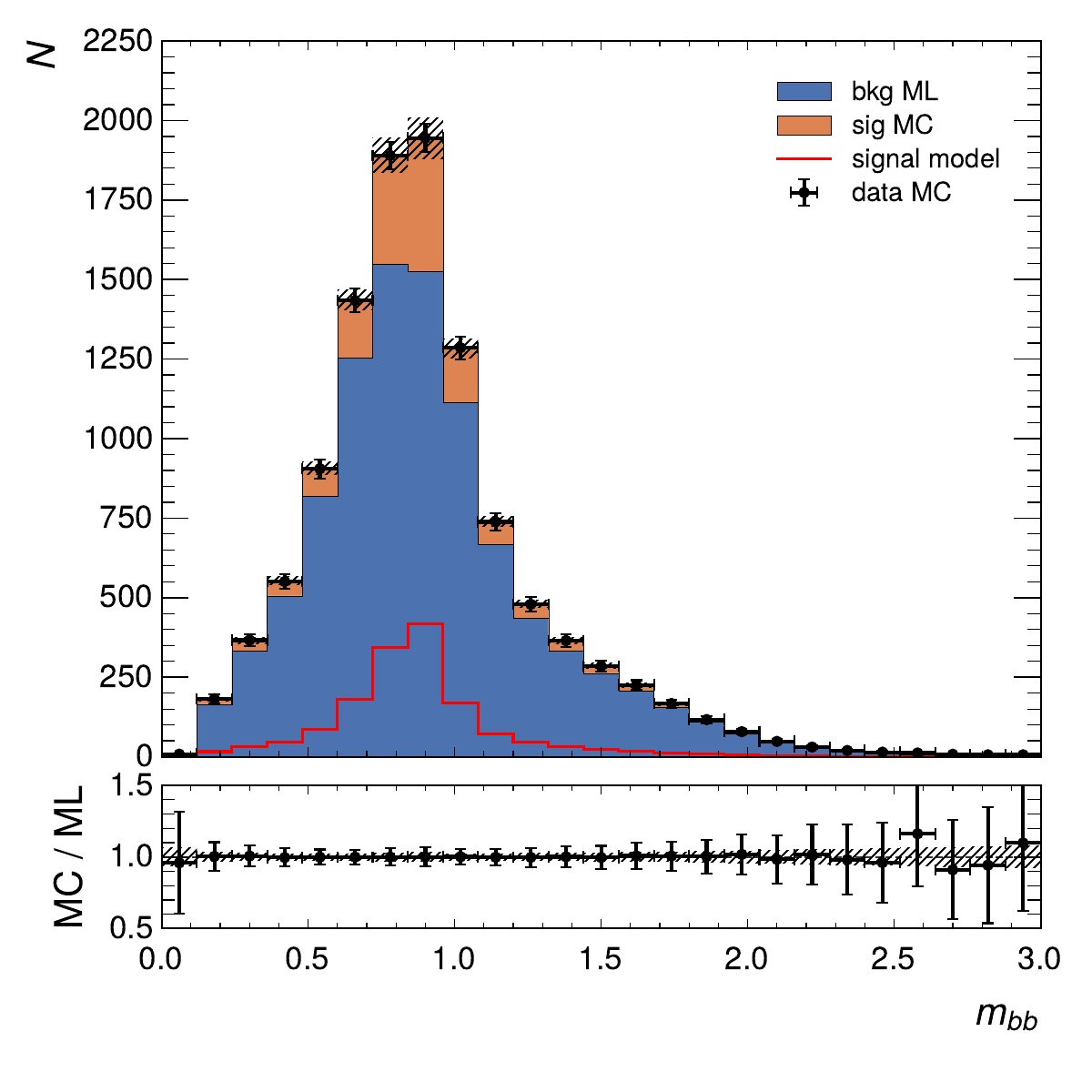}
    \includegraphics[width=0.47\textwidth]{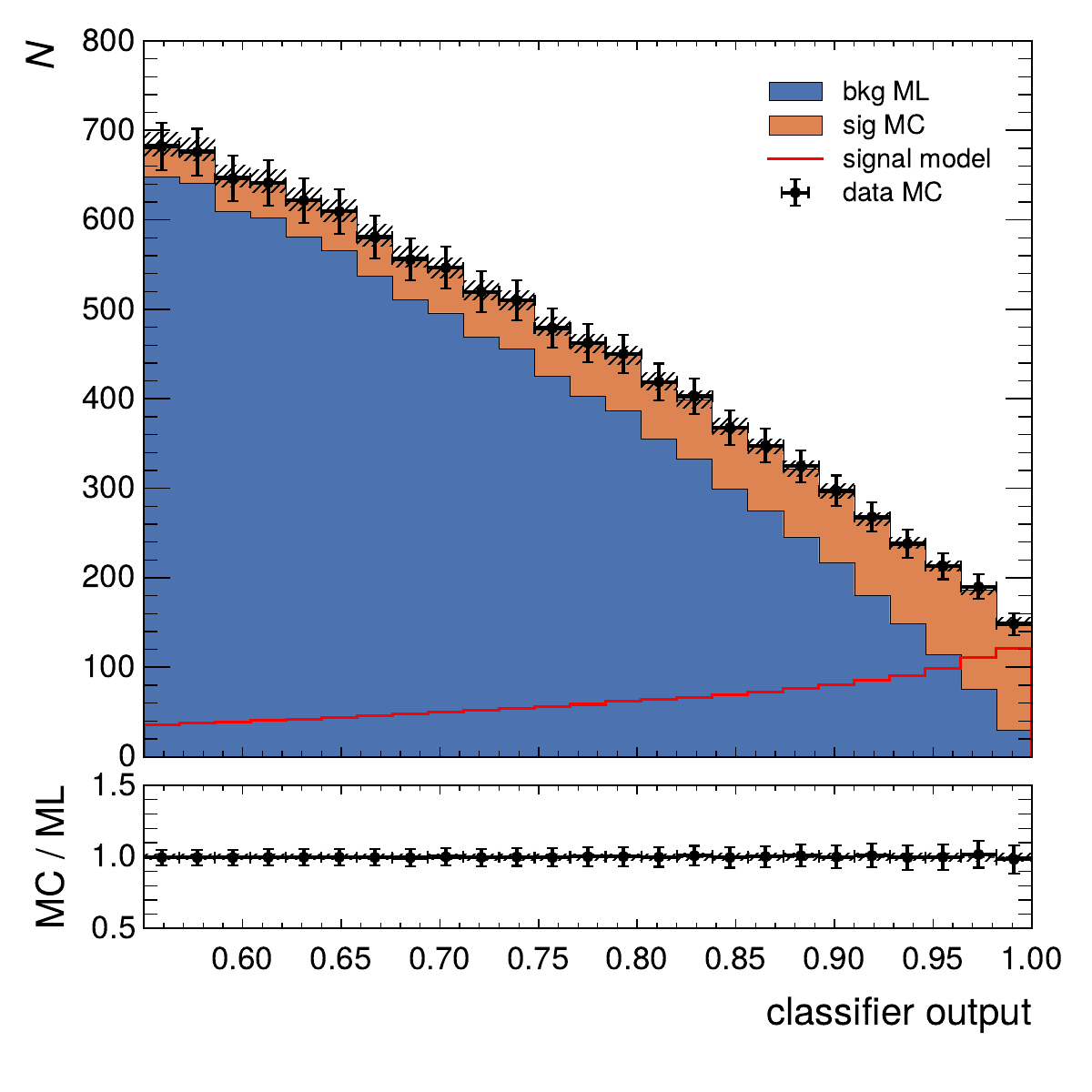}
    \caption{Post-fit distributions for profile likelihood fits to the $m_{bb}$ or the classifier score.  The ``MC data'' (crosses) represents the Asimov data set composed from MC signal and background prediction and matches almost perfectly the  combined MC signal (orange) and
       fitted ML background (blue) prediction. The hatched area on the MC+ML histograms shows the full systematic uncertainty estimated in the fit. The red histogram shows the separate shape of the signal prediction.}
    \label{fig: hists_postfit}
\end{figure}

Examples of event distributions obtained after the profile likelihood fit to the Asimov data, as described above, are shown
in Fig. \ref{fig: hists_postfit} for the two different variable selections. One can observe a very nice agreement
between the fitted prediction and Asimov data.

\begin{figure}[ht!]
    \centering
    \includegraphics[width=0.47\textwidth]{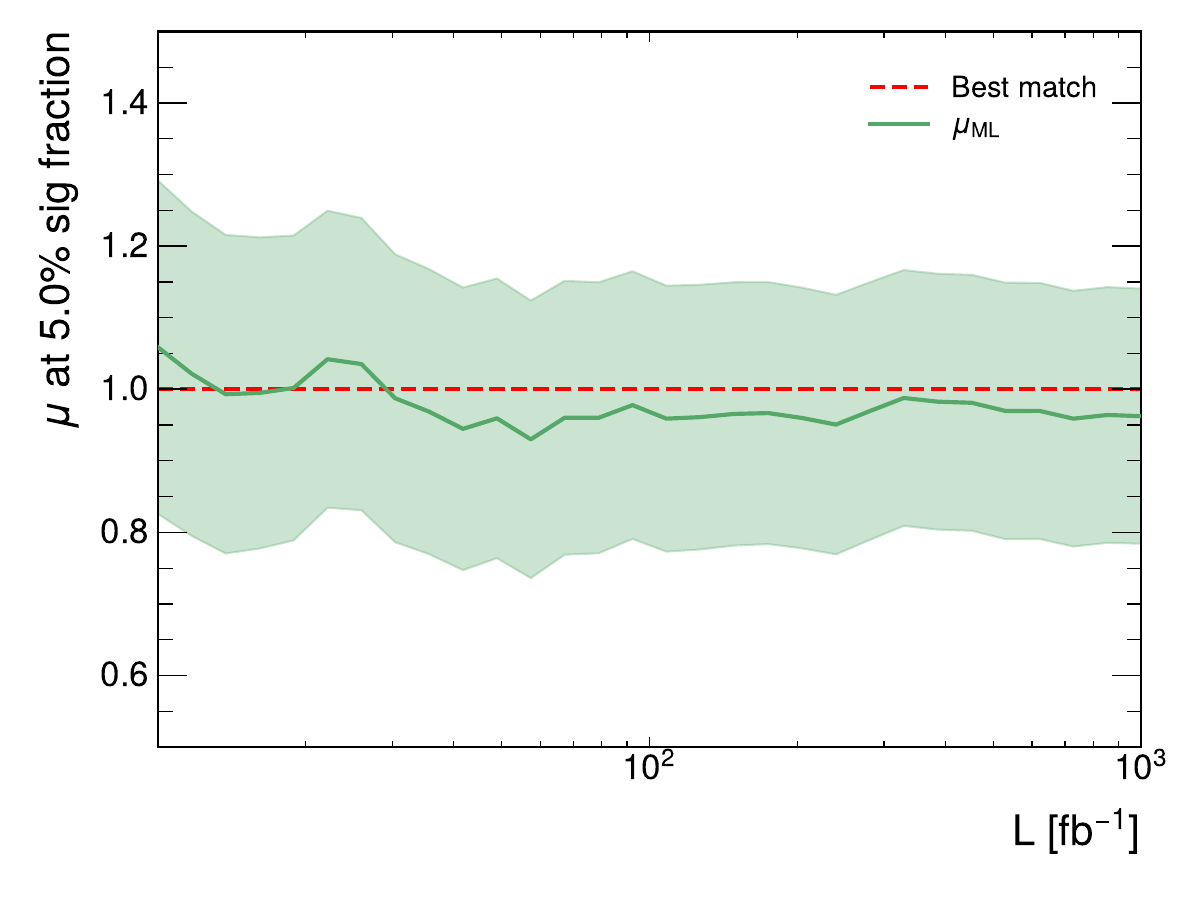}
    \includegraphics[width=0.47\textwidth]{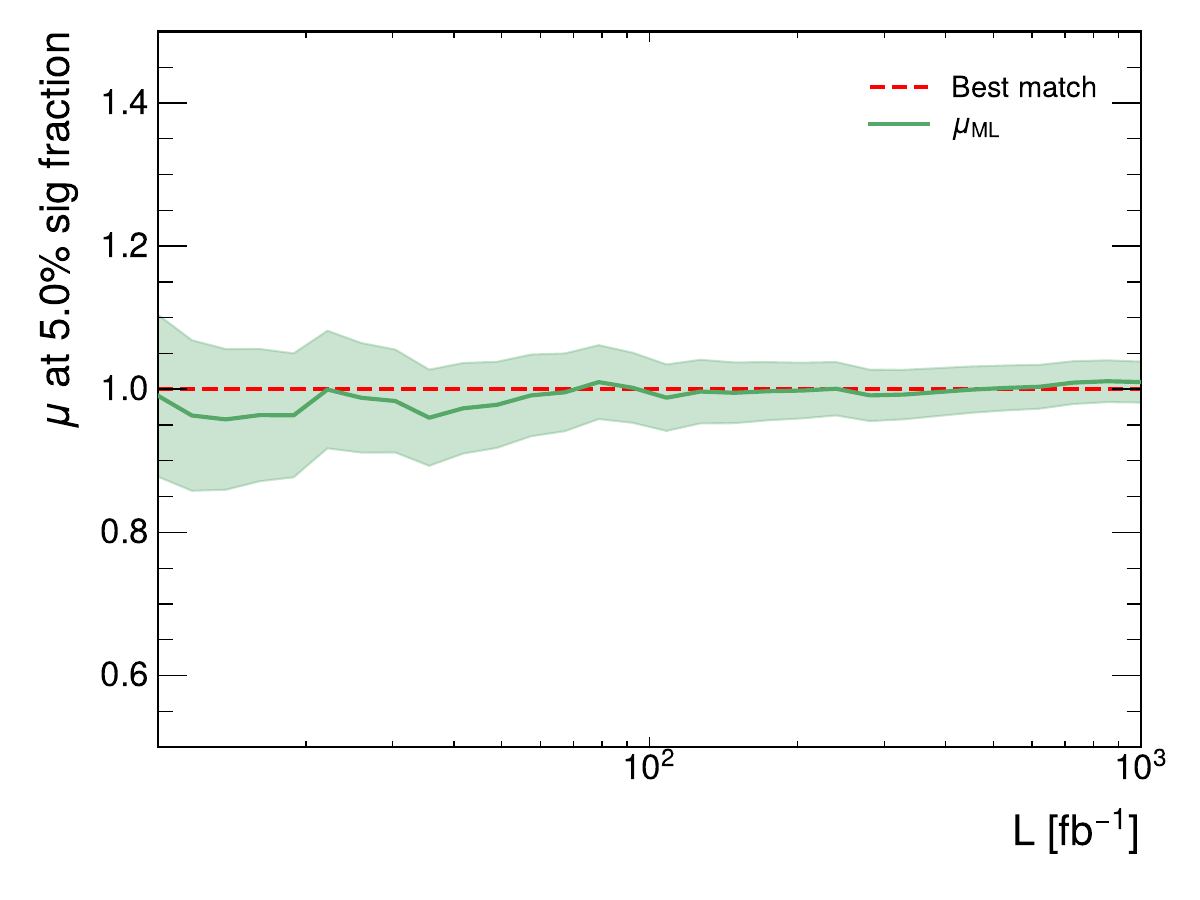}
    \caption{Fitted parameter of interest, the signal scale $\mu$, as a function of integrated luminosity $L$ for ML-generated background is shown for the profile likelihood fit using the
        $m_{bb}$ variable (left) or the classifier score (right). The statistical error on the estimated $\mu$ value is shown with the error band around the central line of best values obtained from the likelihood fit. It is evident that the statistical estimation quite reliably reproduces the expected value of $\mu=1$ for a (small) injected signal at $S/B=5\%$ fraction of the background.}
    \label{fig: mus_injection}
\end{figure}

Fig. \ref{fig: mus_injection} shows how well the expected value of the signal strength $\mu$ is reproduced in the statistical
evaluation. The estimated value of $\mu$ with its uncertainty is
shown w.r.t. the increase in integrated luminosity.  It is evident that the statistical estimation quite reliably reproduces the expected value of $\mu = 1$
for a (small) injected signal at the fraction  $S/B = 5\%$  of the background. One can observe a small bias, which of course persists with increasing integrated luminosity for the
ML-generated sample. The (biased) values are well within the uncertainty, but it is of course clear that background mis-modelling,
present when using ML-generated background, leads to biases and possible  sensitivity loss in an analysis with relatively tiny
signal presence. This evaluated discrepancy between the injected and estimated signal in the final analysis fit can also be interpreted as the presence
of a spurious signal \cite{spur}, which is another common test in the LHC analyses.

\begin{figure}[ht!]
    \begin{minipage}[t]{0.49\textwidth}
        \centering
        \includegraphics[width=0.99\textwidth]{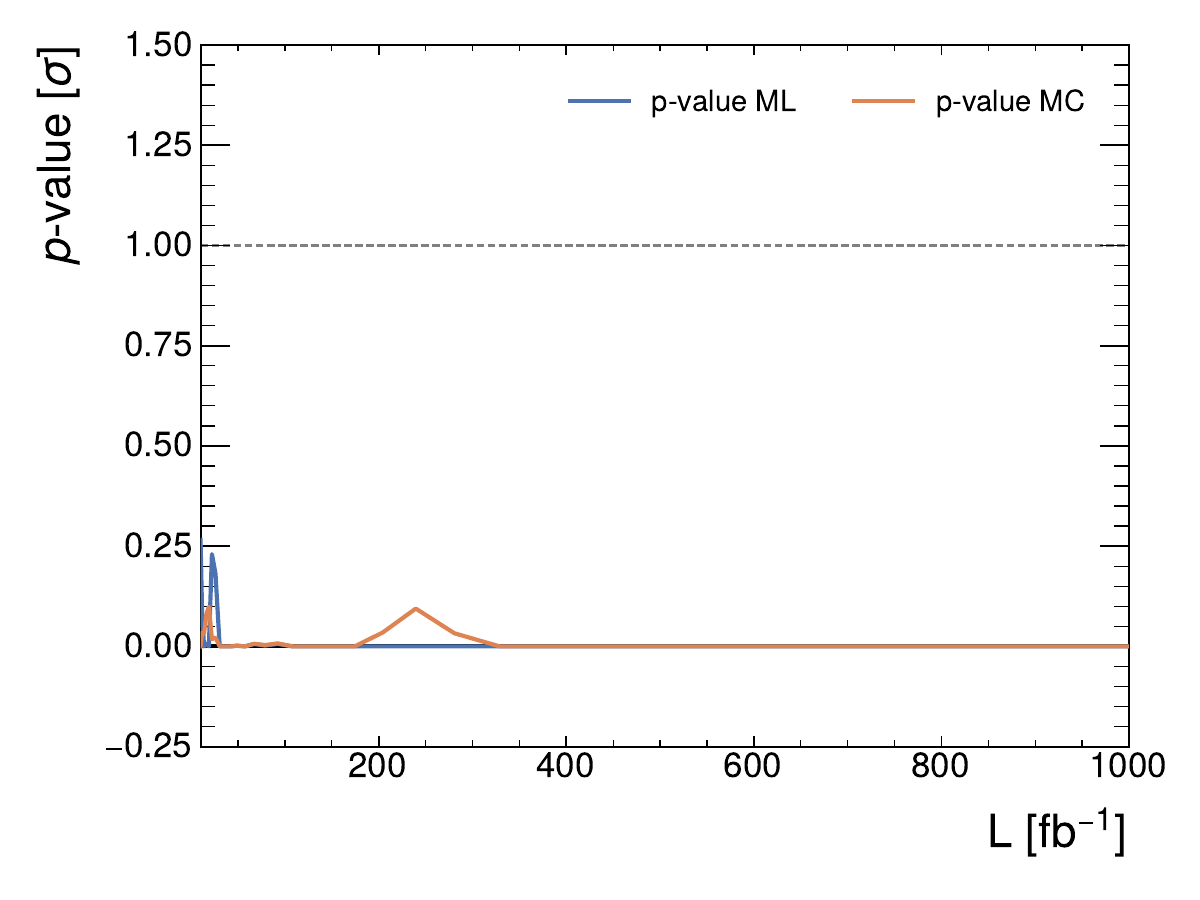}
    \end{minipage}
    \hfill
    \begin{minipage}[t]{0.49\textwidth}
        \centering
        \includegraphics[width=0.99\textwidth]{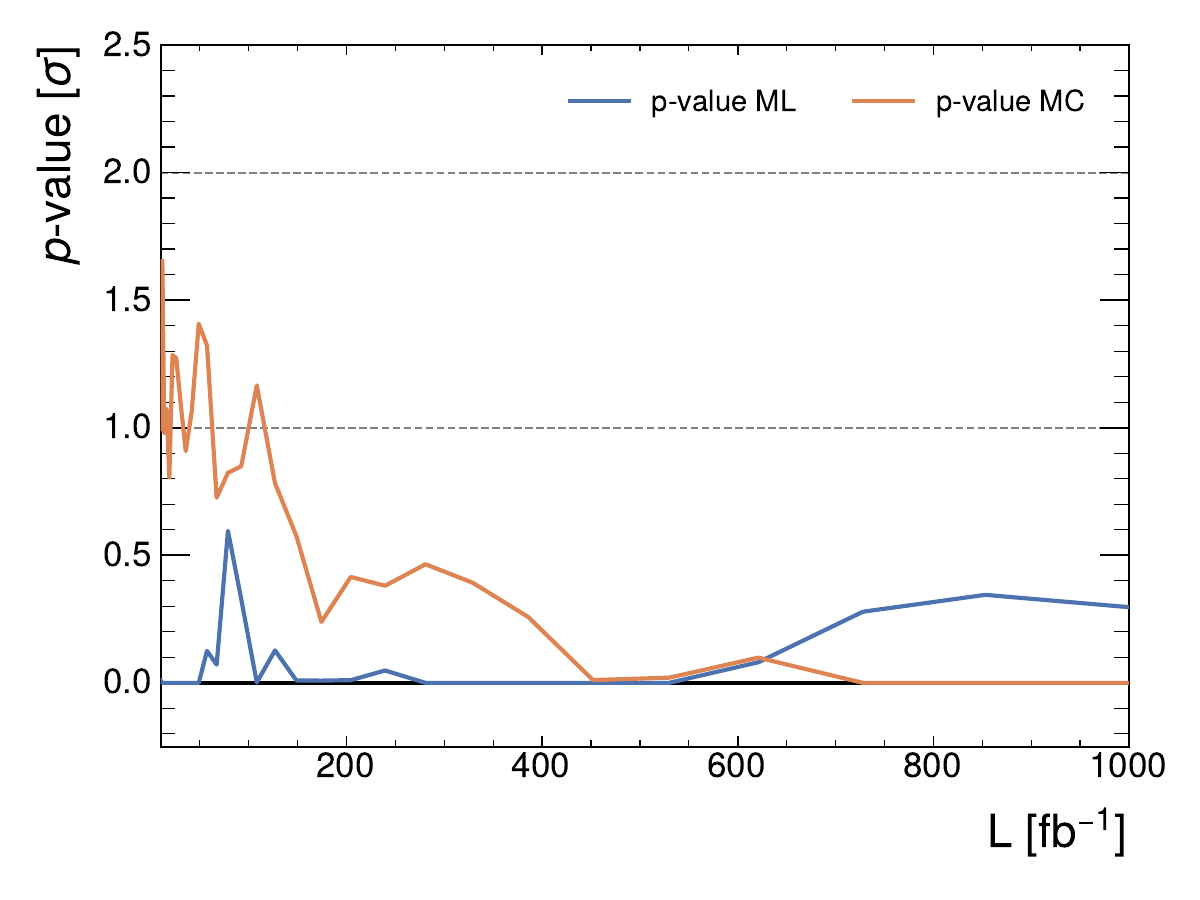}
    \end{minipage}

    \caption{Estimated $p$-values in units of standard deviation ($\sigma$) as the discovery significance for both fitting scenarios ($m_{bb}$ left and classifier score right) at different values of luminosity are shown. One can observe that the use of MC simulated samples converges to full compatibility while the ML-generated samples show a small bias, well below one standard deviation.}
    \label{fig: pvals}
\end{figure}

\begin{figure}[ht!]
    \begin{minipage}[t]{0.49\textwidth}
        \centering
        \includegraphics[width=0.99\textwidth]{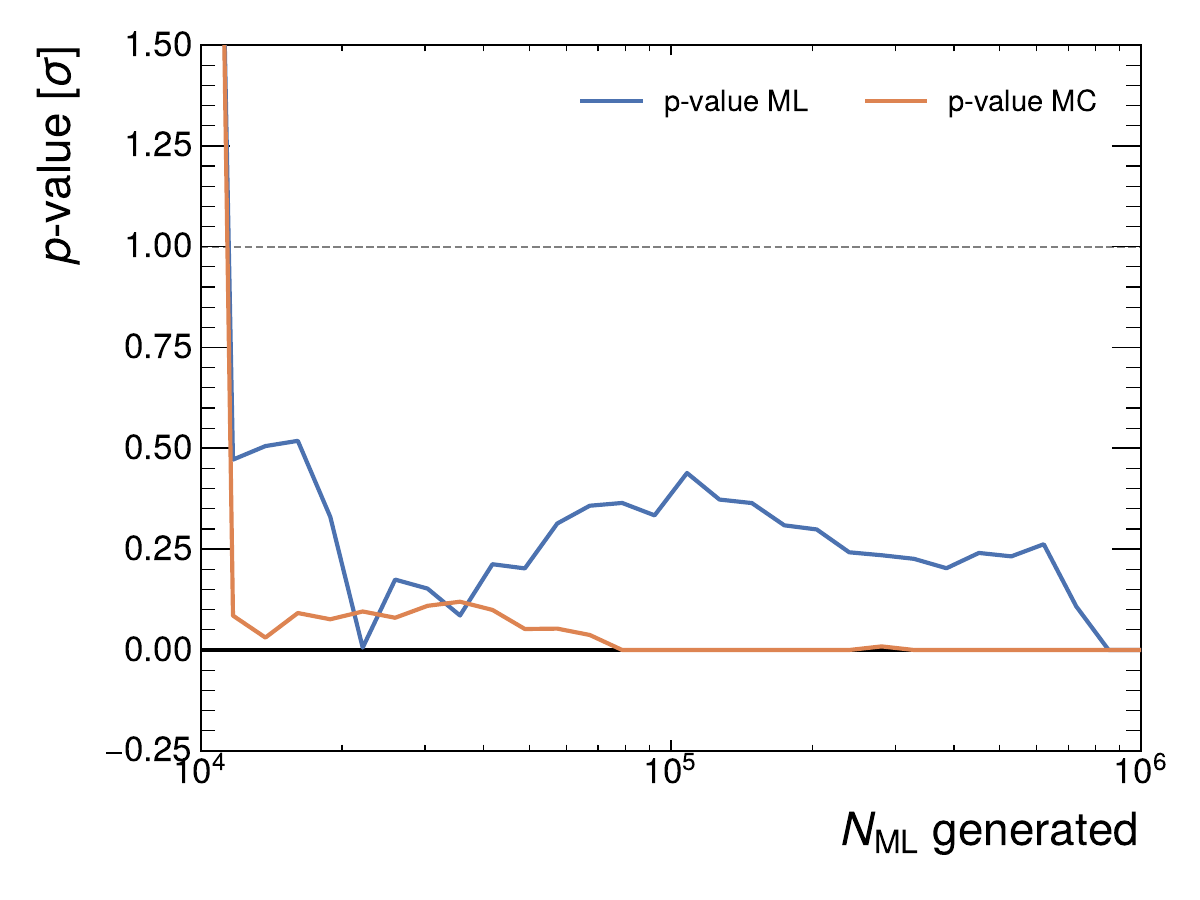}
    \end{minipage}
    \hfill
    \begin{minipage}[t]{0.49\textwidth}
        \centering
        \includegraphics[width=0.99\textwidth]{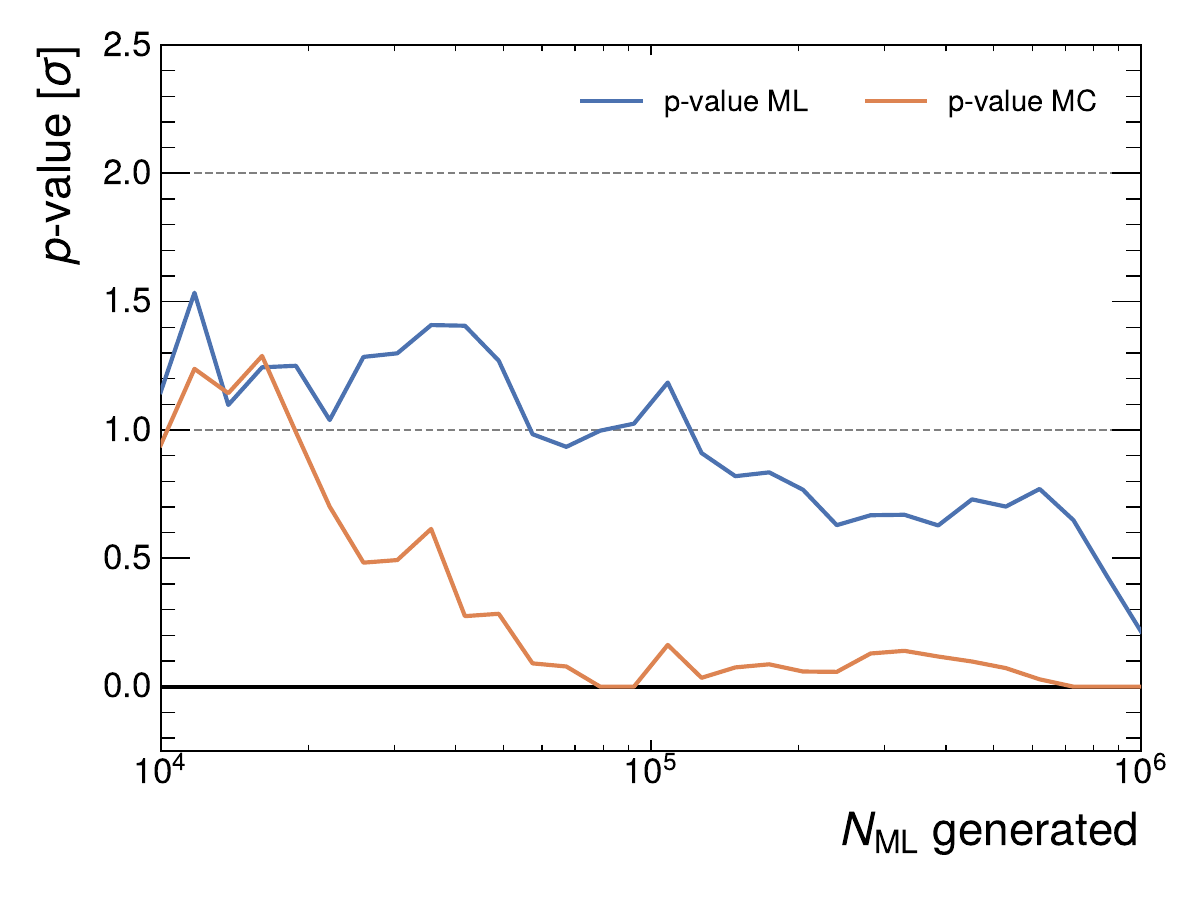}
    \end{minipage}

    \caption{Estimated $p$-values in units of standard deviation ($\sigma$) as the discovery significance for both fitting scenarios ($m_{bb}$ left and classifier score right) at different values of ML-generated events are shown. One can observe that the use of MC simulated samples converges to full compatibility while the ML-generated samples show a small bias, well below one standard deviation. It is evident that the more sensitive observable also exhibits a higher sensitivity to the sample discrepancies.}
    \label{fig: pvals_N}
\end{figure}

A standard statistical test,estimating the background-only hypothesis probability (probability for the $\mu=0$ hypothesis) as described above, i.e. the discovery probability ($p$-value in LHC jargon), is shown in Figures \ref{fig: pvals} and \ref{fig: pvals_N}
as a function of luminosity and ML-generated statistics. Results for the ideal scenario, using only the MC simulated samples,
which match the Asimov dataset, are shown for comparison with the ML-generated background results. As already stated, in this test the signal presence was set to zero in the Asimov data, which means that any significant discrepancy in the p-value when using the ML-generated background would indicate a spurious (fake) signal, reproducing another statistical test at the LHC. For an easier interpretation of results, the p-value is presented in units of standard deviation ($\sigma$) as the discovery significance, as described above: in this representation the ideal agreement is given by zero significance and the higher the value, the worse the agreement - a p-value of 5 ($\sigma$ significance) would mean that the ML-generated background caused a spurious signal statistically corresponding to an experimental observation. As one can observe from the plots, this agreement between the p-values obtained by using ML-generated backgrounds have a very low significance at an appropriate statistics, as desired and anticipated.

\subsection{Upper limit estimation}

\begin{figure}[ht!]
    \centering
    \begin{minipage}[t]{0.49\textwidth}
        \centering
        \includegraphics[width=0.99\textwidth]{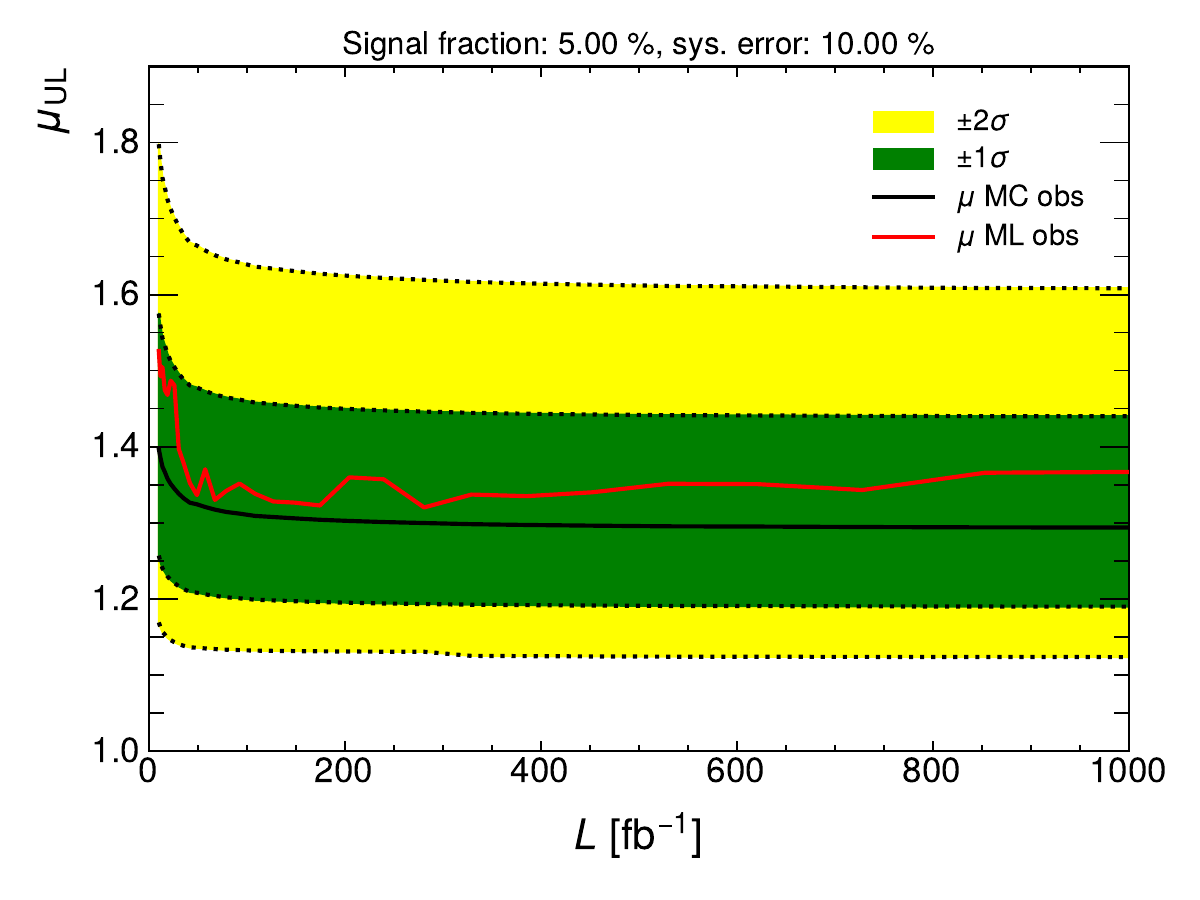}
    \end{minipage}
    \hfill
    \begin{minipage}[t]{0.49\textwidth}
        \centering
        \includegraphics[width=0.99\textwidth]{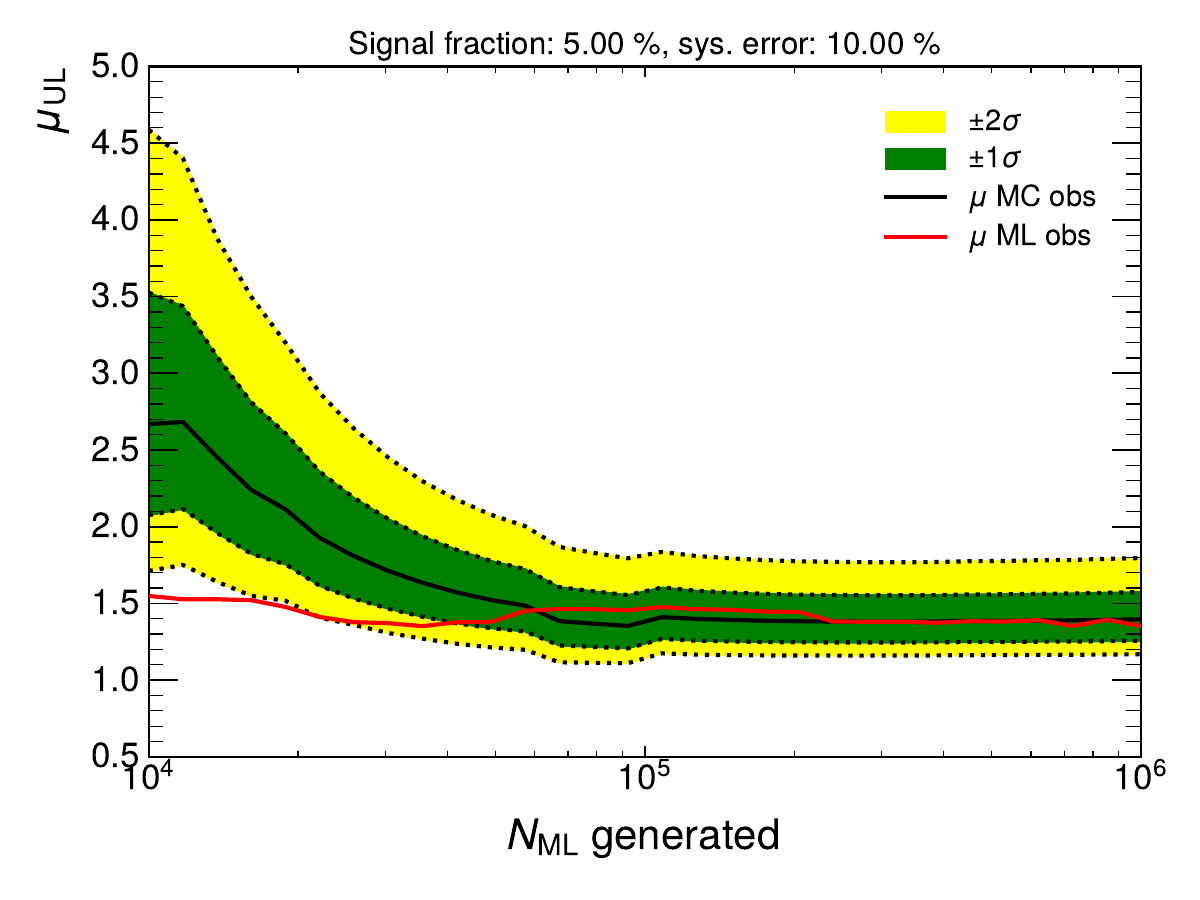}
    \end{minipage}
    \caption{Upper limits on the signal strength $\mu$  for the likelihood fit to the $m_{bb}$ distribution as a function of integrated luminosity $L$ and the number of ML-generated events. The discrepancies (biases) are deemed acceptable and, one can observe that a sufficiently large value of ML-generated events is essential to get an acceptably unbiased result.}
    \label{fig:mbb_ul}
\end{figure}

\begin{figure}[ht!]
    \centering
    \begin{minipage}[t]{0.49\textwidth}
        \centering
        \includegraphics[width=0.99\textwidth]{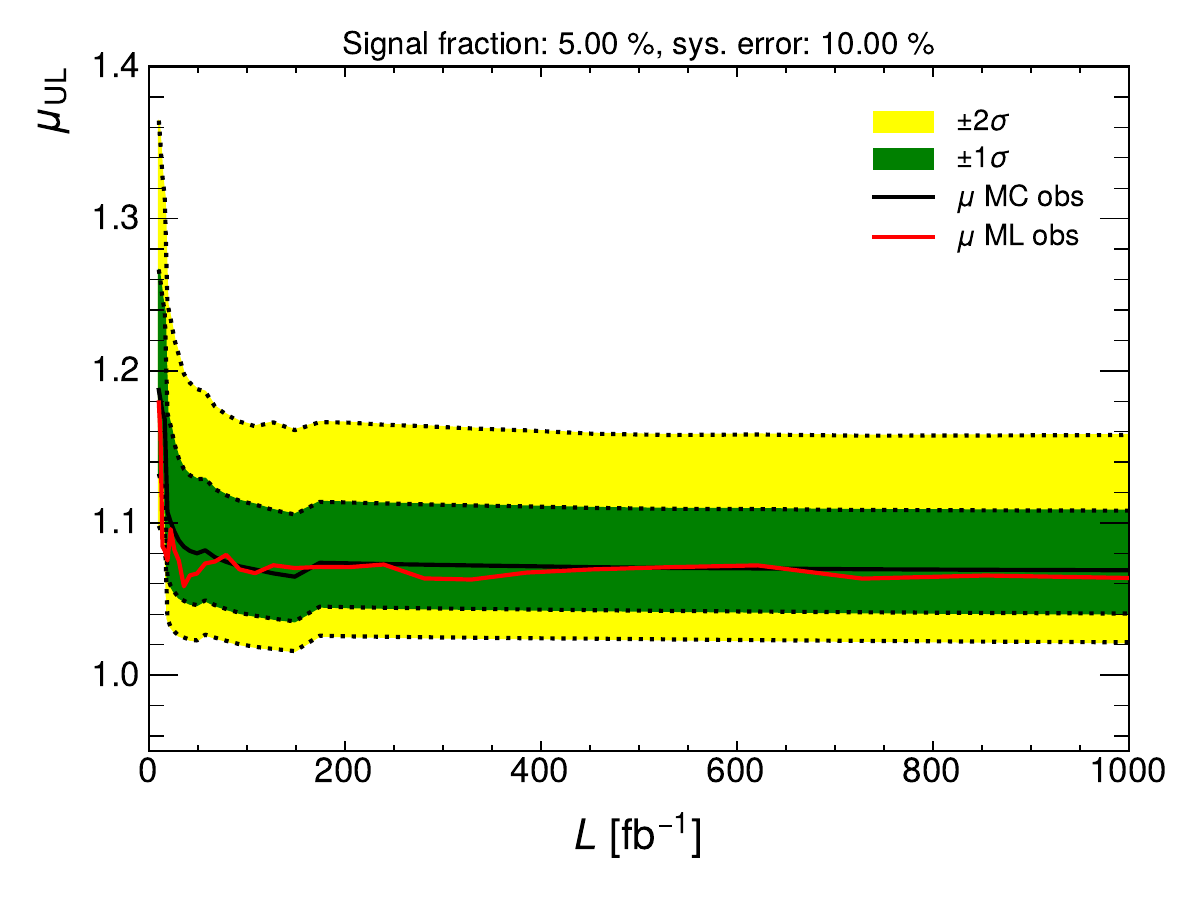}
    \end{minipage}
    \hfill
    \begin{minipage}[t]{0.49\textwidth}
        \centering
        \includegraphics[width=0.99\textwidth]{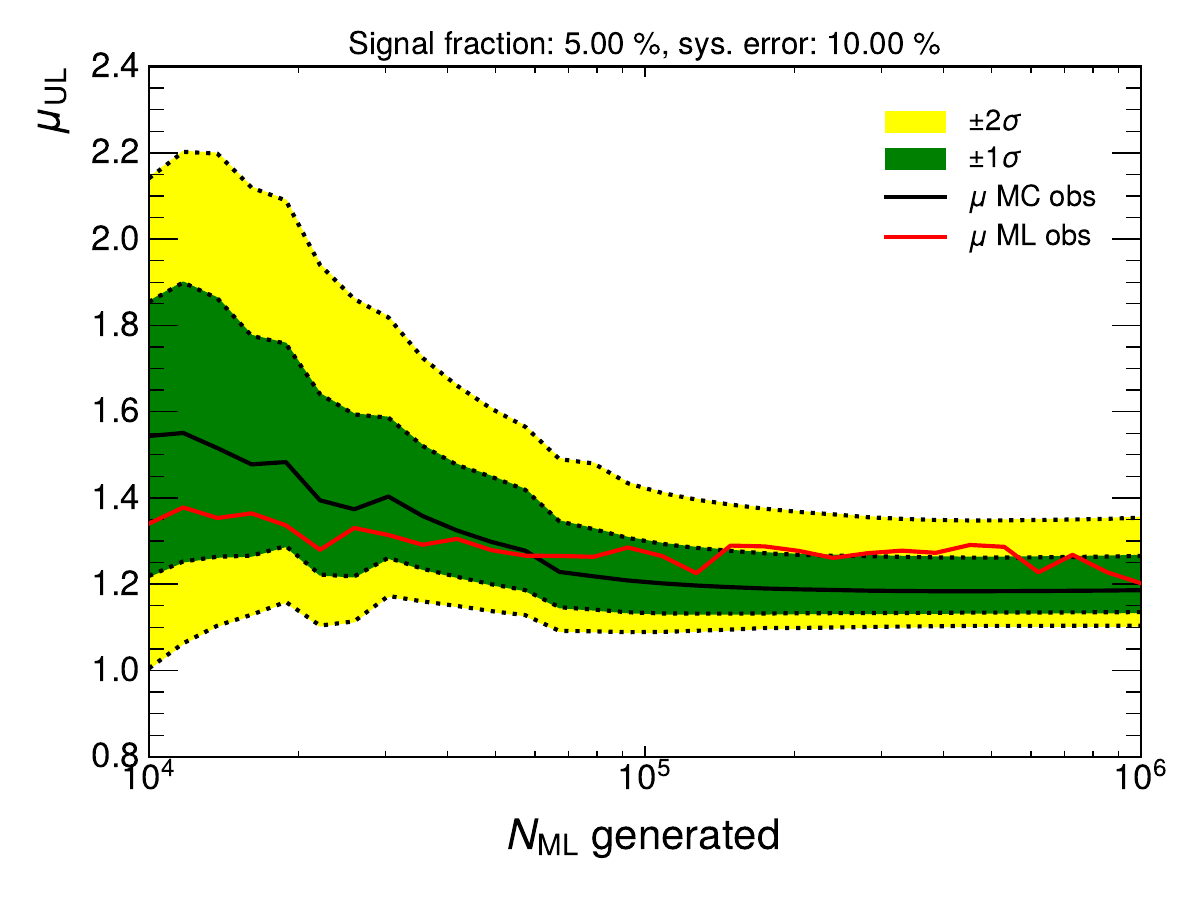}
    \end{minipage}
    \caption{Upper limits on the signal strength $\mu$  for the likelihood fit to the classifier score distribution as a function of integrated luminosity $L$ and the number of ML-generated events. The discrepancies (biases) are deemed acceptable and, one can observe that a sufficiently large value of ML-generated events is essential to get an acceptably unbiased result.}
    \label{fig:class_ul}
\end{figure}

As the final step in this physics analysis study, one aims to evaluate the upper limits on the signal strength ($\mu_{\rm UL}$), together
with the uncertainty estimates using the profile-likelihood-based test statistics, as is done in LHC analyses. The
dependence of the extracted upper limit as a function of integrated luminosity is shown in Fig. \ref{fig:mbb_ul}
and Fig. \ref{fig:class_ul} for different values of integrated luminosity and ML-generated number of events, presented for the two fitting scenarios. When varying the integrated luminosity, the number of ML-generated events is set to a very large value ($N_{\rm ML}=10^6$), while
when varying the ML-generated events the integrated luminosity is set to $L=100~\text{fb}^{-1}$.

Again, the ideal (reference) scenario, using the MC simulated samples both for Asimov data construction and simulated
predictions is used as a reference, and is in the figures shown together with the derived uncertainty bands.
One can observe that the shifts in upper limit estimation are on a scale compatible with the estimated uncertainties for
the reference scenario. The use of the classifier score as the distribution in the profile likelihood fits 
shows a higher sensitivity to the signal presence consistently in all performed tests,  however this also leads to an increased sensitivity to the background
mis-modelling using ML-generated events. The discrepancies (biases) are deemed acceptable in all considered cases, as one can observe from
these statistical tests. One can observe that a sufficiently large value of ML-generated events is essential to get an acceptably 
unbiased result, the production of these large statistics is however very fast, many orders of magnitude faster that the standard
MC simulation, as shown in the Appendix \ref{app: times}.

\begin{figure}[ht!]
    \centering
    \begin{minipage}[t]{0.49\textwidth}
        \centering
        \includegraphics[width=0.99\textwidth]{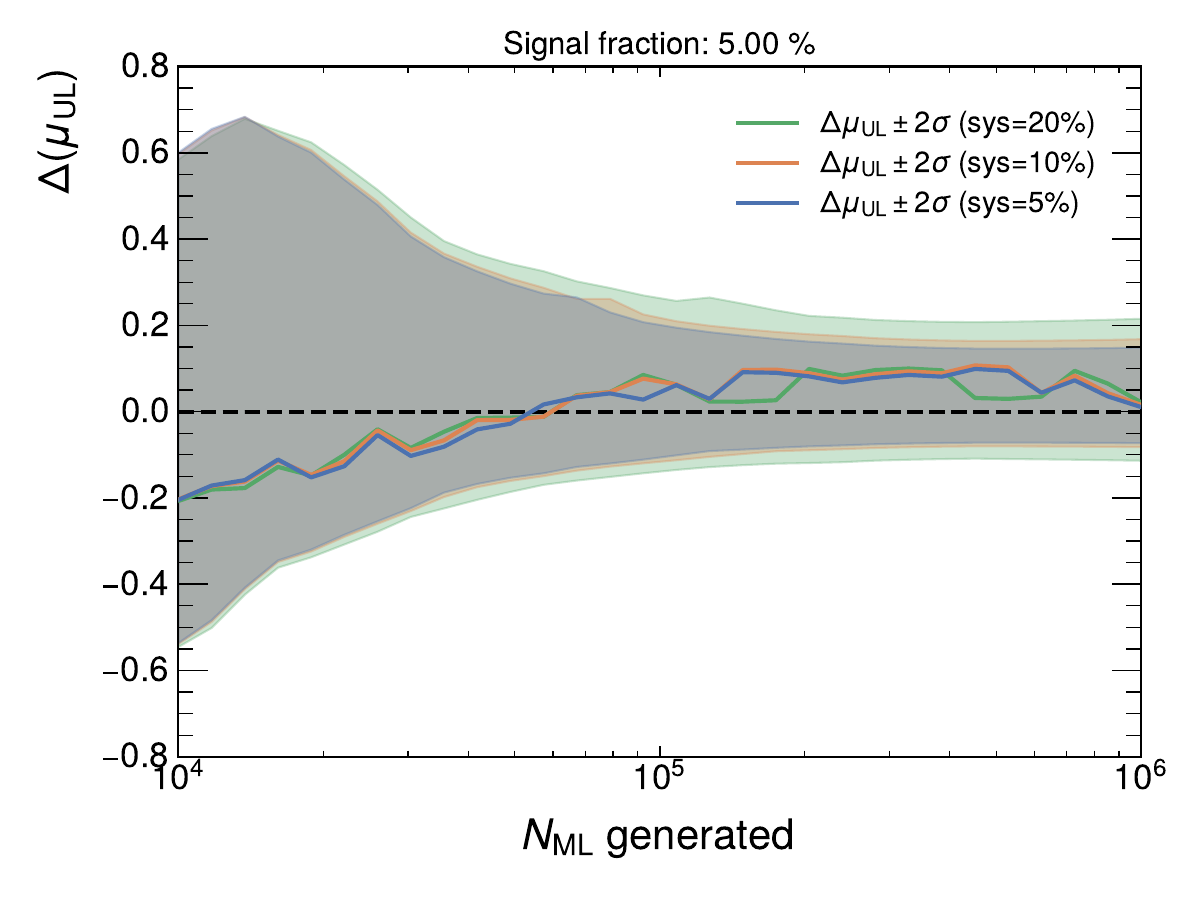}
    \end{minipage}
    \hfill
    \begin{minipage}[t]{0.49\textwidth}
        \centering
        \includegraphics[width=0.99\textwidth]{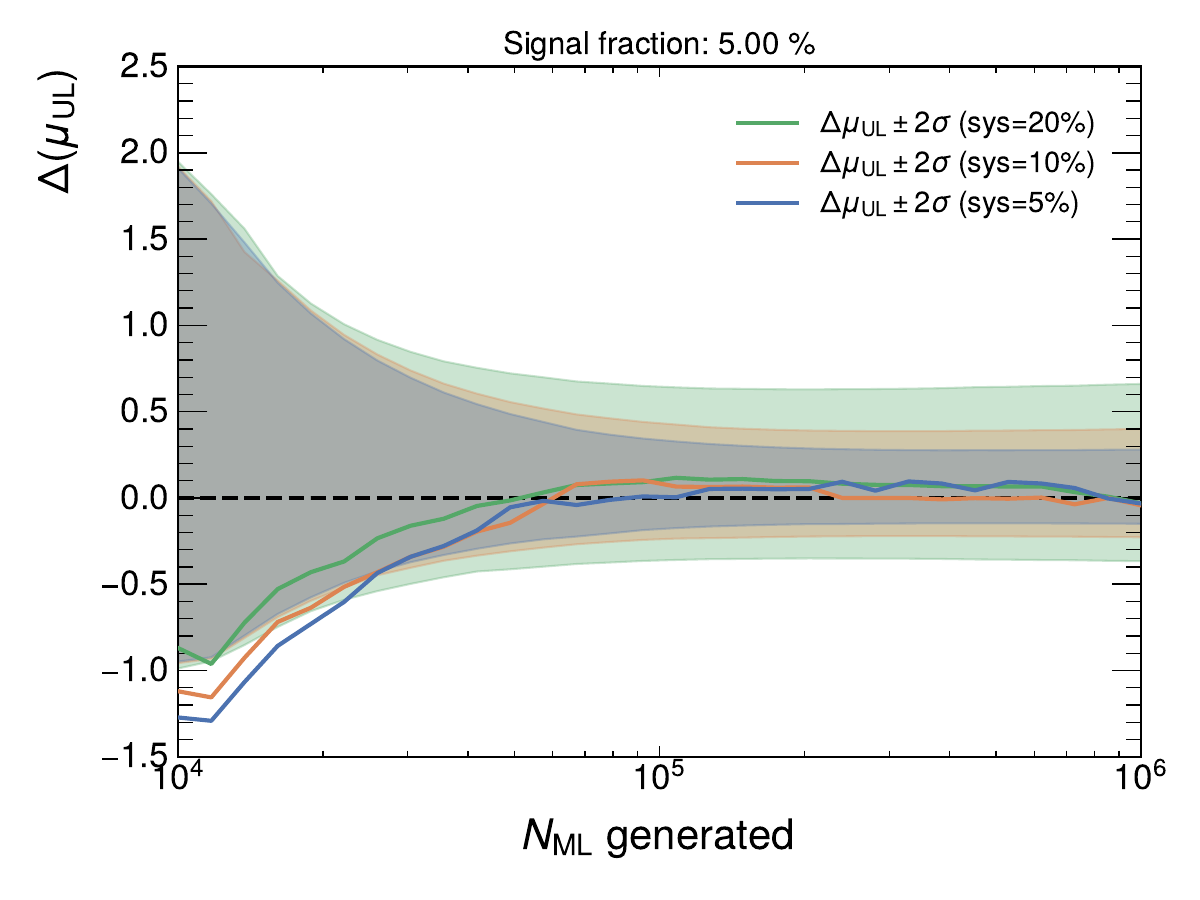}
    \end{minipage}
    \caption{The difference $\Delta\mu_{\rm UL}$ of the upper limits on the signal strength $\mu$, as predicted by using the MC or ML-generated background, for the likelihood fit to the $m_{bb}$ distribution (left) and the classifier score (right), as a function of the number of ML-generated events. In addition, the fixed injected systematics uncertainty is varied between 5\% and 20\%. The discrepancies (biases) are well within the estimated uncertainties and are thus deemed acceptable. One can observe the nice convergence to the fixed injected uncertainties at a sufficiently large value of ML-generated events, when the systematic uncertainty due to finite simulation (ML) statistics becomes negligible.}
    \label{fig:mon_ul}
\end{figure}

An alternative presentation of the estimated dependence of the upper limit $\mu_{\rm UL}$ on the number of ML-generated events, as shown in  Figs. \ref{fig:mbb_ul} and \ref{fig:class_ul}, is given in  Fig. \ref{fig:mon_ul}, which shows the dependence of the difference $\Delta\mu_{\rm UL}$ between the upper limit values when using the MC and ML background predictions on the number of ML-generated events, as well as the injected systematic uncertainty,
which is set to 5\%, 10\% and 20\%. It is nicely visible, how the upper limit uncertainty converges to the fixed injected systematic uncertainty value, when the systematic uncertainty due to finite simulation (ML) statistics becomes negligible.

From these results it is evident that the ML-generated samples can indeed be used in a physics analysis as a surrogate
model for the background prediction. Nonetheless, to further minimize the impact of the background ML mis-modelling, one would
need to work on implementing techniques that go even beyond the current commercial state-of-the-art approaches, similar to the one used in
this paper, and to understand how to optimally adapt them for this use case in high energy physics.

\section{Discussion and outlook}

The paper investigates the possibility of using deep generative models for analysis-specific ML-based generation of events
used by the final stage of a given particle physics analysis, where the state-of-the-art generative ML algorithms are
trained on the available MC-simulated samples. The extended custom event samples can thus serve to extend the MC statistics,
thereby minimizing the analysis uncertainty due to the statistics limitations of MC events or, equivalently, to smooth
and minimize the uncertainties on the predicted kinematic event distributions used in the final statistical analysis of
the data.

The envisaged ML modelling strategy is to learn the $D$-dimensional distributions of kinematic observables used in a
representative physics analysis at the LHC, and produce large amounts of ML-generated  events at a low computing cost
(see Appendix \ref{app: times}). The input kinematic distributions are derived from the MC-simulated samples, which are
statistics-limited by design since they are produced in a very computationally expensive procedure, albeit giving very
accurate predictions of the physics performance of the LHC experiments. Furthermore, the study presented in this paper
replicates the realistic case of the final event selection in a physics analysis using a cut on a ML-derived discriminating parameter,
which defines the final data sets for physics analysis. In this final set, the  MC statistics is usually too low to effectively
train a ML procedure, thus the training needs to happen at the stage before the final filtering and it is essential that
the ML-generated samples reproduce the correlations between the observables, so that the agreement between the ML-generated and original
MC-simulated data is preserved after the final event selection. This paper demonstrates that this can, in fact, be achieved
by using the implemented ML techniques, with reasonable precision.

The generative ML approach described in this paper is inherently analysis-specific, meaning that each analysis would require
a dedicated training setup. As a demonstrator for this paper, a number of different state-of-the-art generative architectures
with different parameters were implemented. The procedures were not fine-tuned for specific analysis
and/or MC dataset to preserve generality, but could potentially achieve even better performance with further optimization
of hyper-parameters in Appendix \ref{app: hyper}. The implemented models were trained on the LHC-specific HIGGS dataset
(beyond the Standard model Higgs boson decay and corresponding backgrounds) with $\mathcal{O}(10)$ observables. Both coupling layer models and autoregressive models were considered as discussed in Section \ref{sec: methods}. All of the models were capable of learning complicated
high-dimensional distributions to some degree of accuracy, with autoregressive models having an advantage at the cost of
somewhat longer sampling times, which does not pose a problem since our distributions are relatively low-dimensional and the
absolute speed-ups are impressive in all cases, i.e. orders of magnitude better with respect to the standard MC generation
procedures used at the LHC.

Performance evaluations using various divergence measures, from $\chi^2$ to Wasserstein distance, classifier two sample test (C2ST)
as well as a simplified statistical analysis, matching the procedures used for upper limit estimation of new physics searches at the
LHC, show that the generally available MC samples of $\mathcal{O}(10^6)$ events are indeed enough to train such state-of-the-art
generative ML models to a satisfactory precision to be used to reduce the systematic uncertainties due to the limited MC statistics.
The generative modelling strategy described in this paper could alleviate some of the high CPU and disk size requirements
(and costs) of generating and storing simulated events. When trained, these models provide not only fast sampling but also
encode all of the distributions in the weights and biases of neural networks, which take up significantly less space than
the full MC datasets and can generate analysis-specific events practically on-demand, which is a functionality that goes
beyond the scope of this paper but should be studied in a dedicated project.

The listed advantages become even more crucial when considering future LHC computing requirements for physics simulation
and analysis, as it is clear that the increase in collision rates will result in much larger data volumes and even more complex
events to analyse. Using generative modelling could thus aid in faster event generation as well as future storage requirements
coming with the HL-LHC upgrade and beyond, however a careful statistical evaluation of the quality of thus produced samples is essential.

\section{Acknowledgements}

The authors would like to acknowledge the support of Slovenian Research and Innovation agency (ARIS) by funding the
research project J1-3010 and programme P1-0135.


\clearpage
\appendix

\section{Code availability}
All networks were implemented in PyTorch \cite{paszke2019pytorch}. The code is available on GitHub (\texttt{\url{https://github.com/j-gavran/MlHEPsim}})
and follows the implementations of \cite{nflows}.

\section{Data partitioning}\label{app: splits}

\begin{figure}[h!]
    \centering
    \includegraphics[width=0.5\textwidth]{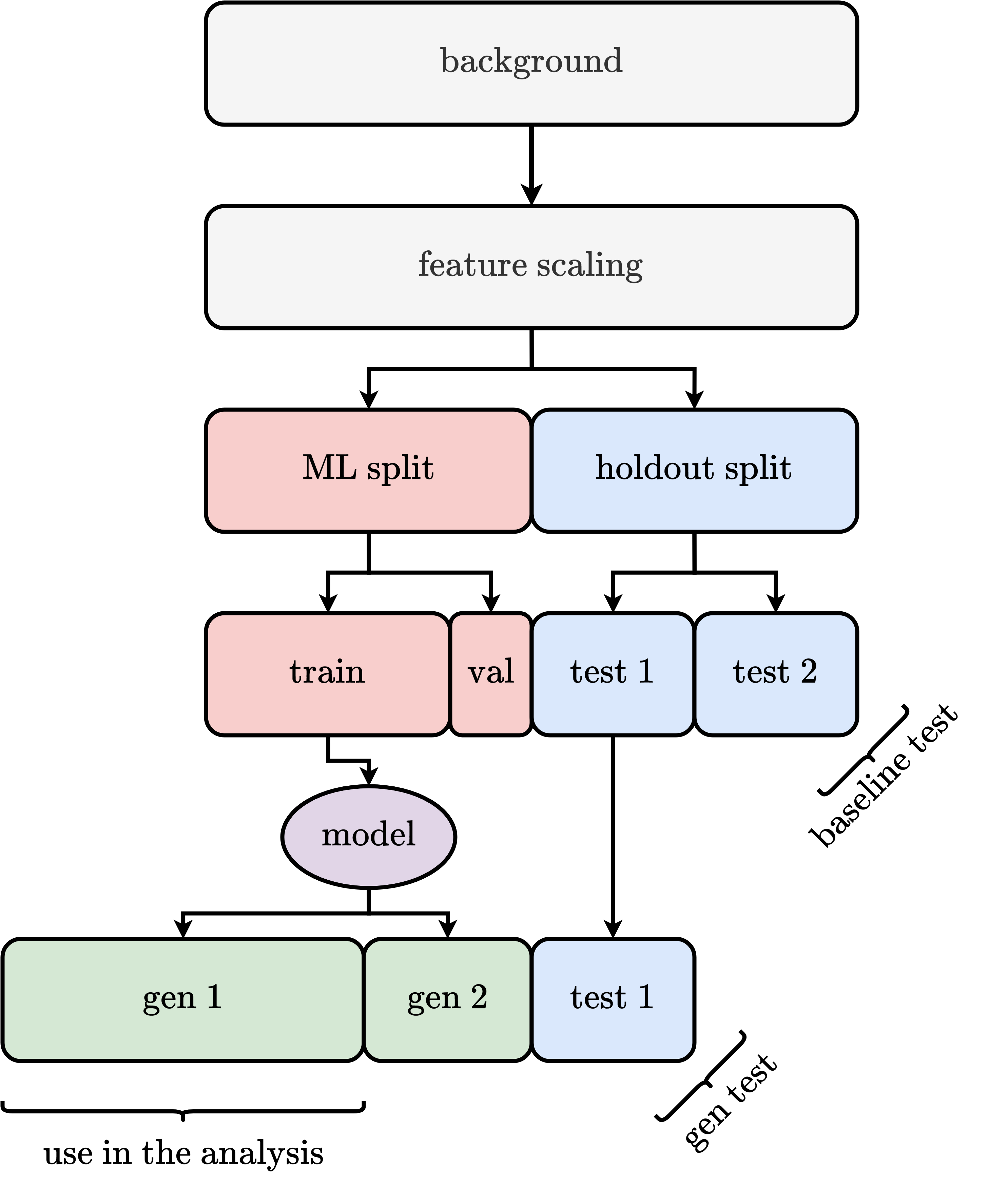}
    \caption{Data split into training, validation and test sets. Splits between the sets are arbitrary and are left to the
        discretion of the method user.}
    \label{fig: splits}
\end{figure}

For simplicity we have used 50\% split for ML and holdout sets, 80\% split for the
training and validation sets, and again 50\% split for both test sets. Equivalent procedure can also be used for the
signal dataset. The procedure is shown in Figure \ref{fig: splits}. After evaluating the generated samples and confirming
they match the MC samples to some accuracy, we can combine the ML-generated background with all the MC events to form the
final enlarged dataset for the analysis.

\section{Event generation times}\label{app: times}

The computational time per event on a GPU is given in Table \ref{tab: times}. Autoregressive models have longer sampling times
because of the modelling constraint that variable $i$ is dependent on all variables preceding variable at index $i$ in an input vector,
giving us $\mathcal{O}(D)$ sampling time.

\begin{table}[h]
    \centering
    \begin{tabular}{ |p{3cm}||p{3cm}|  }
        \hline
        \multicolumn{2}{|c|}{Event generation timing}   \\
        \hline
        Model      & Time $[\mu \text{s}/\text{event]}$ \\
        \hline
        Glow       & $1.33\pm 0.02$                     \\
        MAFMADEMOG & $11.10\pm 0.08$                    \\
        RQS        & $103.59 \pm 0.94$                  \\
        LHC MC Sim & $\sim 10^7$                        \\
        \hline
    \end{tabular}
    \vspace*{0.5cm}
    \caption{Event generation times using for model parameters from Table \ref{tab: hyper1}. For comparison, the event timing for the standard MC simulation of the LHC experiments, which is in the range of 10-20 seconds/event is also stated. }
    \label{tab: times}
\end{table}

\clearpage

\section{Hyper-parameters}\label{app: hyper}

Tables of parameters used in the model testing, model comparison and final analysis stages. In Section \ref{sec: analysis} we used the
MADEMOG model where we have further increased the number of blocks to 30, used GELU activation, and kept the rest of the hyper-parameters the same
as in Table \ref{tab: hyper_comp}. All the models were trained with variable random seeds

\begin{table}[h]
    \centering
    \begin{tabular}{ |p{3.3cm}||p{5.5cm}|  }
        \hline
        \multicolumn{2}{|c|}{Hyper-parameter list}          \\
        \hline
        Parameter           & Value                         \\
        \hline
        Hidden layer size   & 128                           \\
        Blocks              & 10                            \\
        Activation function & ReLU                          \\
        Batch size          & 1024                          \\
        Training size       & $2.5\times 10^5$              \\
        Optimizer           & Adam                          \\
        Learning rate       & $3\times 10^{-4}$             \\
        Weight decay        & $1\times 10^{-6}$             \\
        Max. epochs         & 100                           \\
        Early stopping      & 15                            \\
        Feature scaling     & logit + normal transformation \\
        GPU                 & NVIDIA GeForce RTX 3070       \\
        \hline
    \end{tabular}
    \vspace*{0.5cm}
    \caption{List of used hyper-parameters in the initial model testing stage in Section \ref{sec: methods}.}
    \label{tab: hyper1}
\end{table}

\begin{table}[h]
    \centering
    \begin{tabular}{ |p{3.3cm}||p{5.5cm}|  }
        \hline
        \multicolumn{2}{|c|}{Hyper-parameter list}                                    \\
        \hline
        Parameter               & Value                                               \\
        \hline
        Hidden layer size       & 512                                                 \\
        Blocks                  & 10 (30 for final MADEMOG)                           \\
        Residual connection     & every 2 blocks                                      \\
        Activation function     & ReLU (GELU for final MADEMOG)                       \\
        Batch size              & 1024                                                \\
        Training size           & all split background (see Figure \ref{fig: splits}) \\
        Optimizer               & Adam                                                \\
        Max. learning rate      & $1 \times 10^{-3}$ reached on epoch start           \\
        Min. learning rate      & $0$ reached on epoch end                            \\
        Learning rate scheduler & cosine annealing with warm restarts                 \\
        Weight decay            & $1\times 10^{-7}$                                   \\
        Max. epochs             & 150                                                 \\
        Early stopping          & 15                                                  \\
        Feature scaling         & Gauss scaler                                        \\
        GPU                     & NVIDIA GeForce RTX 4090                             \\
        \hline
    \end{tabular}
    \vspace*{0.5cm}
    \caption{List of used hyper-parameters used for model comparison in Section \ref{sec: performance} and in the final
        physics analysis in Section \ref{sec: analysis}.}
    \label{tab: hyper_comp}
\end{table}

\clearpage

\section{Generated events in linear scale}\label{app: nolog}

\begin{figure}[h!]
    \centering
    \includegraphics[width=0.77\textwidth]{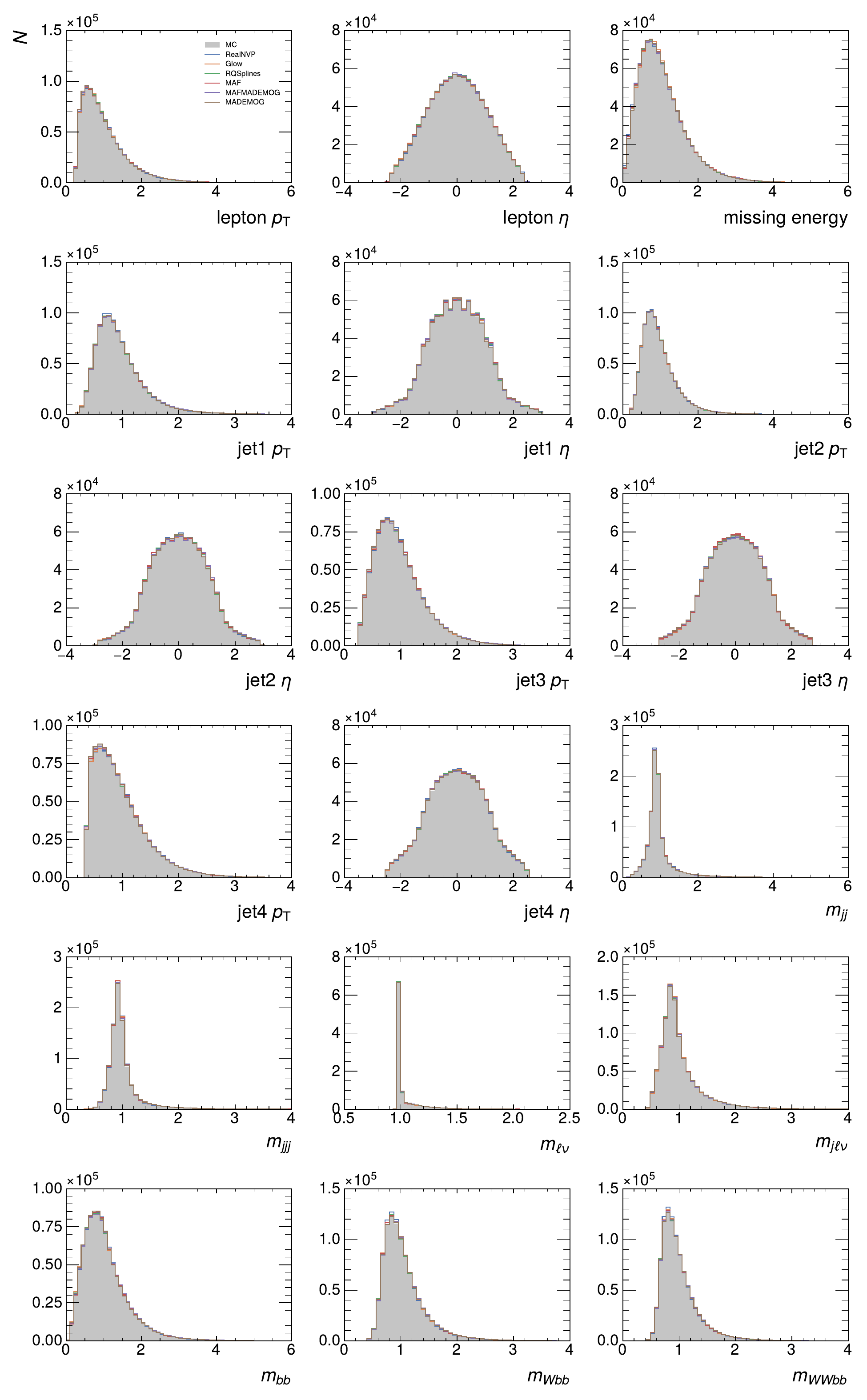}
    \caption{Distributions of generated events in liner scale. Original MC distribution is shown in grey.}
\end{figure}

\section{Corner plot}\label{app: corner}

\begin{figure}[h!]
    \centering
    \includegraphics[width=0.99\textwidth]{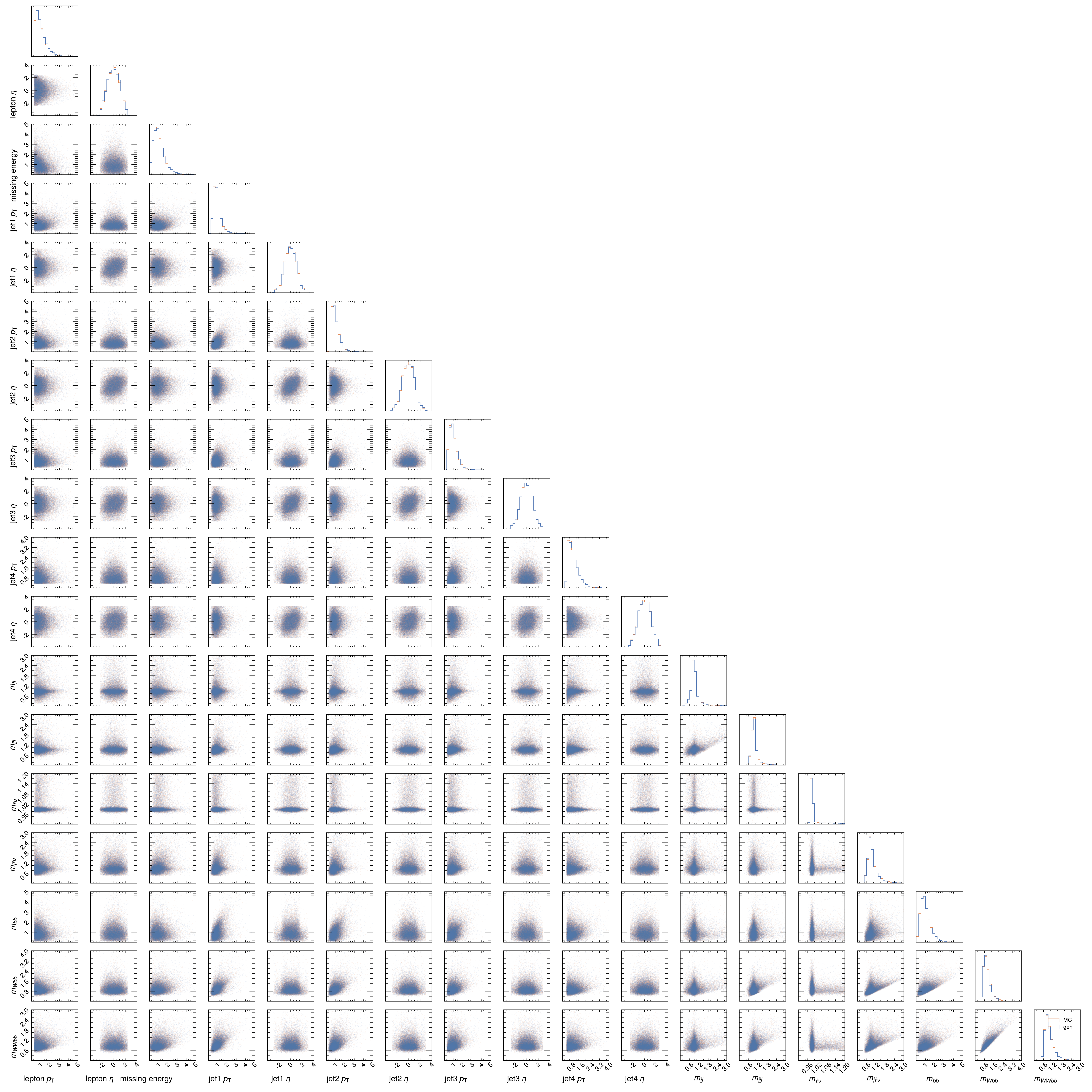}
    \caption{Corner plot of generated events with the MADEMOG model.}
    \label{fig: corner}
\end{figure}

Figure \ref{fig: corner} shows the corner plot of the generated events. Each one and two dimensional projection of the
sample is plotted. The plot shows the correlations between the observables in the generated and MC events.

\clearpage

\section{Effects of increasing the parameter count}\label{app: paramas}

\begin{figure}[h!]
    \centering
    \includegraphics[width=0.9\textwidth]{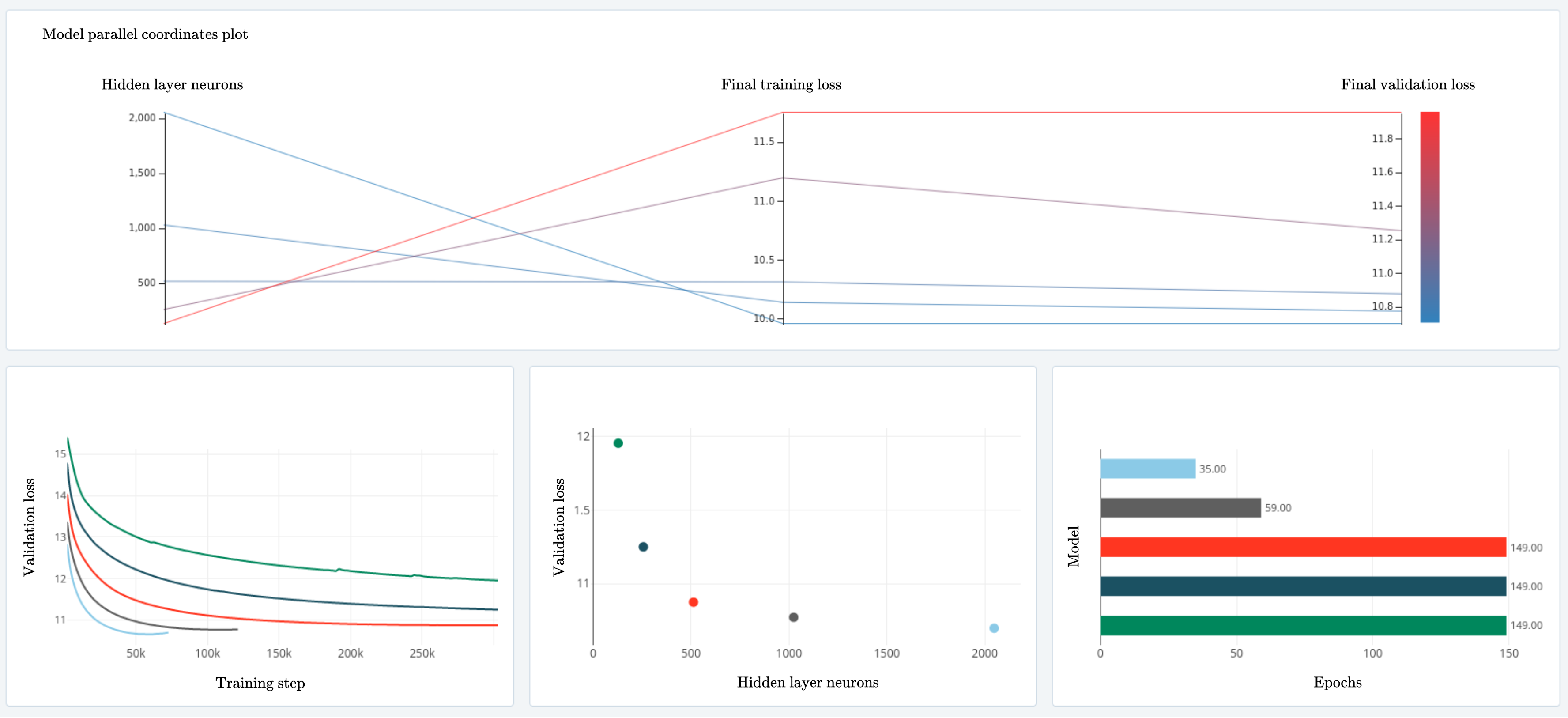}
    \caption{Model losses with varying number of neurons in the hidden layers of the MADEMOG model.}
    \label{fig: model_params}
\end{figure}

\begin{figure}[ht!]
    \centering
    \begin{minipage}[t]{0.46\textwidth}
        \centering
        \includegraphics[width=0.95\textwidth]{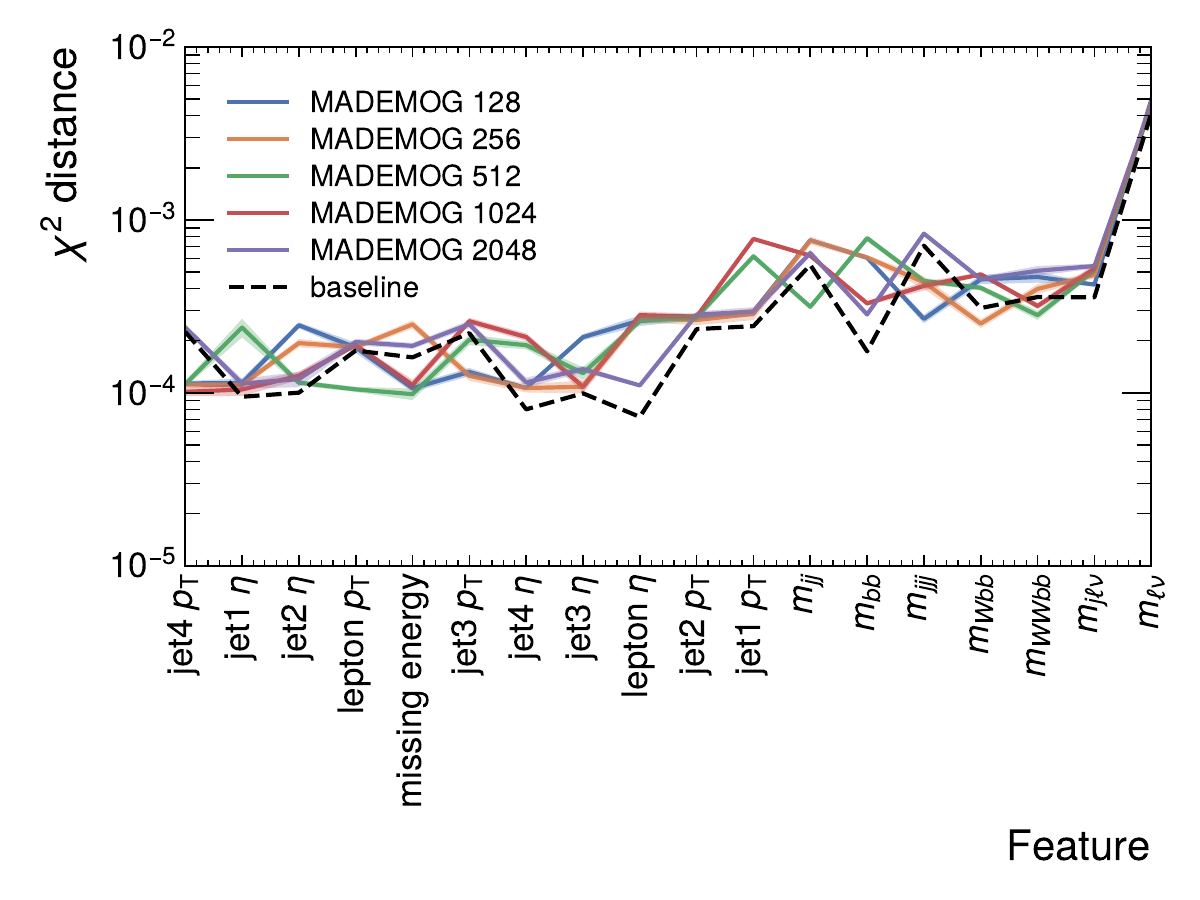}
    \end{minipage}
    \hfill
    \begin{minipage}[t]{0.46\textwidth}
        \centering
        \includegraphics[width=0.95\textwidth]{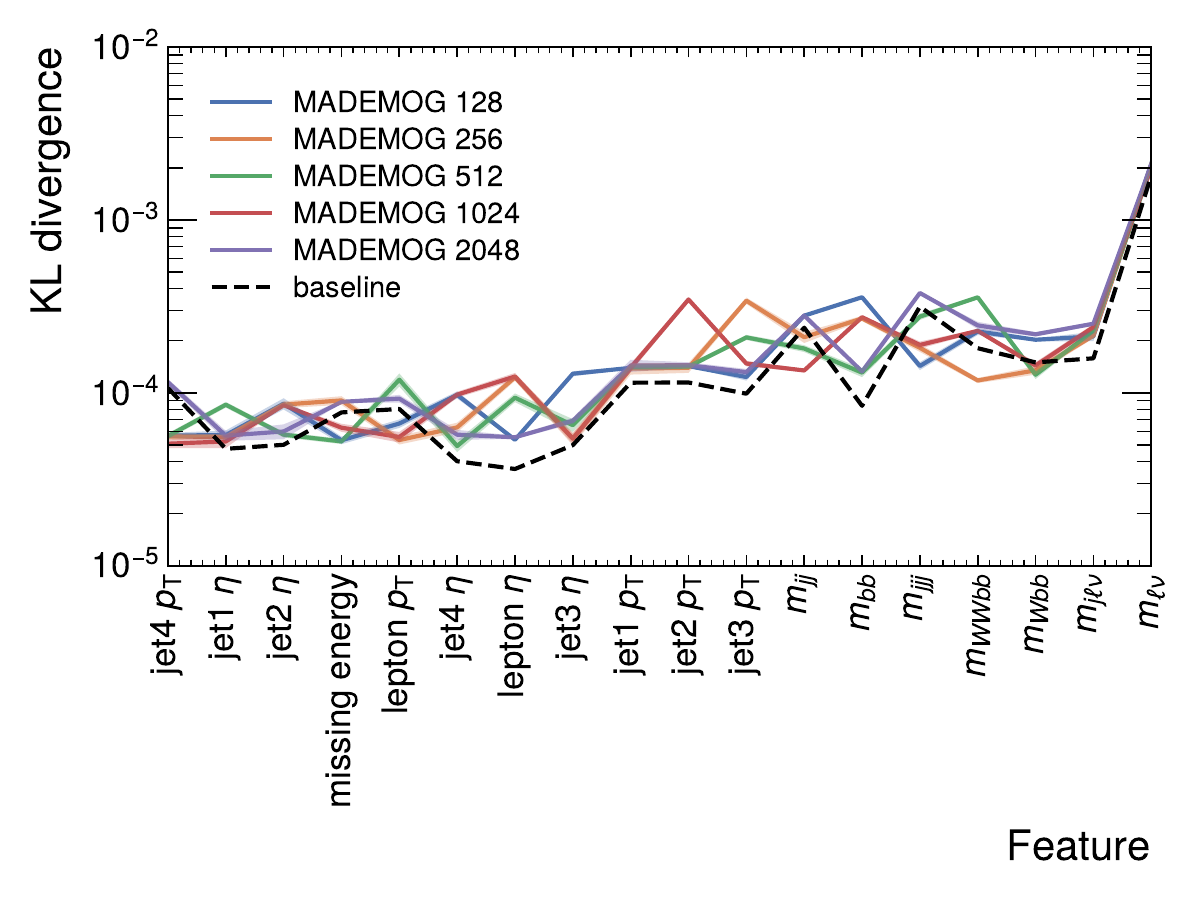}
    \end{minipage}
    \hfill
    \begin{minipage}[t]{0.46\textwidth}
        \centering
        \includegraphics[width=0.95\textwidth]{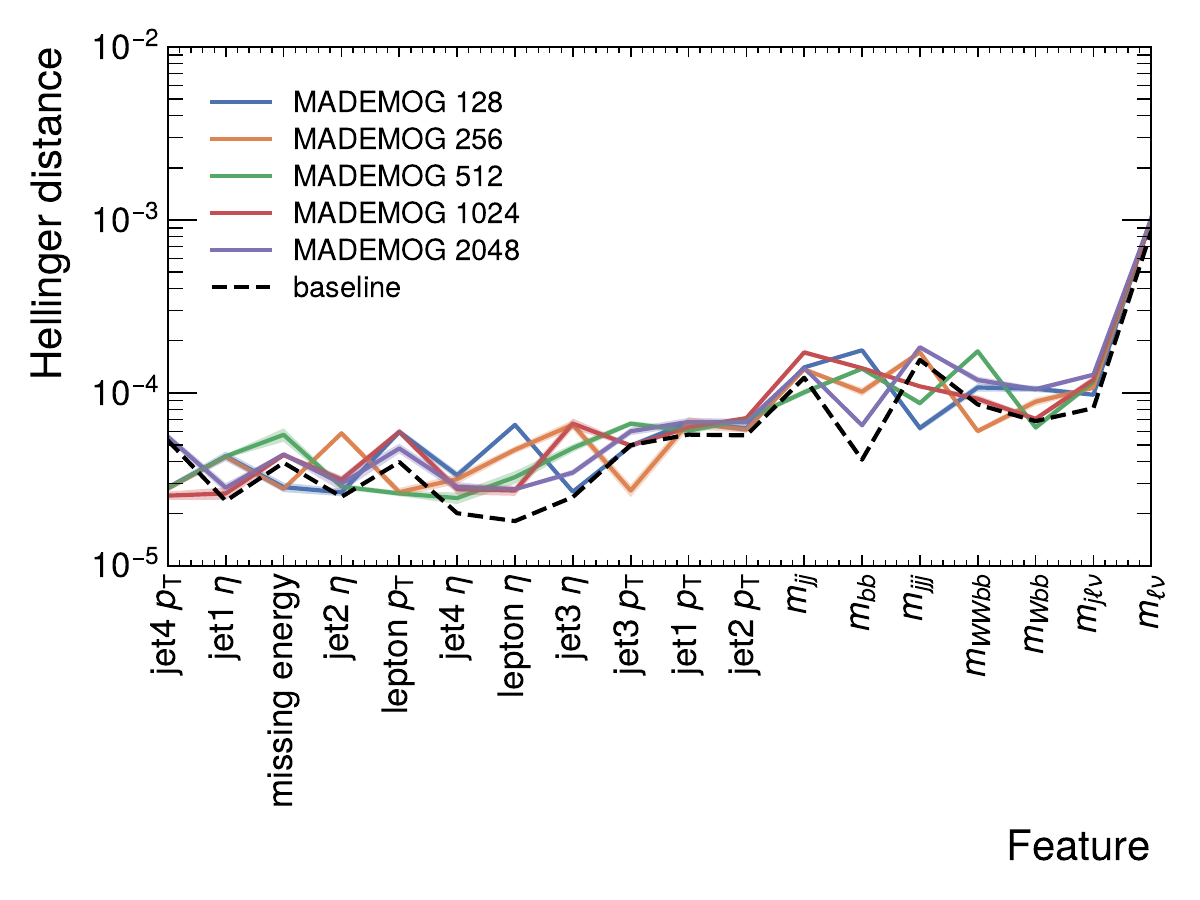}
    \end{minipage}
    \hfill
    \begin{minipage}[t]{0.46\textwidth}
        \centering
        \includegraphics[width=0.95\textwidth]{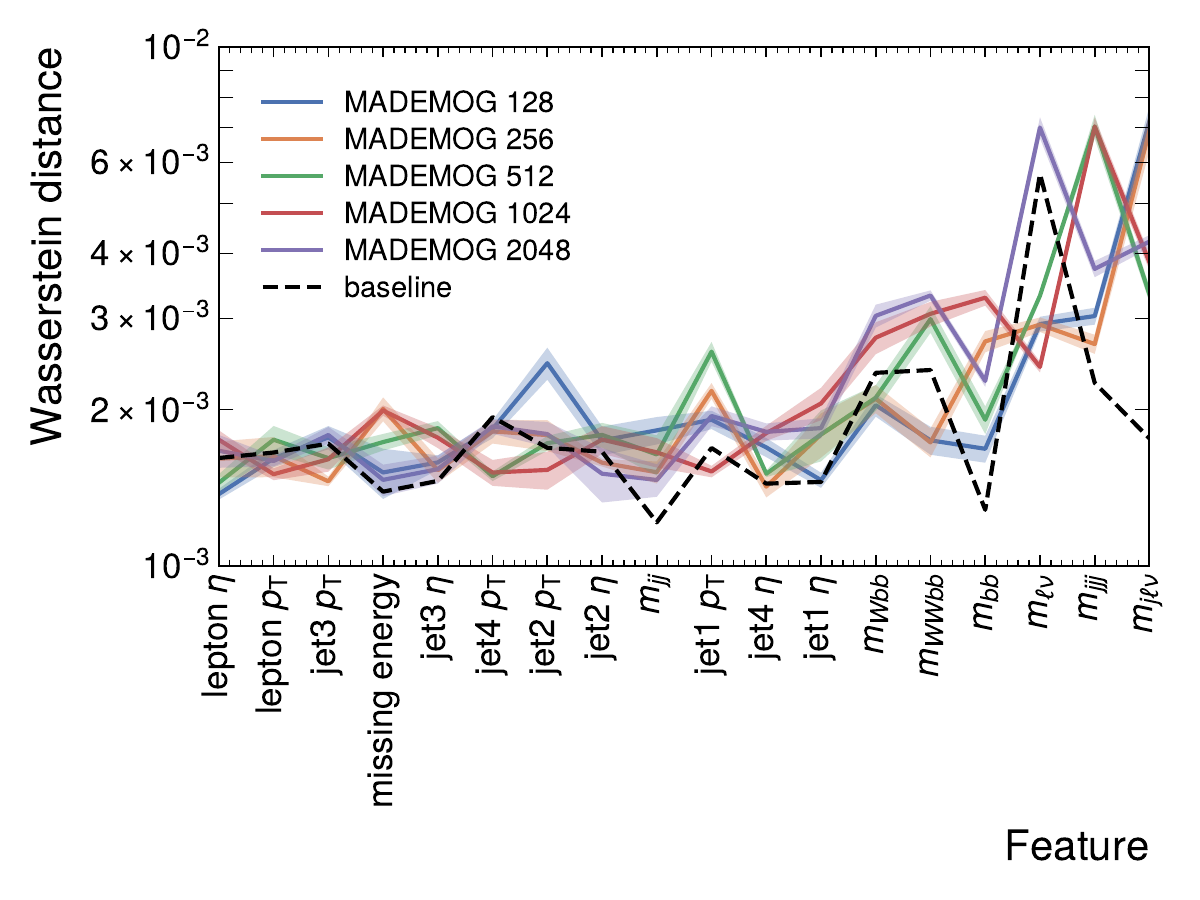}
    \end{minipage}
    \caption{Comparison of different MADEMOG models where we vary the number of hidden neurons in all the hidden layers
        of the model. The distances are calculated between the generated and MC samples.}
    \label{fig: distances_dim}
\end{figure}

\begin{figure}[h!]
    \centering
    \begin{minipage}[t]{0.44\textwidth}
        \centering
        \includegraphics[width=0.99\textwidth]{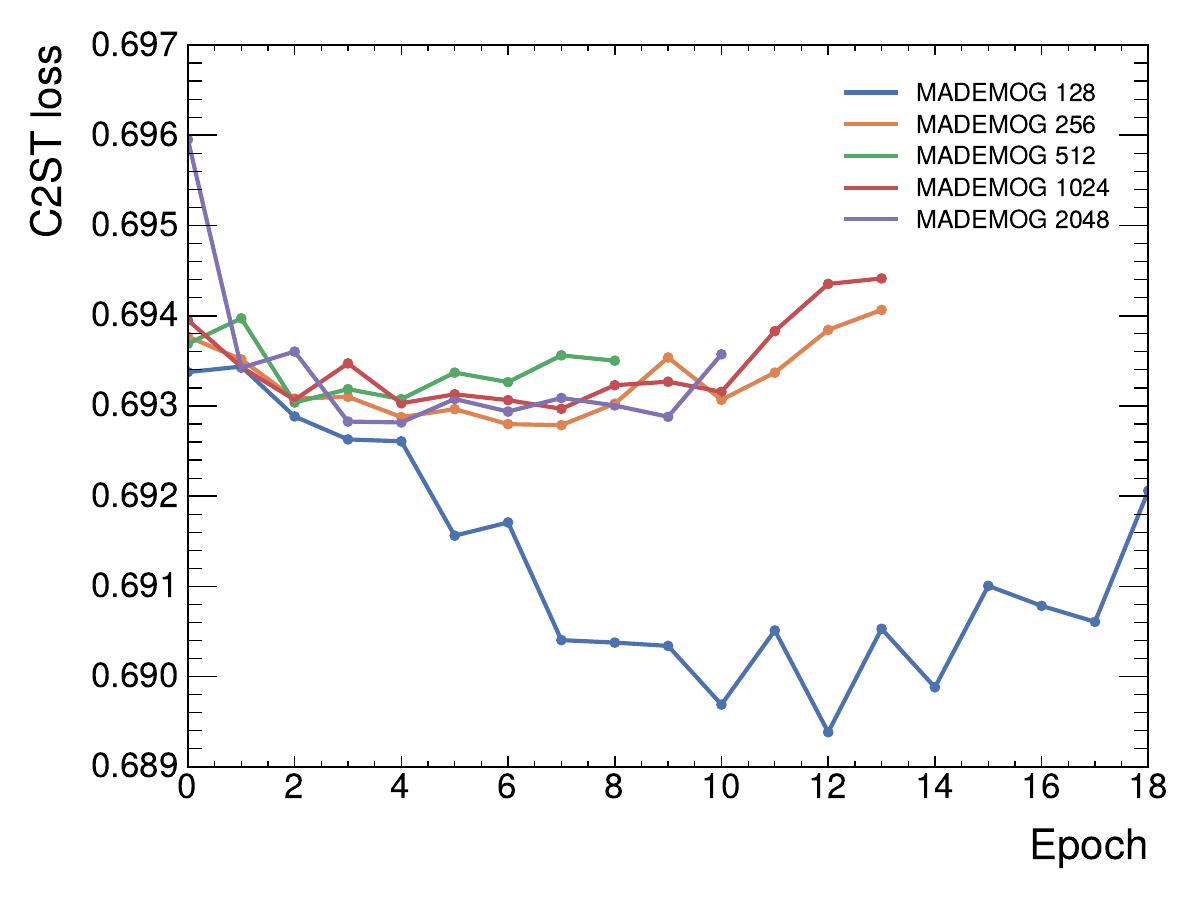}
        \caption{Validation loss of the C2ST test for different MADEMOG model sizes.}
        \label{fig: c2st_dim_loss}
    \end{minipage}
    \hfill
    \begin{minipage}[t]{0.44\textwidth}
        \centering
        \includegraphics[width=0.99\textwidth]{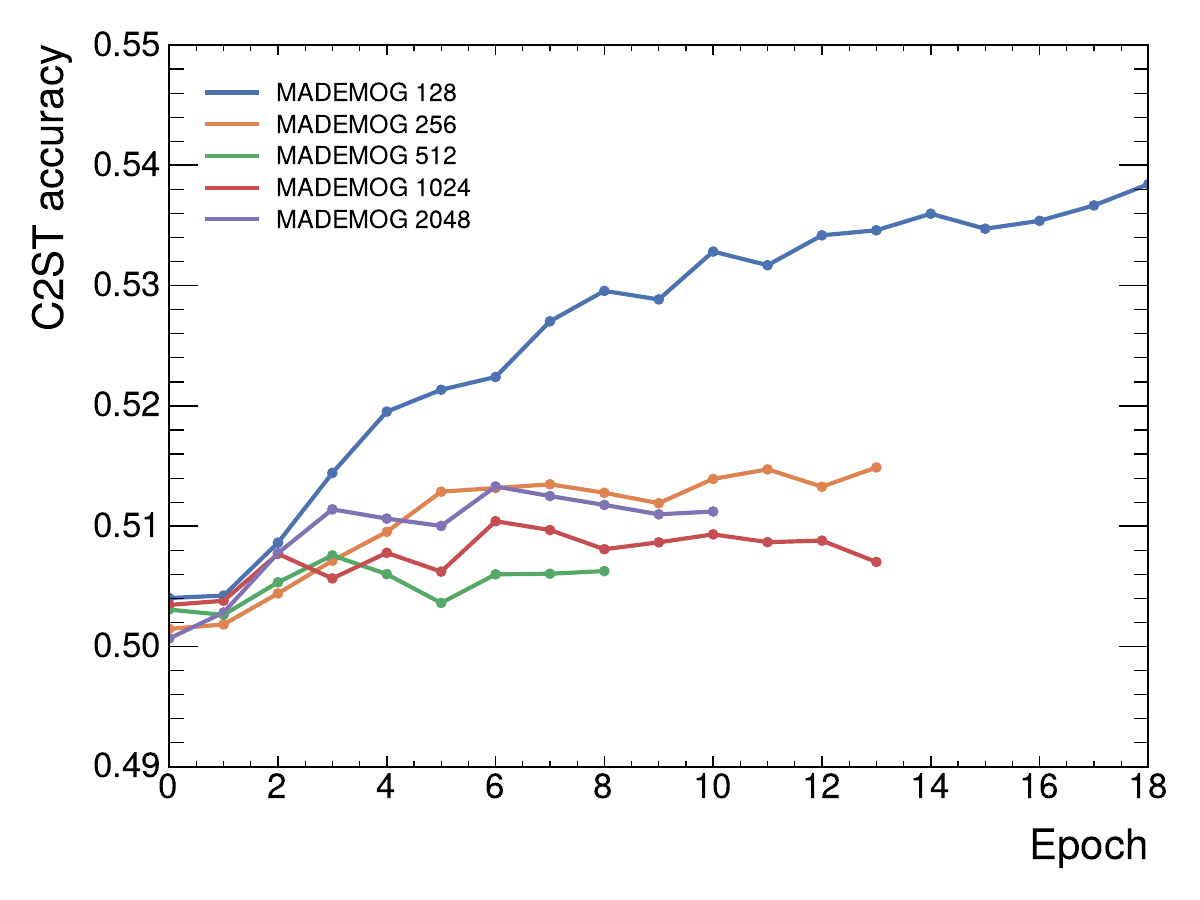}
        \caption{Validation accuracy of the C2ST test for different MADEMOG model sizes.}
        \label{fig: c2st_dim_acc}
    \end{minipage}
\end{figure}

Figures \ref{fig: model_params}, \ref{fig: distances_dim} and \ref{fig: c2st_dim_loss} show the effects of increasing the
number of parameters in the model. This is done by increasing the number of neurons in the hidden layers and keeping all
other parameters from Table \ref{tab: hyper_comp} fixed. The model with 512 hidden neurons in the hidden layers was chosen
for the final analysis.

\clearpage

\section{Classifier density ratio}\label{app: c2st}

Using the Bayes' rule with a prior $p(\vy=1)=\pi=\frac{1}{2}$, and taking into account the complementarity of the two
hypotheses, i.e. $p(\vy=1|\vx) = 1- p(\vy=0|\vx)$, we can write the density ratio as
\begin{equation}
    \begin{aligned}
        r(\vx) & = \frac{P(\vx)}{Q(\vx)} = \frac{p(\vx|\vy=1)}{p(\vx|\vy=0)} =
        \frac{p(\vy=1|\vx)p(\vx)}{p(\vy=1)} \cdot \frac{p(\vy=0)}{p(\vy=0|\vx)p(\vx)} \\
               & = \frac{p(\vy=1|\vx)}{p(\vy=0|\vx)} \cdot \frac{\pi}{1-\pi}
        = \frac{p(\vy=1|\vx)}{1 - p(\vy=1|\vx)} = \exp \left[ \log \frac{p(\vy=1|\vx)}{1 - p(\vy=1|\vx)}  \right] \>,
    \end{aligned}
\end{equation}
which we can estimate with a classifier trained on a two-sample dataset from Eq. (\ref{eq: c2st}) as:
\begin{equation}
    r(\vx) = \exp \left\{ \sigma^{-1} \left[ p(\vy=1|\vx) \right] \right\}
    \approx \exp \left\{ \sigma^{-1} \left[ f(\vx;\vtheta) \right] \right\} \>,
\end{equation}
where $\sigma$ is the sigmoid function and $\sigma^{-1}$ is it's inverse (the logit function). For the classifier we have
used a simple neural network with a sigmoid output, and a binary cross entropy loss function.

\bibliography{references}
\bibliographystyle{unsrtnat}

\end{document}